\documentclass[usenatbib,useAMS]{mn2e}
\bibpunct{(}{)}{;}{a}{}{,}

\usepackage{enumerate}

\usepackage{aas_macros}

\usepackage{amsmath}
\usepackage{mathtools}
\usepackage{IEEEtrantools}

\usepackage{txfonts}
\usepackage[utf8]{inputenc}
\usepackage[english]{babel}
\usepackage{booktabs}
\usepackage{multirow}

\usepackage{savesym}
\savesymbol{iint}
\savesymbol{iiint}
\savesymbol{iiiint}
\savesymbol{idotsint}

\usepackage{amssymb,amsmath}

\usepackage{amstext}
\usepackage{subfigure}
\usepackage{graphicx}
\usepackage[np,autolanguage]{numprint}

\usepackage{bigstrut}

\usepackage[usenames,dvipsnames]{xcolor}
\usepackage{hyperref}
\usepackage{breakurl}

\hypersetup{
colorlinks=false,
citecolor = blue,
linkcolor = red
}

\usepackage{pdfsync}
\bibpunct{(}{)}{;}{a}{}{,} % to follow the A&A style

\usepackage{threeparttable}

\newcommand{\LargecolWidth}{0.28\hsize}

\newcommand{\colWidth}{0.35\hsize}
\newcommand{\colhspace}{0.02\hsize}

\def\mathbi#1{\textbf{\em #1}}

\renewcommand{\bmath}[1]{\mbox{ \boldmath $\!#1\!$ \unboldmath}}
\renewcommand{\mathbfss}[1]{\mathsf{\mathbf{#1}}}

\newcommand{\txn}[1]{\textnormal{#1}}

\newcommand{\taul}{\hbox{$\hat{\tau}_\lambda$}}
\newcommand{\tauLth}{\hbox{$\hat{\tau}_\lambda(\theta)$}}
\newcommand{\tauli}{\hbox{$\hat{\tau}_{\lambda,i}$}}
\newcommand{\tauVismTh}{\hbox{$\hat{\tau}^\txn{ISM}_V(\theta)$}}
\newcommand{\tauHismTh}{\hbox{$\hat{\tau}^\txn{ISM}_H(\theta)$}}
\newcommand{\tauvP}{\hbox{$\tau^\txn{mid}_{V\perp}$}}

\newcommand{\meantauvISM}{\hbox{$\langle\hat{\tau}^\txn{ISM}_V\rangle_\theta$}}
\newcommand{\meantaulISM}{\hbox{$\langle\hat{\tau}^\txn{ISM}_\lambda\rangle_\theta$}}

\newcommand{\Ha}{$\txn{H}\alpha$}
\newcommand{\Hb}{$\txn{H}\beta$}
\newcommand{\Hd}{$\txn{H}\delta_A$}

\newcommand{\logOH}{\hbox{$12+\log(\txn{O}/\txn{H})$}}

\newcommand{\Dn}{$\txn{D}_\txn{n}4000$}
\newcommand{\DnHd}{$\txn{D}_\txn{n}4000-\txn{H}\delta_A$}

\newcommand{\Tthin}{\hbox{$\txn{t}_\txn{thin}$}}
\newcommand{\Tthick}{\hbox{$\txn{t}_\txn{thick}$}}
\newcommand{\Tbulge}{\hbox{$\txn{t}_\txn{bulge}$}}
\newcommand{\lgTthin}{\hbox{$\log({\txn{t}_\txn{thin}/\txn{Gyr}})$}}
\newcommand{\lgTthick}{\hbox{$\log({\txn{t}_\txn{thick}/\txn{Gyr}})$}}
\newcommand{\lgTbulge}{\hbox{$\log({\txn{t}_\txn{bulge}/\txn{Gyr}})$}}

\newcommand{\taubP}{\hbox{$\tau_{B\perp}$}}
\newcommand{\tauLbc}{\hbox{$\hat{\tau}^\txn{BC}_\lambda$}}
\newcommand{\tauVbc}{\hbox{$\hat{\tau}^\txn{BC}_V$}}
\newcommand{\tauVism}{\hbox{$\hat{\tau}^\txn{ISM}_V$}}
\newcommand{\tauLismTh}{\hbox{$\hat{\tau}^\txn{ISM}_\lambda(\theta)$}}
\newcommand{\tauLismTt}{\hbox{$\hat{\tau}^\txn{ISM}_\lambda(\theta,t)$}}
\newcommand{\tauLthinth}{\hbox{$\hat{\tau}^\txn{thin}_\lambda(\theta)$}}
\newcommand{\tauLthickth}{\hbox{$\hat{\tau}^\txn{thick}_\lambda(\theta)$}}
\newcommand{\tauLbulgeth}{\hbox{$\hat{\tau}^\txn{bulge}_\lambda(\theta)$}}
\newcommand{\nVism}{\hbox{$n^\txn{ISM}_V$}}
\newcommand{\nVismTh}{\hbox{$n^\txn{ISM}_V(\theta)$}}
\newcommand{\nHismTh}{\hbox{$n^\txn{ISM}_H(\theta)$}}
\newcommand{\nLism}{\hbox{$n^\txn{ISM}_\lambda$}}
\newcommand{\nLismTh}{\hbox{$n^\txn{ISM}_\lambda(\theta)$}}

\newcommand{\xiVthin}{\hbox{$\xi^\txn{thin}_V$}}
\newcommand{\xiVthick}{\hbox{$\xi^\txn{thick}_V$}}
\newcommand{\xiVbulge}{\hbox{$\xi^\txn{bulge}_V$}}

\newcommand{\fSFH}{$f_{\txn{SFH}}$}
\newcommand{\psiS}{\hbox{$\psi_{\txn{S}}$}}

\newcommand{\muS}{\hbox{{$\mu_{\ast}$}}}
\newcommand{\Mstar}{\hbox{{$\txn{M}_{\ast}$}}}
\newcommand{\Msun}{\hbox{$\txn{M}_{\sun}$}}

\newcommand{\Mfib}{\hbox{$\txn{M}^\txn{fib}_\ast$}}

\newcommand{\MdMs}{\hbox{$\txn{M}_\txn{dust} / \txn{M}_{\ast}$}}

\newcommand{\HaHb}{\hbox{$\txn{H}\alpha/\txn{H}\beta$}}

\newcommand{\Nii}{[\mbox{N\,{\sc ii}}]}
\newcommand{\Oiii}{[\mbox{O\,{\sc iii}}]}

\renewcommand{\bar}{\overline}
\renewcommand{\deg}{^\circ}

\newcommand{\MultiNest}{\textsc{MultiNest}}
\newcommand{\CosmoMC}{\textsc{CosmoMC}}
\newcommand{\LibAGF}{\textsc{LibAGF}}

\newcommand{\sn}{\hbox{\textrm{S/N}}}

\title[Content and spatial distribution of dust in galaxies]{Insights into the content and spatial distribution of dust from the integrated spectral properties of galaxies}
\author[J. Chevallard, S. Charlot, B. Wandelt,
V. Wild]{J. Chevallard$^{1}$\thanks{e-mail:chevalla@iap.fr},
  S. Charlot$^{1}$, B. Wandelt$^{1}$, V. Wild$^{2,3}$\\
$^{1}$UPMC-CNRS, UMR7095, Institut d'Astrophysique de Paris, F-75014, Paris, France \\
$^{2}$School of Physics and Astronomy, University of St Andrews, North Haugh, St Andrews, KY16 9SS, U.K. \\
$^{3}$Institute for Astronomy, University of Edinburgh, SUPA, Royal Observatory, Blackford Hill, Edinburgh, EH9 3HJ, U.K.
}

\begin{document}

\date{Accepted for publication on MNRAS on 22/03/2013}

\maketitle

\label{firstpage}

% Abstract
\begin{abstract}
We present a new approach to investigate the content and spatial distribution of dust in structurally unresolved star-forming galaxies from the observed dependence of integrated spectral properties on galaxy inclination. Motivated by the observation that different stellar populations reside in different spatial components of nearby star-forming galaxies, we develop an innovative combination of generic models of radiative transfer in dusty media with a prescription for the spectral evolution of galaxies, via the association of different geometric components of galaxies with stars in different age ranges. We start by showing that a wide range of radiative transfer models all predict a quasi-universal relation between slope of the attenuation curve at any wavelength, from the ultraviolet to the near-infrared, and $V$-band attenuation optical depth in the diffuse interstellar medium (ISM), at all galaxy inclinations. This relation predicts steeper (shallower) dust attenuation curves than both the Calzetti and Milky-Way curves at small (large) attenuation optical depths, which implies that geometry and orientation effects have a stronger influence on the shape of the attenuation curve than changes in the optical properties of dust grains. We use our new, combined radiative transfer and spectral evolution model to interpret the observed dependence of the \HaHb\ ratio and $ugrizYJH$ attenuation curve on inclination in a sample of about 23 000 nearby star-forming galaxies, which we correct for systematic biases by developing a general method based on importance sampling. From the exploration of the model parameter space by means of a Bayesian Markov Chain Monte Carlo technique,  we measure the central face-on $B$-band optical depth of this sample to be $\taubP\approx1.8\pm0.2$ (corresponding to an angle-average $\meantauvISM \approx 0.3$). We also quantify the enhanced optical depth towards newly formed stars in their birth clouds, finding this to be significantly larger in galaxies with bulges than in disc-dominated galaxies, while \taubP\ is roughly similar in both cases. This can arise if, for example, galaxies with significant bulges have higher central star formation efficiencies than their disc-dominated counterparts at fixed specific star formation rate, and dustier stellar birth clouds because of the higher metallicity. We find that over 80 percent of the attenuation in galaxies in our sample is characteristic of that affecting thin-disc stars in radiative transfer models. The median unattenuated $V$-band luminosity ratio of thick-disc to thin-disc stars is 0.1--0.2, in good agreement with the results from spatially resolved studies of nearby edge-on disc galaxies. Finally, we show that neglecting the effect of geometry and orientation on attenuation can severely bias the interpretation of galaxy spectral energy distributions, as the impact on broadband colours can reach up to 0.3--0.4\,mag at optical wavelengths and 0.1\,mag at near-infrared ones. This paper also contains an original application of Gaussian Random Processes to extend the wavelength range of dust attenuation curves.

\end{abstract}

%Keywords

% Inizio testo

\section{Introduction}

A small but significant mass fraction of the interstellar medium (ISM) in galaxies is in the form of dust grains, which absorb and scatter the light emitted by stars at all wavelengths \citep[e.g.][]{Spitzer1978}. Dust grains are produced during late phases of stellar evolution -- mainly supernova explosions and winds of asymptotic-giant-branch (AGB) stars \citep[e.g.][]{Dwek1998,Hofner2009,Cherchneff2010} -- and are destroyed by energetic photons and shocks \citep[e.g.][]{Jones2004, Tielens2005,Jones2011}. This makes the amount, composition and spatial distribution of dust in a galaxy depend in a complex way on the star formation history and geometry, and hence, the history of hierarchical merging, gas infall and outflow. All these factors influence the wavelength dependence of the attenuation of starlight by dust, which we must understand to retrieve star formation and chemical enrichment histories from observed spectral energy distributions of galaxies.

Much information has been gathered on the shape of the dust attenuation curve in galaxies. In the Milky Way, the Large and the Small Magellanic Clouds (LMC and SMC), comparisons of observed spectral energy distributions of stars of fixed spectral type and luminosity class along different lines of sight have provided estimates of the extinction\footnote{We follow the standard nomenclature that `extinction' refers to the absorption of photons along, and their scattering out of, a line of sight, while `attenuation' (or `effective absorption') refers to the combined effects of absorption and scattering in and out of the line of sight caused by both local and global geometric effects \citep[e.g.][]{Calzetti2001}.} curve along individual lines of sight \citep[e.g.][]{Savage1979,Prevot1984,Bouchet1985,Clayton1985,Cardelli1988,Cardelli1989}. By extending this approach to the analysis of the integrated light in a sample of nearby starburst galaxies, for which the unattenuated spectral energy distribution is known to be dominated by blue massive stars,  \citet[see also \citealt{Calzetti2001}]{Calzetti1994} derived constraints on the mean attenuation curve in these galaxies. This is significantly greyer than the Milky Way, LMC and SMC extinction curves at ultraviolet wavelengths. Theoretically, the greyness of the \citet{Calzetti2001} attenuation curve can be accounted for by a simple angle-averaged, two-component dust model, in which young stars in giant molecular clouds suffer more attenuation than older stars in the diffuse ISM, for standard (i.e. LMC-like) optical properties of dust grains \citep{CF00}. Models of this type are useful because they can be implemented easily in any spectral synthesis analysis of galaxies \citep[e.g.][]{Brinchmann2004,daCunha2008,Pacifici2012}. Recently, however, the identification of systematic changes of the dust attenuation curve as a function of both inclination and star formation history in nearby galaxies has shown the limitation of angle-averaged dust models \citep{Wild2011b}. Investigating the physical origin of these changes requires more sophisticated models to compute the transfer of photons through the dusty ISM of spatially resolved galaxies \citep[e.g.][]{Silva1998,Tuffs2004,Pierini2004,Jonsson2010}. In practice, such models are computationally expensive and cannot be easily applied to the detailed analysis of large samples of galaxies.

In this paper, we present a new approach to account in a simple yet physically consistent way for the dependence of dust attenuation on orientation in spectral analyses of structurally unresolved galaxies. This approach is based on an original exploitation of existing sophisticated models of radiative transfer in dusty media. We consider four types of models relying on Monte-Carlo \citep[hereafter P04]{Pierini2004}, analytic (\citealt[hereafter T04]{Tuffs2004}; \citealt[hereafter S98]{Silva1998}), smoothed particle hydrodynamics \citep[SPH,][hereafter J10]{Jonsson2010} calculations of the scattering and absorption of photons through the diffuse ISM of galaxies with multiple-disc and bulge stellar components. We show that these different types of models all predict a quasi-universal relation between slope of the attenuation curve at any wavelength from the ultraviolet to the near infrared and $V$-band attenuation optical depth of the dust in the diffuse ISM, at all galaxy inclinations. We further show that appropriate choices of the dust optical depths and relative luminosities of the various disc and bulge components in the flexible T04 model can reproduce the dependence of attenuation on galaxy inclination predicted by the P04 and J10 models. On these stable grounds, we combine the T04 model with a semi-analytic post-treatment of the Millennium cosmological simulation \citep{Springel2005,DeLucia2007}, the latest version of the \citet{BC03} stellar population synthesis code and a prescription for dust attenuation in stellar birth clouds \citep{CF00}, to investigate the physical origin of the observed changes of the dust attenuation curve in the \citet{Wild2011b} sample of nearby galaxies. We achieve this by relating the geometric components of galaxies to stellar age ranges in spatially unresolved models of spectral evolution. In practice, we build a comprehensive library of model spectral energy distributions encompassing wide ranges of stellar and dust parameters for galaxies with multiple-disc and bulge components. Then, we use this library to constrain the dust content and stellar age ranges in the different geometric components of the galaxies in the \citet{Wild2011b} sample by performing a Bayesian analysis of the observed \HaHb\ ratio and broadband (optical and near-infrared) spectral energy distribution. The results of this analysis provide a calibration of the dependence of attenuation on both orientation and star formation history in structurally unresolved galaxies. We show that accounting for this effect has important implications for the interpretation of galaxy spectral energy distributions.

In Section 2 below, we briefly recall the formalism to describe the attenuation of starlight by dust in galaxies and compare the radiative transfer models of \citet{Tuffs2004}, \citet{Pierini2004}, \citet{Jonsson2010} and \citet{Silva1998}. We show that the predictions of these different models are consistent with one another and yield a quasi-universal relation between slope of the attenuation curve at any wavelength and attenuation optical depth of the dust. In Section 3, we combine the T04 model with models of the spectral evolution of galaxies in a cosmological context to build a comprehensive library of synthetic spectral energy distributions of dusty galaxies. We use this library to interpret the observed line and continuum emission of the \citet{Wild2011b} sample of nearby star-forming galaxies in terms of constraints on the content and distribution of the dust between the different stellar components. We discuss the implications of these results for the analysis of structurally unresolved galaxies in Section 4, where we also summarise our conclusions.

\section{General properties of dust attenuation models}\label{sec:models}

The intensity $I_{\lambda} (\theta)$ emerging at wavelength $\lambda$  in a direction $\theta$ from the normal to the equatorial plane of a galaxy can always be related to the intensity $I^{\,0}_{\lambda}$  produced by stars, assumed isotropic, by an expression of the form
\begin{equation} \label{eq:lumin_atten}
I_{\lambda} (\theta)= I^{\,0}_{\lambda} \exp\left [ -\taul(\theta) \right ]\,,
\end{equation}
where \tauLth\ is the `attenuation' (or `effective absorption') optical depth of the dust affecting those photons emitted in all directions by all stars in the galaxy that emerge in the direction $\theta$. Expression~\ref{eq:lumin_atten} implicitly assumes azimuthal symmetry. In practice, the quantity  \tauLth\ depends on the geometry of the galaxy and the optical properties and spatial distribution of dust relative to the stars. This can make the intensity $I_{\lambda} (\theta)$ either smaller or larger than $I^{\,0}_{\lambda}$, depending on the influence of scattering on the paths of photons before these escape from the galaxy.

In practice, the attenuation optical depth \tauLth\ incorporates contributions from the different geometric components of the galaxy (i.e. bulge, thin and thick discs), each influencing in a different way the dependence of \taul\ on $\theta$. We write
\begin{equation}\label{eq:dust_cmpn}
\hat{\tau}_{\lambda} (\theta)  = - \ln{ \displaystyle\sum_{i} \xi_{\lambda}^i \ \exp \left [-\hat{\tau}_{\lambda,i} (\theta) \right ]  } \,,
\end{equation}
where $\tauli(\theta)$ is the contribution by the $i^{\mathrm th}$ component to the absorption optical depth of the dust affecting photons emerging at wavelength $\lambda$ in the direction $\theta$, and 
\begin{equation}\label{eq:xi}
\xi_{\lambda}^{\,i}   = I^{\,0,i}_{\lambda}\,\big/\,I^{\,0}_{\lambda}
\end{equation}
is the contribution by this component to the intensity produced by stars at wavelength $\lambda$ (with $\sum_i{I^{\,0,i}_{\lambda}}=I^{\,0}_{\lambda}$). 
The intensity at wavelength $\lambda$ can be expressed as
\begin{equation}\label{eq:intensity}
I^{\,0}_{\lambda}(t) = \int_0^t dt' \psi(t-t')S_{\lambda}[t',Z(t-t')]   \,,
\end{equation}
where $\psi(t-t')$ is the star formation rate at time $t-t'$ and $S_{\lambda}[t',Z(t-t')]$ is the intensity produced per unit wavelength and per unit mass by a stellar population of age $t'$ and metallicity $Z(t-t')$.

For some purposes, it is useful to compute the mean of \tauLth\ over all solid angles. In the case of azimuthal symmetry ($d\Omega = 2\pi\: d\txn{cos}\,{\theta}$), and if we assume that the radiation is symmetric with respect to the equatorial plane of the galaxy, this is given by
\begin{equation}\label{eq:ang_aver}
\langle \taul \rangle_{\theta}   = \int_0^{\pi/2} d\txn{cos}\,{\theta} \;\tauLth\,.
\end{equation}

\subsection{Radiative Transfer models}\label{sec:RT}

\begin{figure}
	\centering
	\resizebox{\hsize}{!}{\includegraphics{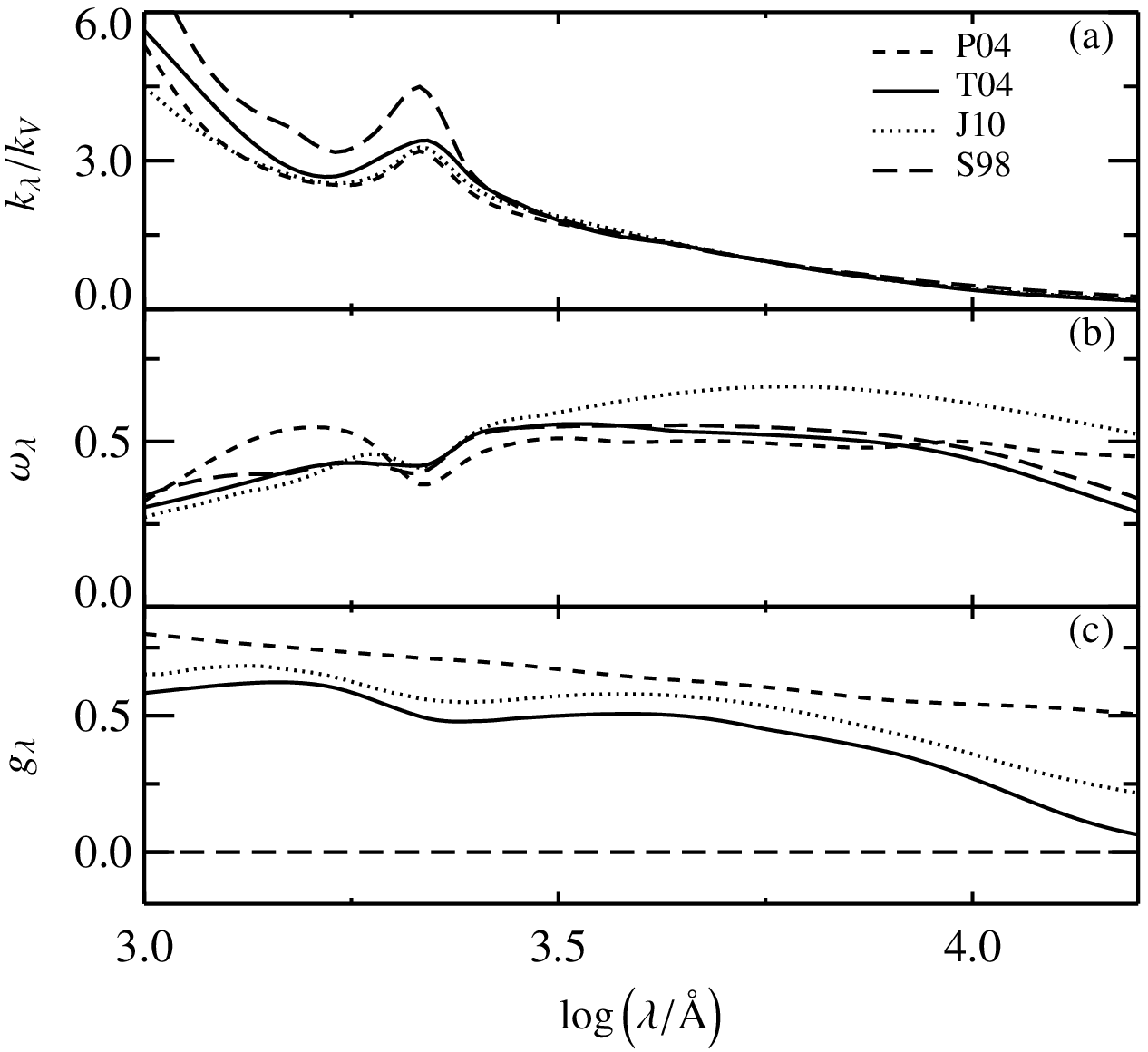}}
	\caption{Dust-absorption coefficient in units of the $V$-band value, $\kappa_\lambda/\kappa_V$, asymmetry parameter of the scattering phase function, $g_\lambda$, and single-scattering albedo, $\omega_\lambda$, plotted against wavelength $\lambda$, as adopted for Milky-Way type dust in the P04 (short-dashed line; from \citealt{Witt2000}), T04 (solid line; from \citealt{Laor1993,Mathis1977}), J10 (dotted line; from \citealt{Weing2001,Draine2007}) and S98 (long-dashed line; from \citealt{Draine1984, Laor1993}, with modifications by \citealt{Silva1998}) models.}
	\label{fig:ext_curves}
\end{figure}

The attenuation optical depth \tauLth\ can be computed analytically for simple geometries, such as a uniform screen of dust in front of the stars and a uniform mixture of stars and dust  \citep[e.g.][]{Yip2010}. The study of more complex geometries requires a numerical approach to follow the interactions between photons and dust grains in the interstellar medium. We focus here on four recent, popular models of radiative transfer in dusty media:  \citet{Tuffs2004}, \citet{Pierini2004}, \citet{Jonsson2010} and \citet{Silva1998}. Fig.~\ref{fig:ext_curves} shows that the dependence on wavelength of the dust-absorption coefficient, single-scattering albedo and asymmetry parameter of the scattering phase function adopted in these different models for Milky-Way type dust are roughly consistent with one another, except for the fact that S98 assume isotropic scattering at all wavelengths (i.e. $g_\lambda=0$). The small apparent discrepancies between the dust optical properties of the P04, T04 and J10 models have a negligible impact on our analysis. In fact, we show in Sections~\ref{sec:S_A_rel} and \ref{sec:tau_theta} below that the attenuation curves obtained from these radiative transfer models have similar global properties at fixed geometry and inclination, while the adoption of isotropic scattering in the S98 model introduces significant differences in the shape of the attenuation curve at low attenuation optical depths. In the next paragraphs, we briefly review the T04, P04, J10 and S98 models and establish their common behaviours and mutual coherence. We focus here on the predictions of these models for attenuation by dust in the ambient (i.e. diffuse) ISM, which we refer to as \tauLismTh. The characteristics of these models are summarised in Table~\ref{tab:RT_models}.

\begin{table}
	\caption{Summary of the parameters defining the geometric configuration of the different radiative transfer models described in Section~\ref{sec:RT}.}
\begin{threeparttable}
%\caption{Summary of the parameters determining the geometric configuration of the different radiative transfer models used in this work.}
	\centering
	\begin{tabular}{ l l c c c c }

\toprule

{\bf Model}	           & {\bf Type}           & \multicolumn{2}{c}{scale length (kpc )}    & \multicolumn{2}{c}{scale height (kpc )}     \\

\cmidrule{3-4} \cmidrule{5-6} 

                                              && stars ($h_\ast$) & dust ($h_d$) &  stars ($z_\ast$) & dust ($z_d$)  \\
\midrule

\multirow{2}*{P04}	& Bulge 	& 1.0 	&             &                   &             \\ 
                                & Disc	& 3.0 	& 3.0    &  0.06\,--\,0.37 & 0.11 \\ 

\midrule

\multirow{3}*{T04\tnote{a}}	& Bulge	& 0.69	&             &                                   \\ 
                          & Thin Disc	& 3.0 	& 3.0   &  0.048    & 0.048 \\ 
                          & Thick Disc	& 4.2	& 4.2   &  0.22    & 0.14 \\ 

\midrule

\multirow{2}*{J10\tnote{b}}	& G-series	& 1.1\,--\,2.8 	& $3\ h_\ast$    &  0.125\,--\,$0.2\ h_\ast$                                   \\ 
                                & Sbc-series	& 4.0\,--\,7.0 	& $3\ h_\ast$    &  0.125\,--\,$0.2\ h_\ast$                          \\

\midrule

\multirow{2}*{S98}	& S-series	& 5 	& $ h_\ast$    &  0.5        & $z_\ast$                                   \\ 
                                & N-series	& 5\,--\,8 	& $ h_\ast$    &  0.4\,--\,0.1     & $z_\ast$                      \\

\bottomrule

	\end{tabular}
	\begin{tablenotes}
\item [a] T04 express all parameters in units of the thin-disc scale length, while P04 and J10 adopt physical units. For the sake of comparison, we re-normalise here the T04 parameters to the scale length of the P04 disc.
	\item [b] J10 do not provide any measure of the dust disc scale height, which is very sensitive to the resolution of the adopted SPH simulation for each galaxy.
\end{tablenotes}
\end{threeparttable}
	\label{tab:RT_models}
\end{table}

\subsubsection{The P04 model \citep{Pierini2004}}\label{sec:P04}

\begin{figure}
	\centering
	\resizebox{\hsize}{!}{\includegraphics{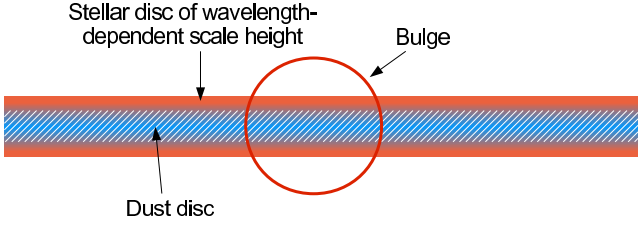}}
	\caption{Geometric components of the radiative transfer model of P04: a stellar disc with a double exponentially declining light profile and wavelength-dependent scale height (blue-to-red rectangle); a stellar bulge with an exponentially declining radial light profile (red circle);  both penetrated by a dust disc with a double exponentially declining density profile (white diagonal hatching). See Section~\ref{sec:P04} and Table~\ref{tab:RT_models} for detail.}
	\label{fig:Pierini}
\end{figure}

The P04 model relies on the DIRTY radiative transfer code of \citet{Gordon2001}. This follows the multiple scattering and absorption of photons through an analytic distribution of dust and stars by means of a Monte Carlo approach. P04 use DIRTY to compute the attenuation  optical depth \tauLismTh\ for galaxies consisting of a stellar bulge and disc pervaded by a dust disc, as depicted in Fig.~\ref{fig:Pierini}.

The `stellar bulge' is modelled as a sphere with an exponentially declining radial light profile of scale length 1\,kpc truncated at 4\,kpc. The `stellar disc' has a double exponentially declining light profile of the form 
\begin{equation}\label{eq:expon}
\rho(r,|Z|) \propto \exp\left ({ -\frac{r}{h} -\frac{|Z|}{z} } \right )\,,
\end{equation}
where $r$ and $|Z|$ indicate the radial and polar coordinates and $h$ and $z$ the scale length and scale height of the disc, respectively. P04 adopt a fixed scale length for the stellar disc, $h_\ast=3$ kpc. To account for the fact that young (blue) stars tend to have a  lower scale height than older (redder) stars in observed spiral galaxies \citep[e.g.][]{Yoachim2008A,Yoachim2012}, P04 take the scale height of the stellar disc to increase with wavelength, from $z_\ast=60$\,pc at $\lambda=0.1\:\micron$ to 370\:pc at 3\:\micron.
The `dust disc' also has a double exponentially declining density profile, with the same scale length as the stellar disc ($h_d=3$\,kpc) and a fixed scale height $z_d=110$\,pc, which equals that of the stellar disc at wavelength $\lambda=1850$\:\AA. The stellar and dust discs both extend to a maximum radius of 12\,kpc and maximum height of 2.1\:kpc. P04 compute \tauLismTh\ for two types of dust distribution: homogeneous and clumpy.\footnote{They describe the clumpy medium by associating in a stochastic way each resolution element (corresponding to a cubic cell of $L=44$\:pc on the side) to either a high or a low density state. The high-to-low density contrast is set to 100 and the filling factor of the high density clumps to 0.15.}  We consider only the clumpy distribution here because it better mimics the patchy and filamentary structure of the ISM \citep[see, e.g. the recent observations with the {\it Herschel}/PACS camera in][]{Mookerjea2011, Arzoumanian2011}.  

P04 conveniently parametrize their results in terms of the central face-on $V$-band optical depth of the dust from the equatorial plane to the surface of a galaxy, which we denote by \tauvP. They consider values of this parameter in the range $0.25 \leq \tauvP \leq 8.00$ and compute \tauLismTh\ at angles in the range $0\,^{\circ} \leq \theta \leq 90\,^{\circ}$ and wavelengths in the range $0.1\leq \lambda \leq 3.0$\:\micron. We note that, even for $\tauvP=8.0$, the average over angles \meantauvISM\ amounts to only 1.6 because of the exponentially declining density profile of the dust disc.

\subsubsection{The T04 model \citep{Tuffs2004}}\label{sec:T04}

The T04 model relies on the radiative transfer code of \citet{Kylafis1987}. As in the case of the P04 models, this code includes multiple anisotropic scattering and can be applied to analytic distributions of stars and dust. Unlike the Monte Carlo approach of P04, the code adopted by T04 is based on an approximate solution to the radiative transfer equation, which can be easily computed \citep[for detail, see][]{Kylafis1987}.

\begin{figure}
	\centering
	\resizebox{\hsize}{!}{\includegraphics{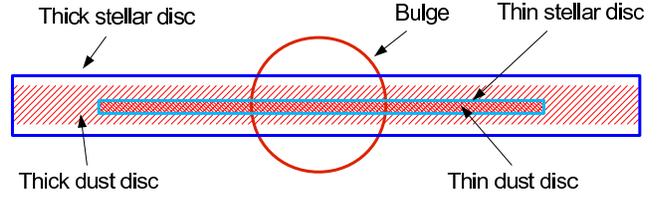}}
	\caption{Geometric components of the radiative transfer model of T04: a `thin' and a `thick' stellar disc with a double exponentially declining light profile (cyan and blue rectangles, respectively); a stellar bulge with a de Vaucouleurs $\exp(-r^{1/4})$ declining radial light profile (red circle); each penetrated by a `thin' and a `thick' dust discs with a double exponentially declining density profile (red diagonal hatching). The aspect ratios of the different components reproduce those adopted in the model. See Section~\ref{sec:T04} and Table~\ref{tab:RT_models} for detail.}
	\label{fig:Tuffs}
\end{figure} 

T04 describe galaxies as the sum of three stellar components (a thin and a thick disc, and a bulge) and two dust components (associated with the stellar discs), as illustrated by Fig.~\ref{fig:Tuffs}. The two stellar discs, along with the associated dust discs, are described as double exponentially declining light and density profiles, respectively (equation~\ref{eq:expon}). T04 normalise all the scale parameters of a galaxy to the $B$-band scale length of the thin stellar disc, which is by construction $h_\ast^\txn{thin}=1.0$. The `thin stellar disc' has a scale height $z_\ast^\txn{thin} = 0.016$ and is pervaded by a `thin dust disc' of same scale height $z_\ast^\txn{thin}$ and scale length $h_\ast^\txn{thin}$.   The `thick stellar disc' has larger scale length and scale height than the thin disc ($h_\ast^\txn{thick} = 1.406$, $z_\ast^\txn{thick} = 0.074$ $\approx 4.6 \, z_\ast^\txn{thin}$) and it is penetrated by a dust disc with the same scale length and smaller scale height than the thick stellar disc ($z_d^\txn{thick}\approx 0.65 \, z_\ast^\txn{thick}$).  The `stellar bulge' has a de Vaucouleurs $\exp(r^{1/4})$ light profile, at variance with the exponentially declining light profile in the P04 model. We stress that each stellar component in the T04 model (bulge, thick and thin disc) is attenuated by both the thin and the thick dust discs, but with different implications for \tauLismTh\ because of the different relative distributions of stars and dust. 

T04 parametrize the dust content of a galaxy by means of the central face-on $B$-band optical depth, which includes the contributions from the two dust discs, $\taubP=\tau^{\txn{thick}}_{B\perp} + \tau^{\txn{thin}}_{B\perp}$. They fix $\tau^{\txn{thick}}_{B\perp} / \tau^{\txn{thin}}_{B\perp} = 0.387$ and explore the range $0.1 \leq \taubP \leq 8.0$. T04 compute \tauLismTh\ at angles in the range $0\,^{\circ} \leq \theta \leq 90\,^{\circ}$ and wavelengths in the range $0.09 \leq \lambda \leq 2.2 \: \micron$ for the thin disc, but only in the range $0.45\leq \lambda \leq 2.2 \:\micron$ for the thick disc and bulge. While stars in the thick disc and bulge are not expected to dominate the radiation at wavelengths  $0.09 \leq \lambda \leq 0.45\:\micron$, the lack of information on \tauLismTh\ in this range limits the exploration of the attenuation of potentially interesting populations of evolved stars in these two components. To overcome at least partially this limitation, we can try to guess the attenuation at $\lambda \leq 0.45\:\micron$ based on that at $\lambda \geq 0.45\:\micron$ for the thick stellar disc and bulge, using the information available at all wavelengths for the thin stellar disc.\footnote{Unfortunately, it was not possible for the authors of T04 to compute \tauLismTh\  in the missing wavelength domain for our purpose.} This is possible because, as noted above, the same dust discs determine the attenuation of all three stellar components.

To best-guess the shape of the attenuation curve \tauLismTh\ at wavelengths $\lambda \leq 0.45\:\micron$ from that at $\lambda \geq 0.45\:\micron$ for the thick stellar disc and bulge, we do not choose extrapolation methods invoking parametric functions (e.g. polynomials, power laws). Instead, we use a `Gaussian random process'  to learn the shape of the attenuation curves at $\lambda \leq 0.45\:\micron$ from that at $\lambda \geq 0.45\:\micron$ based on the calculations available for the thin-disc attenuation (see Appendix \ref{app:GRP} for detail). This approach is valid under two main assumptions: firstly, that the attenuation curves red-ward and blue-ward of 0.45\:\micron\ be strongly enough correlated that the knowledge of one allows the prediction of the other; the finding by \citet{Cardelli1988} that the ultraviolet extinction law correlates well with the optical/near-infrared one for stars in various Galactic environments supports this hypothesis. Secondly, that our learning sample (the thin-disc  attenuation curves for various $\taubP$ and $\theta$) well cover the parameter space sampled by the incomplete attenuation curves (of the thick disc and bulge). In our case, the thin-disc calculations of T04 do encompass the full range of optical depths sampled by the thick-disc and bulge calculations. In Appendix \ref{app:GRP}, we describe and thoroughly test our algorithm to extend the thick-disc and bulge attenuation curves computed by T04. From the analysis of different datasets, we show that \tauLismTh\  at wavelengths $0.25 \la \lambda \leq 0.45\:\micron$ can be retrieved reliably from the attenuation curve at $\lambda \geq 0.45\:\micron$, but that the recovery worsens at wavelengths $\lambda < 0.25\:\micron$.

We note that T04 account in a separate way for the attenuation of newly born stars in stellar birth clouds by introducing an analytic component of clumpy dust, which absorbs a fraction of the ultraviolet radiation. This component is not shown in Fig.~\ref{fig:Tuffs} and it is not used as part of the T04 model in this work, since we independently model the attenuation of newly born stars following the prescription of \citet[see Section~\ref{sec:dust} below]{CF00}.

\subsubsection{The J10 model \citep{Jonsson2010}}\label{sec:J10}

The J10 model relies on the SUNRISE radiative transfer code of \citet{SUN2006}. J10 appeal to a set of smoothed particle hydrodynamics simulations to describe the input distributions of stars and gas in SUNRISE, which differ from the analytic profiles used by P04 and T04. Also, J10 adopt a Monte Carlo method to solve the equation of radiative transfer, as P04, but considering polychromatic rays instead of monochromatic ones.\footnote{This is achieved by expressing the probability for photons of any wavelength to be absorbed or scattered by dust along a ray in terms of that of a photon of reference wavelength $0.9\:\micron$ (see J10 for detail).} With this approach, single rays can be used to trace the paths of photons of different wavelengths, which enables high spectral resolution across the whole range from ultraviolet to infrared wavelengths \citep[for detail and a discussion of drawbacks, see][]{SUN2006}. 

The SPH simulations in the J10 model were performed with the GADGET code of \citet[\citealt{GADGET2}]{Springel2001} and include supernova feedback and metal enrichment \citep{Jonsson2006, Cox2006, Cox2008}. J10 compute the attenuation curve \tauLismTh\ for two sets of SPH simulations: the `G' series, with stellar and interstellar properties typical of nearby galaxies with stellar masses in the range $1\times10^{9} \leq \Mstar/M_{\odot} \leq 6  \times 10^{10}$ in the Sloan Digital Sky Survey \citep[SDSS,][]{SDSS}; and the `Sbc' series, with properties typical of local late-type spiral galaxies with masses in the range  $5 \times 10^{10} \leq \Mstar/M_{\odot} \leq 1\times10^{11}$ \citep[see figs~6 and 7 of][]{Jonsson2010}. All simulations include the self-consistent treatment of the evolution of dark matter, stellar and gas particles. Attenuation by dust is added a posteriori, assuming that a constant fraction of 40 percent of the mass of metals in the gas is in form of dust grains. In the original simulations of J10, the attenuation of newly born stars in their birth clouds is computed using the photo-ionisation code MAPPINGSIII \citep{Groves2004,Groves2008}. In the present study, for the purpose of comparison with the P04 and T04 models, we use a set of  J10 simulations which do not include the attenuation of young stars in stellar birth clouds (we are grateful to B.~Groves and P.~Jonsson for kindly making these calculations available to us).

A main advantage of the J10 model over the T04 and P04 ones is to rely on a physically motivated spatial distribution of stars and dust. The drawback is that this approach is computationally expensive: the simulations of only 7 galaxies published to date do not allow proper statistical analyses. Moreover, J10 show that the results of their calculations depend on the resolution of the SPH simulations used as input in SUNRISE. An order-of-magnitude increase in this resolution makes the face-on ultraviolet attenuation smaller and the edge-on near-infrared one larger (both by about 10 percent) for their model Sbc galaxy. J10 conclude that high-resolution simulations are required to better explore the origin of these effects.

\subsubsection{The S98 model \citep{Silva1998}}\label{sec:S98}

The S98 model relies on the GRASIL radiative transfer code \citep{Silva1998, Bressan2002, Panuzzo2003}. This solves the radiative transfer equation in any location of model galaxies described by analytic distributions of dust and stars, using the iterative algorithm of \citet{Granato1994}. S98 compute in a consistent way the properties of stellar populations and the masses of interstellar gas and dust using an analytic model of galaxy evolution, which includes infall of pristine gas and the enrichment of the ISM by successive generations of stars. They assume that galaxies are composed of a stellar bulge and disc, penetrated by a dust disc. For the bulge, they adopt a King profile of the form $\rho(r) \propto [1 + (r/h_\ast)^2]^{-3/2}$ truncated at $\log(r/h_\ast) = 2.2$, which differs from the exponentially declining and de Vaucouleurs profiles adopted by P04 and T04, respectively.

As P04 and T04, S98 adopt double exponentially declining light and density profiles to describe the stellar and dust discs, respectively (equation~\ref{eq:expon}), assuming that both discs have equal scale-lengths and scale-heights (Table~\ref{tab:RT_models}). We note that S98 also include a component of dust in stellar birth clouds, which attenuates the radiation from newly born stars. As in the case of the T04 and J10 models above, we ignore this component here and consider only the attenuation of starlight by dust in the diffuse ISM. 

We consider the attenuation curves \tauLismTh\ computed by S98 for five model galaxies: an Sb and an Sc galaxy (the S-series) available from the GRASIL website\footnote{\url{http://adlibitum.oat.ts.astro.it/silva/grasil/modlib/modlib.html}.}, and three models obtained from the spectral fits of late-type spiral galaxies (M\,100, M\,51 and NGC\,6946; the N-series) presented in \citet{Silva1998}. All models were evolved for  13 Gyr. The Sb and Sc galaxies differ by their timescales for gas infall and efficiencies of star formation, which make the gas reservoir be depleted on a longer timescale in the Sc model than in the Sb one. The N-series galaxies are described as bulge-less, with roughly similar timescales for gas infall and efficiencies of star formation. 

\subsection{Relation between slope of the attenuation curve and dust attenuation optical depth}\label{sec:S_A_rel}

We have described the different prescriptions of P04, T04, J10 and S98 to calculate the attenuation optical depth \tauLismTh\ of dust in the ambient ISM at an angle  $\theta$ between the line of sight and the normal to the equatorial plane of a galaxy. We now show that these prescriptions are consisted with one another and yield a quasi-universal relation between slope of the attenuation curve at any wavelength and $V$-band attenuation optical depth of the dust. To achieve this, we fit the results of the four models by expressing the dependence of \tauLismTh\ on $\lambda$ at fixed $\theta$ as  
\begin{equation}\label{eq:pow_law}
\tauLismTh = \tauVismTh \left ( \frac{\lambda}{0.55\,\micron}\right )^{-n^\mathrm{ISM}_\lambda(\theta)} \, ,
\end{equation} 
where the exponent of the power law is a linear function of wavelength,
\begin{equation}\label{eq:n_fit}
\nLismTh= A(\theta) + B(\theta)\,(\lambda/\micron -0.55) \,.
\end{equation} 
This parametrisation provides excellent fits to the radiative transfer calculations of P04, T04, J10 and S98. In practice, we select the best-fit $A(\theta)$ and $B(\theta)$ corresponding to every combination of input parameters of each of the P04, T04, J10 and S98 models. We focus below on the slope of the attenuation curve estimated in this way at $\lambda=0.55$ and $1.60\,\mu$m, which we refer to as \nVismTh\ and \nHismTh.
\begin{figure*}
	\centering
	\resizebox{\hsize}{!}{\includegraphics{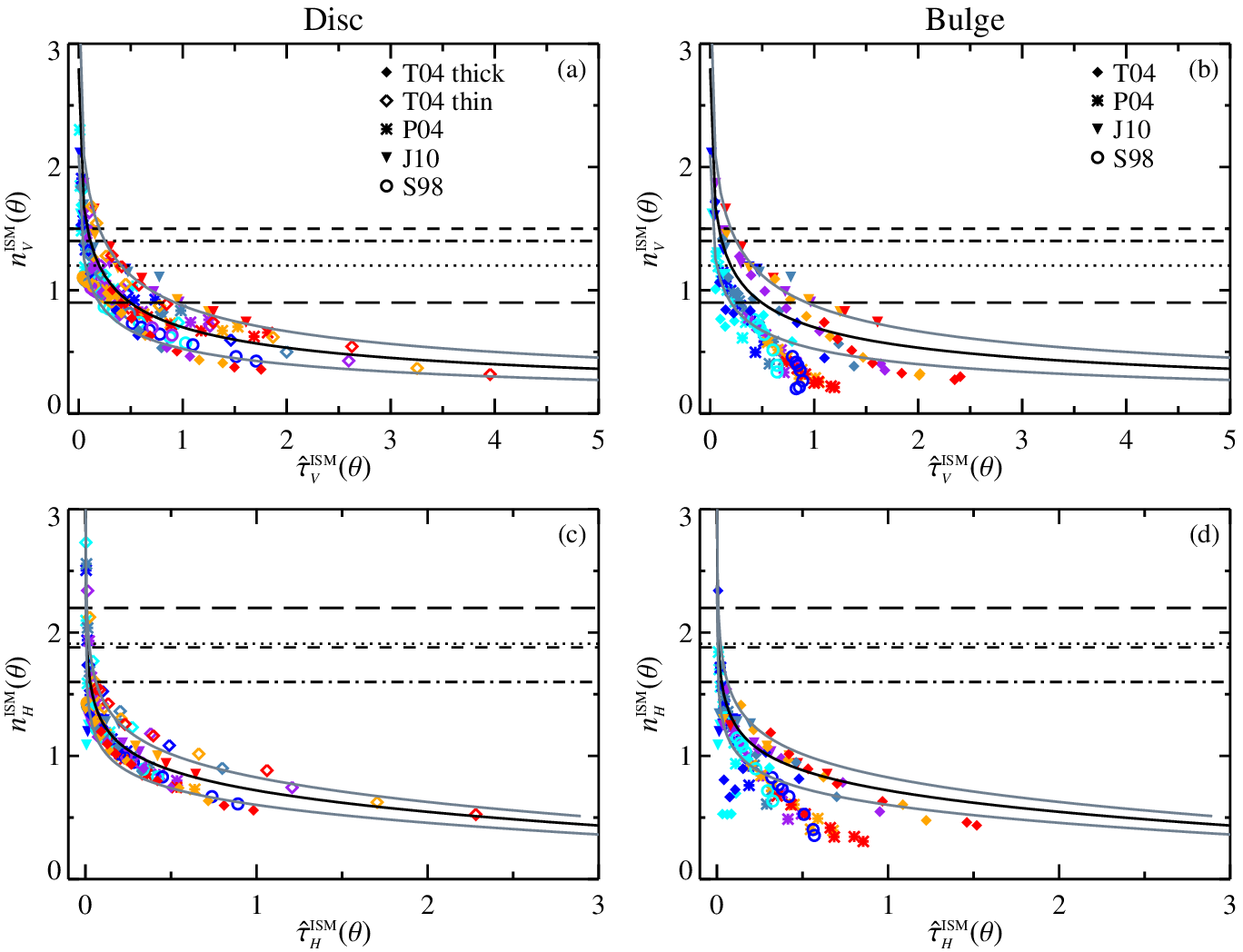}}
	\caption{({\it a}) Slope of the attenuation curve \nLismTh\ at galaxy inclination $\theta$ plotted against dust attenuation optical depth \tauLismTh\ (equations~\ref{eq:pow_law}--\ref{eq:n_fit}) , at $\lambda = 0.55$ \micron, as predicted for the stellar disc components of the P04, T04 and S98 models  and the (inseparable) disc+bulge components of the J10 SPH simulations. For the P04 and T04 models, different colours refer to different face-on optical depths \tauvP\ and $\taubP$, respectively:  0.25 (cyan), 0.50 (blue), 1.0 (steel blue), 2.0 (purple), 4.0 (orange) and 8.0 (red).  For the J10 model, different colours refer to different simulations: G0 (cyan:  $\meantauvISM = 0.06$), G1 (blue: $\meantauvISM = 0.11$), G2 (steel blue: $\meantauvISM = 0.14$), G3 (purple: $\meantauvISM = 0.24$),  Sbc$-$ (orange: $\meantauvISM= 0.31$), and Sbc$+$ (red: $\meantauvISM = 0.48$). For the S98 model, different colours refer to different simulated galaxies: Sb (cyan:  $\meantauvISM = 0.65$), Sc (blue: $\meantauvISM = 0.97$), M100 (steel blue: $\meantauvISM = 0.15$), M51 (purple: $\meantauvISM = 0.31$), and NGC\,6946 (orange: $\meantauvISM= 0.12$).  At fixed dust optical depth, the same symbol plotted at different \tauVismTh\ corresponds to the same model seen at different inclinations, $\theta=0\,^{\circ}$, $36\,^{\circ}$, $50\,^{\circ}$, $63\,^{\circ}$, $74\,^{\circ}$, $85\,^{\circ}$ and $90\,^{\circ}$ from left to right. ({\it b})~Same as~({\it a}), but for the bulge component of the P04, T04 and S98 models  [the J10 simulations are the same as in ({\it a})]. ({\it c})~Same as~({\it a}), but at $\lambda = 1.6$ \micron. ({\it d})~Same as~({\it b}), but at $\lambda = 1.6$ \micron\ [the J10 simulations are the same as in ({\it c})]. For reference, in panels ({\it a}) and ({\it b}), horizontal lines indicate the slopes in the range $0.4 \leq \lambda/\micron \leq 0.5$ of the mean Milky-Way \citep[{\it dotted line}]{ODonnell1994}, SMC \citep[{\it dashed line}]{Pei1992}, LMC \citep[{\it dot-dashed line}]{Pei1992} extinction curves and of the \citet[{\it long-dashed line}]{Calzetti2001} attenuation curve. Horizontal lines in panels ({\it c}) and ({\it d}), indicate the slopes in the range $1.4 \leq \lambda/\micron \leq 1.6$ of the mean Milky-Way \citep[{\it dotted line}]{Pei1992}, SMC \citep[{\it dashed line}]{Pei1992}, LMC \citep[{\it dot-dashed line}]{Pei1992} extinction curves and of the \citet[{\it solid line}]{Calzetti2001} attenuation curve. The solid black line in ({\it a}) shows the mean relation given by  equation~(\ref{eq:n_tau_fit}), and the solid grey lines show the associated 25-percent dispersion [these lines are repeated in ({\it b}) for reference]. The solid black line in  ({\it c}) shows the mean relation given by equations~(\ref{eq:n_lambda})--(\ref{eq:b_tau_fit}), and the solid grey lines show the associated 25+10-percent dispersion [these lines are repeated in ({\it d}) for reference].}
	\label{fig:V_slope}
\end{figure*} 
	 
Fig.~\ref{fig:V_slope} shows \nVismTh\ as a function of \tauVismTh\ (top panels), and \nHismTh\ as a function of \tauHismTh\ (bottom panels), as predicted by the P04, T04, J10 and S98  models over the full explored ranges in face-on dust optical depth and galaxy inclination. The left-hand panels show the attenuation of the stellar disc components in the P04, T04 and S98 models  along with the 7 SPH simulations (with inseparable dic+bulge components) of J10. The right-hand panels show the attenuation of the bulge component in the P04, T04 and S98 models and the same composite J10 models as in the left-hand panels. Different colours refer to different face-on dust optical depths. At fixed face-on dust optical depth, the same symbol plotted at different abscissae shows the same model seen at different inclinations (see caption for detail). Also shown for reference as horizontal lines in all panels in Fig.~\ref{fig:V_slope} are the slopes of the mean Milky-Way \citep{ODonnell1994}, SMC and LMC \citep{Pei1992} extinction curves and the \citet{Calzetti2001} attenuation curve. 

\begin{figure}
	\centering
	\resizebox{\hsize}{!}{\includegraphics[angle=90]{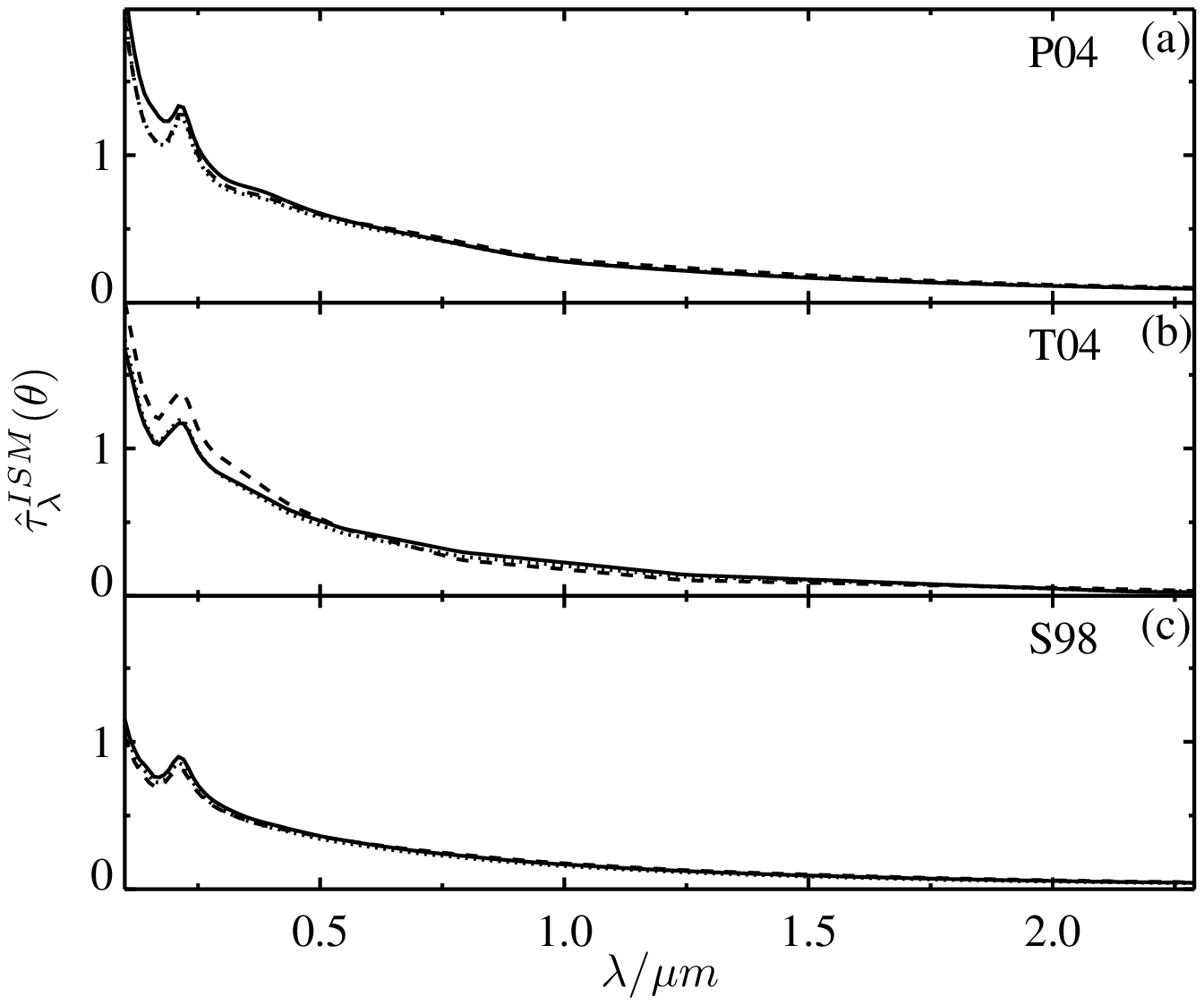}}
	\caption{Attenuation curves \tauLismTh\  of pure-disc models with different dust contents seen at different inclinations, for which the $V$-band attenuation optical depth \tauVismTh\  differ by less than 0.05. ({\it a}) For three P04 models with $\tauVismTh\approx0.55$: $(\tauvP,\theta )=(0.5,85\deg)$, $(2.0, 70\deg)$ and $(8.0,35\deg)$. ({\it b}) For three T04 models with $\tauVismTh\approx0.45$: $(\taubP,\theta)=(0.5,85\deg)$, $(2.0, 75\deg)$ and $(8.0,36\deg)$. ({\it c}) For three S98 models with $\tauVismTh\approx0.33$: $(\txn{disc model},\theta)=(\txn{Sb disc},48\deg)$, $(\txn{M\,51},70\deg)$ and $(\txn{NGC\,6946},90\deg)$. In each panel, the dashed, dotted and solid lines refer to the smallest, intermediate and largest inclination angles, respectively.}
	\label{fig:ex_curves}
\end{figure}

Fig.~\ref{fig:V_slope}a shows that, in all models, the slope \nVismTh\ of the optical attenuation curve decreases (i.e. the curve becomes shallower) when the $V$-band attenuation optical depth \tauVismTh\ increases. The attenuation curve is predicted to be steeper than the reference \citet{Calzetti2001} (Milky-Way) curve at $\tauVismTh\la0.5$ ($\tauVismTh\la0.25$) and shallower at larger \tauVismTh. Differences in the disc geometries adopted by P04, T04, J10 and S98 induce small vertical offsets between the different models (symbols), without  affecting the general trend. It is worth stressing that the overlap of identical symbols with different colours in Fig.~\ref{fig:V_slope}a implies that galaxies with different dust content and inclination but analogous \tauVismTh\ have similar attenuation curves. To better illustrate this effect, we plot in Fig.~\ref{fig:ex_curves} the attenuation curves of pure-disc models with different dust contents seen at different inclinations, for which the $V$-band attenuation optical depth \tauVismTh\ differs by less than 0.05, as computed by P04, T04 and S98. We choose the widest possible dynamic range in $\theta$ that can be sampled at fixed \tauVismTh\ by combining pure-disc S98 models with different dust contents (i.e. from $48\deg$ for the Sb-disc model to $90\deg$ for the NGC\,6946 model). Then, we select similar ranges in $\theta$ from the model grids of P04 and T04. Fig.~\ref{fig:ex_curves} confirms that models with different dust content seen at different inclination but with similar \tauVismTh\ have similar attenuation curves across the whole range from ultraviolet, to optical, to near-infrared wavelengths. This implies that the quasi-universal relation between \nVismTh\  and  \tauVismTh\ of Fig.\ref{fig:V_slope} is not purely the result of orientation effects. 

\begin{figure}
	\centering
	\resizebox{.8\hsize}{!}{\includegraphics{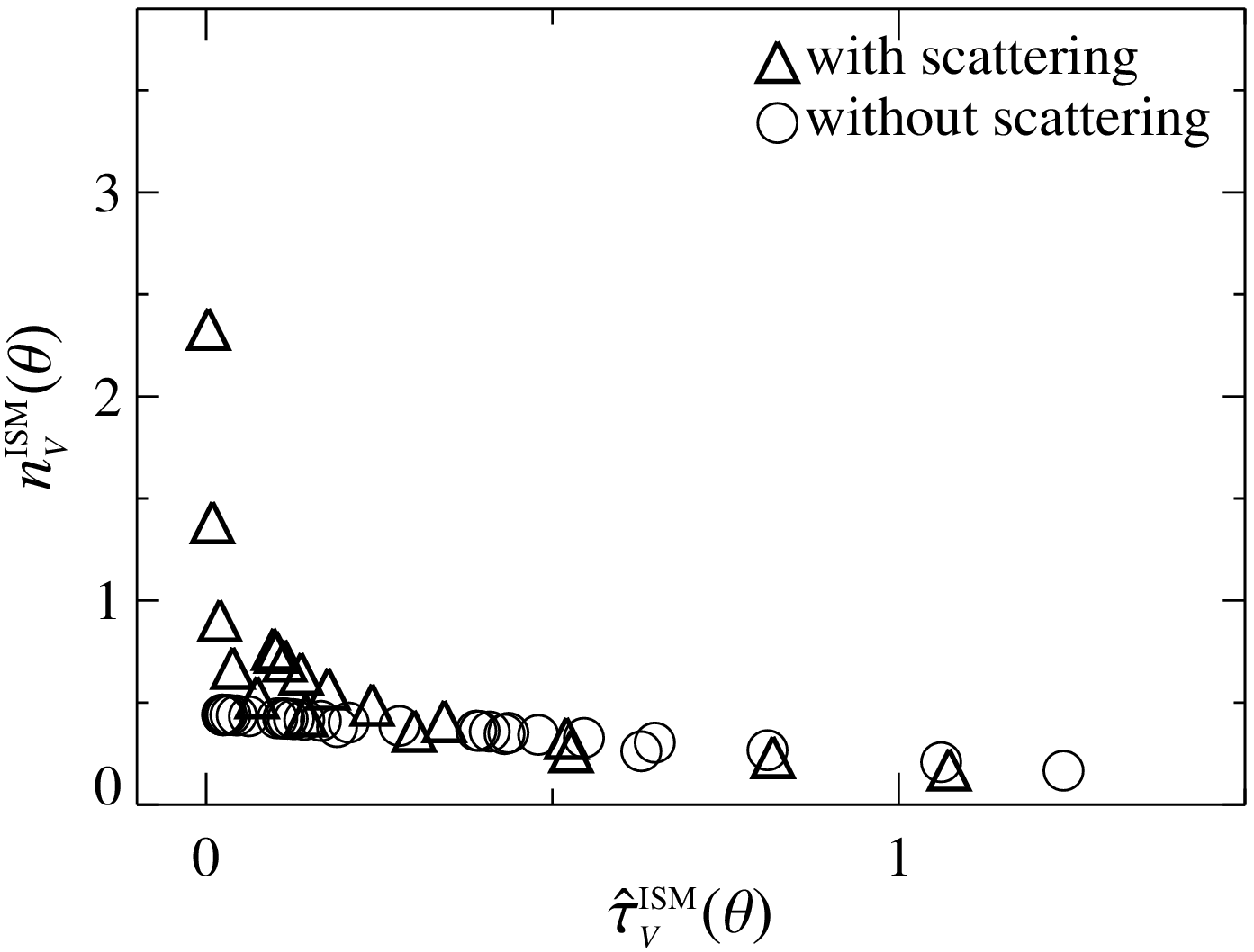}}
	\caption{Slope of the attenuation curve \nLismTh\ at galaxy inclination $\theta$ plotted against dust attenuation optical depth \tauLismTh\ (equations~\ref{eq:pow_law}--\ref{eq:n_fit}), at $\lambda = 0.55$ \micron, as predicted by the \citet{MacLachlan2011} radiative transfer code for disc galaxies. Open triangles show the results for standard calculations, while open circles show identical calculations in which the single-scattering albedo has been set to zero (MacLachlan, private communication). In each case, the different points correspond to galaxies with different face-on $V$-band optical depths, $\tauvP= 0.1$, 0.5 and 2.0, and inclinations, from $\theta=0\,^{\circ}$ to $90\,^{\circ}$ in steps of  $10\,^{\circ}$.}
	\label{fig:scattering}
\end{figure}

We interpret the relation between \nVismTh\  and  \tauVismTh\  in Fig.~\ref{fig:V_slope}a as the combination of two effects, one dominating at low attenuation optical depths, the other at large ones.  At small attenuation optical depths, scattering is the likely reason for the steepness of the attenuation curve. Fig.~\ref{fig:ext_curves} shows that, for the Milky-Way type dust in the P04, T04 and J10 models, the single-scattering albedo is roughly constant across the optical wavelength range $3.6 \la \log({\lambda}/\txn{\AA}) \la 3.9$, while the asymmetry parameter of the scattering phase function decreases monotonically. Thus, scattering is more forward in the blue and more isotropic in the red. This implies that blue photons emitted along the equatorial plane of a galaxy will have less chances than red photons to be scattered away from the plane and escape from the galaxy before they are absorbed. To a face-on observer, this will appear as a steepening of the attenuation curve. We highlight that this effect is important only at small \tauVismTh\ (corresponding also to small $\theta$; see Fig.~\ref{fig:V_slope}a), since at large \tauVismTh, photons are more likely to be absorbed even if they are first scattered, which reduces the imprint of scattering on the shape of the attenuation curve. Also, we note that the attenuation curve of the S98 model, in which scattering is assumed to be isotropic at all wavelengths (Fig.~\ref{fig:ext_curves}c), does not exhibit any steepening at low \tauVismTh, reaching at most $\nVismTh \approx1$. We can illustrate the effect of non-isotropic scattering on the attenuation curve by appealing to the publicly available radiative transfer code of \citet{MacLachlan2011}, which relies on a Monte Carlo algorithm to solve the equation of transfer for analytic distributions of stars and dust. As in the P04 and T04 models, \citet{MacLachlan2011} adopt Milky-Way type dust and double exponentially declining profiles to describe the distributions of stars and dust. Fig.~\ref{fig:scattering} shows  \nVismTh\  versus  \tauVismTh\ for two sets of disc calculations kindly made available to us by J.~MacLachlan (private communication): a standard set with ranges in $\tauvP$ and $\theta$ similar to those of the P04, T04 and J10 models in Fig.~\ref{fig:V_slope}a, in qualitative agreement with these models (open triangles in Fig.~\ref{fig:scattering}); and an identical set in which the single-scattering albedo has been set to zero (open circles). The results in Fig.~\ref{fig:scattering} shows that scattering is the likely reason for the steepening of the attenuation curve at low \tauVismTh\ in Fig.~\ref{fig:V_slope}a. 

At large attenuation optical depths, the shallowness of the attenuation curves in Fig.~\ref{fig:V_slope}a is a well-known signature of mixed distributions of stars and dust \citep[e.g.][]{CF00,Calzetti2001}: at any wavelength, the radiation reaching an observer from such distributions is dominated by stars located at an attenuation optical depth less than unity from the edge of the disc (which corresponds to deeper physical locations for red photons than for blue photons). This causes the attenuation curve of the disc component(s) to be greyer than the input (Milky-Way) extinction curve at large \tauVismTh\ in the models of Fig.~\ref{fig:V_slope}a. Fig.~\ref{fig:V_slope}b shows that, in the P04, T04 and S98 models, the attenuation curve of the bulge component at small and large \tauVismTh\ behaves in a way similar to that of the disc component(s) in Fig.~\ref{fig:V_slope}a, as expected from the fact that the same double exponentially declining dust disc(s) attenuate bulge and disc stars in these models (see Figs~\ref{fig:Pierini} and \ref{fig:Tuffs}). At fixed \tauVismTh , the bulge attenuation curves predicted by the P04 and S98 models in Fig.~\ref{fig:V_slope}b are systematically shallower than that predicted by the T04 model. This is likely to be a consequence of the different light profiles adopted for the stellar bulge in these models, which are much shallower in the P04 and S98 models than in the T04 model, as illustrated by Fig.~\ref{fig:bulge_prof}. As we shall see in Section~\ref{sec:dust_constraints} below, bulge stars in nearby star-forming galaxies account for typically only a few percent of the total optical emission (Table~\ref{tab:MCMC}). Thus, in practice, the global dust attenuation curve in these galaxies should be well represented by that pertaining to the disc component(s) in Fig.~\ref{fig:V_slope}a.

 \begin{figure}
	\centering
	\resizebox{.85\hsize}{!}{\includegraphics{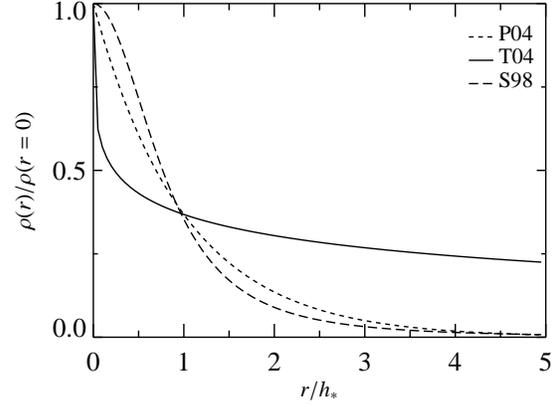}}
	\caption{Light profiles adopted for the stellar bulge component in the P04 [exponentially declining profile: $\rho(r) \propto \exp(-r/h_\ast)$, where $h_\ast$ is the bulge scale length], T04 \{de Vaucouleurs profile: $\rho(r) \propto \exp[(r/h_\ast)^{1/4}]$\} and S98 \{King profile: $\rho(r) \propto [1 + (r/h_\ast)^2]^{-3/2}$\} models.}
	\label{fig:bulge_prof}
\end{figure}

Figs~\ref{fig:V_slope}c and \ref{fig:V_slope}d show the analog of Figs~\ref{fig:V_slope}a and \ref{fig:V_slope}b for the near-infrared attenuation curve. Again, all models in Figs~\ref{fig:V_slope}c appear to follow a quasi-universal relation between \nHismTh\  and  \tauHismTh. The steepening and flattening of the attenuation curve at small and large \tauHismTh, respectively, is qualitatively similar to the behaviour of the optical attenuation curve at small and large \tauVismTh. In Fig.~\ref{fig:V_slope}d, the behaviour of the bulge attenuation curve at near-infrared wavelengths is also qualitatively similar to that found in Fig.~\ref{fig:V_slope}b at optical wavelengths. Again, the different trends followed by the P04 and S98 models on the one hand, and the T04 model on the other hand, are likely to result from the different light profiles adopted for bulge stars in these models (Fig.~\ref{fig:bulge_prof}). As before, we note that this discrepancy should have a negligible implication for measurements of the global attenuation curve in nearby star-forming galaxies, which is expected to be dominated by the disc component(s) (Section~\ref{sec:dust_constraints}).

For practical applications, it is of interest to fit the quasi-universal relation between slope of the attenuation curve and attenuation optical of the dust in the ambient ISM in Figs~\ref{fig:V_slope}a and \ref{fig:V_slope}c by means of a simple analytic function. Adopting for this purpose the simplified notation $\tau_V\equiv\tauVismTh$, we find that rewriting equation~(\ref{eq:n_fit})  as
\begin{equation}\label{eq:n_lambda}
\nLism(\tau_V) = \nVism(\tau_V)+ b(\tau_V)\,(\lambda/\micron -0.55) \,,
\end{equation} 
with
\begin{equation}\label{eq:n_tau_fit}
\nVism(\tau_V) =  \frac{ 2.8 }{ 1+3 \sqrt{\tau_V} }\hspace{10mm} (\pm25\,\txn{percent})
\end{equation}
and
\begin{equation}\label{eq:b_tau_fit}
b(\tau_V) =  0.3-0.05 \, \tau_V\hspace{12mm} (\pm10\,\txn{percent})\,,
\end{equation}
provides an adequate representation of the relations both between \nVismTh\  and  \tauVismTh\ and between \nHismTh\  and  \tauHismTh, as shown by the black solid lines in Figs~\ref{fig:V_slope}a and \ref{fig:V_slope}c (grey solid lines indicate the dispersion around these mean relations). We emphasise that equations~(\ref{eq:pow_law}) and (\ref{eq:n_lambda})--(\ref{eq:b_tau_fit}) summarise the results of four sophisticated radiative transfer models including realistic spatial distributions of stars and dust. These equations can be easily implemented in any spectral analysis of structurally unresolved star-forming galaxies to include in a simple yet physically consistent way for the dependence of dust attenuation on inclination (as exemplified recently by \citealt{Pacifici2012}). The fact that the P04, T04, J10 and S98  models all rely on standard optical properties of Milky Way-type dust should be kept in mind, but we do not expect this to strongly limit applications of equation~(\ref{eq:n_lambda}) to spectral studies of external galaxies. In fact, the relation in Fig.~\ref{fig:V_slope}a extends over a much wider range in \nVismTh\  than the dispersion between the Milky-Way, LMC, SMC and \citet{Calzetti2001} curves. Also, the optical properties of Milky Way- and SMC-type dust are similar at optical wavelengths, exhibiting larger differences in the ultraviolet (see fig.~10 of \citealt{Gordon2003} and fig.~4 of \citealt{Draine2003}). This suggests that, in normal star-forming (disc) galaxies, changes in the shape of the optical attenuation curve induced by geometry and orientation effects are likely to dominate over those induced by differences in the optical properties of dust grains. Finally, we note that the dispersion about the mean relations in equations~(\ref{eq:n_tau_fit}) and (\ref{eq:b_tau_fit}) accounts to some extent for  potential systematic uncertainties in these relations arising from the different assumptions of the P04, T04, J10 and S98 models (i.e. different dust properties, different scale-lengths and scale-heights of the stellar and dust discs; see Table~\ref{tab:RT_models}). A more thorough investigation of the assumptions behind radiative-transfer models and their implications for the systematic uncertainties affecting the results presented in this paper would require running the original models for different values of their main adjustable parameters. This is beyond the scope of the present study.

\subsection{Relation between dust attenuation optical depth and galaxy inclination}\label{sec:tau_theta}

In the previous section, we found that the sophisticated radiative transfer models of P04, T04, J10 and S98 all predict a similar dependence of the slope of the attenuation curve on the dust attenuation optical depth of the diffuse ISM in star-forming galaxies. Here, we show that, at fixed dust content, these models also predict a similar dependence of \tauLismTh\  on galaxy inclination $\theta$. We demonstrate this by using the more versatile T04 model, which includes a thin-disc, a thick-disc and a bulge components, to reproduce the attenuation curves of different types of galaxies computed with the P04 and J10 models (we do not consider here the S98 model, in which the approximation of isotropic scattering makes the attenuation curves much shallower than in the other models at low optical depths, i.e. at face-on inclinations, which would bias the comparisons). To proceed, we explore a wide collection of attenuation curves generated using the T04 model by considering full ranges of central face-on $B$-band optical depth $\taubP$ (Section~\ref{sec:T04}) and of relative intensities $\xi^{\,i}_{\lambda}$ of the thin-disc, thick-disc and bulge components. In practice, for a given star formation law $\psi(t)$, we compute $\xi^{\,i}_{\lambda}$ from equations~(\ref{eq:xi}) and (\ref{eq:intensity}) by relating stars in different age ranges to different geometric components of the T04 model. This is achieved by introducing three age parameters, \Tthin, \Tthick\ and \Tbulge, such that $\Tthin \leq \Tthick \leq \Tbulge$. We associate stars younger than \Tthin\  with the thin-disc component, those in the age range $ \Tthin < t \leq \Tthick$ with the thick-disc component and those in the age range $ \Tthick < t \leq \Tbulge$ with the bulge component (we consider the ranges in $\taubP$,  \Tthin, \Tthick\ and \Tbulge\ listed in Table~\ref{tab:prior}). To compute the spectral evolution $S_{\lambda}[t',Z(t-t')]$ of stellar populations in equation~(\ref{eq:intensity}), we appeal to the \citet{BC03} stellar population synthesis code.

\begin{figure}
	\centering
	\subfigure{
		\resizebox{\hsize}{!}{\includegraphics{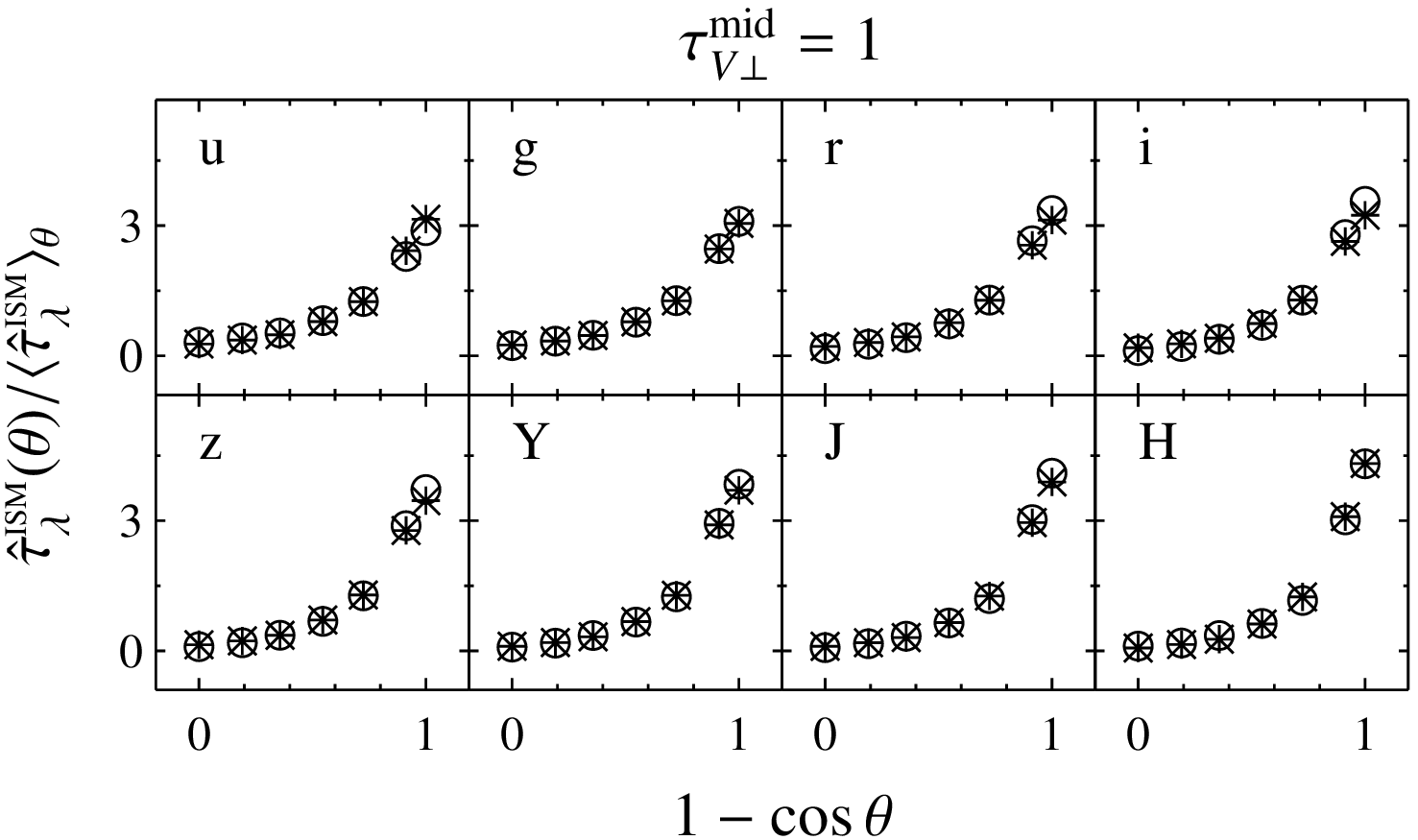}}
               \label{subfig:P04_a}}
	\subfigure{
		\resizebox{\hsize}{!}{\includegraphics{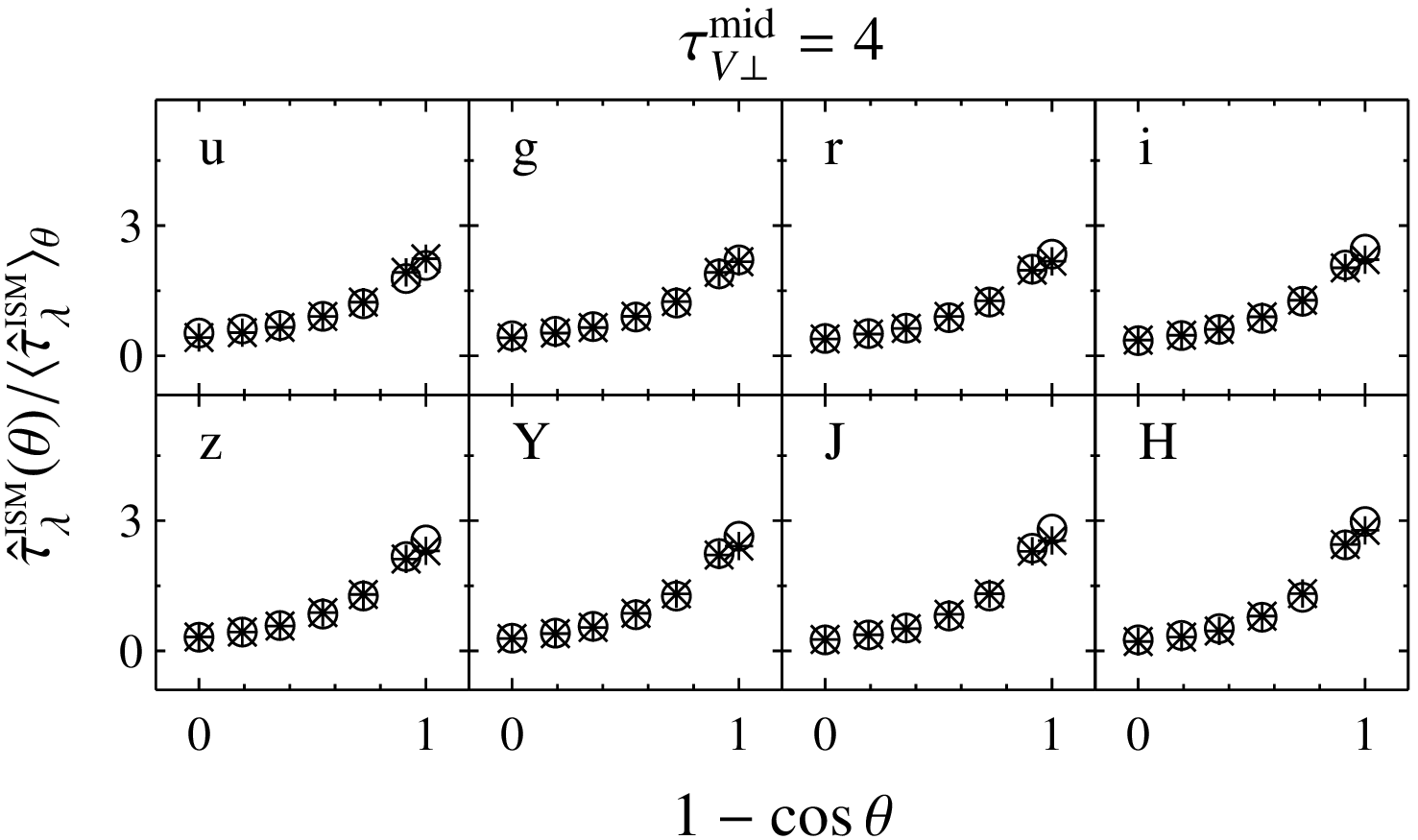}} 
                \label{subfig:P04_b}}
	\caption{Dust attenuation optical depth \tauLismTh\ at galaxy inclination $\theta$, in units of the angle-averaged value \meantaulISM, plotted against $1-\cos\theta$, in the $ugrizYJH$ photometric bands, as indicated. {\it Top}: reproduction of a pure-disc P04 model with $\tauvP=1$ (stars) by a T04 model with $\taubP = 2.2$ and age parameters $\Tthin = 0.01$\,Gyr and $\Tthick =  \Tbulge=6.2$\,Gyr (open circles). {\it Bottom}: reproduction of a pure-disc P04 model with $\tauvP=4$ (stars) by a T04 model with $\taubP = 5.7$ and age parameters $\Tthin = 0.003$\,Gyr and $\Tthick =  \Tbulge=9.7$\,Gyr (open circles). See Section~\ref{sec:tau_theta} for detail.}
	\label{fig:Fit_P04_disc}
\end{figure}

We focus on the dependence of the attenuation optical depth \tauLismTh\ on inclination $\theta$ at the effective wavelengths of the $ugrizYJH$ photometric bands ($\lambda_\txn{eff}=3546$, 4670, 6156, 7471, 8918, 10\,305, 12\,483, and 16\,313\,\AA, respectively). To efficiently explore the T04 parameter space and find the combination of $\taubP$,  \Tthin, \Tthick\ and \Tbulge\ that can best reproduce calculations of \tauLismTh\ by P04 and J10, we appeal to a Markov Chain Monte Carlo (MCMC) algorithm of the type described in Section~\ref{sec:bayes_approach} below (to which we refer for detail; see also Appendix~\ref{app:MCMC}). In Fig.~\ref{fig:Fit_P04_disc}, we show examples of two combinations of T04 model parameters selected in this way to reproduce P04 predictions for two pure-disc models with face-on $V$-band dust optical depth $\tauvP=1$ and 4, respectively (Section~\ref{sec:P04}). In practice, we fit the P04 pure-disc model with the T04 thin- and thick-disc components, excluding the T04 bulge component (i.e. we fix $\Tthick =  \Tbulge$).
For simplicity, since the P04 model does not include any star formation history, we have adopted a constant star formation law, $\psi(t)=$\,const, to compute the relative intensities $\xi^{\,i}_{\lambda}$ from equation~(\ref{eq:intensity}) in the T04 model. The excellent agreement between the stars and circles in all panels of Fig.~\ref{fig:Fit_P04_disc} demonstrates that two different combinations of the T04 thin- and thick-disc components can reproduce the dependence of \tauLismTh\ on $\theta$ predicted at all wavelengths by P04 bulge-less models with $\tauvP=1$ and 4. We note that the T04 best-fit parameters (listed in the figure caption) depend on the adopted star formation law $\psi(t)$, which cannot be constrained independently using the P04 attenuation curves.

\begin{figure}
	\centering
	\subfigure{
		\resizebox{\hsize}{!}{\includegraphics{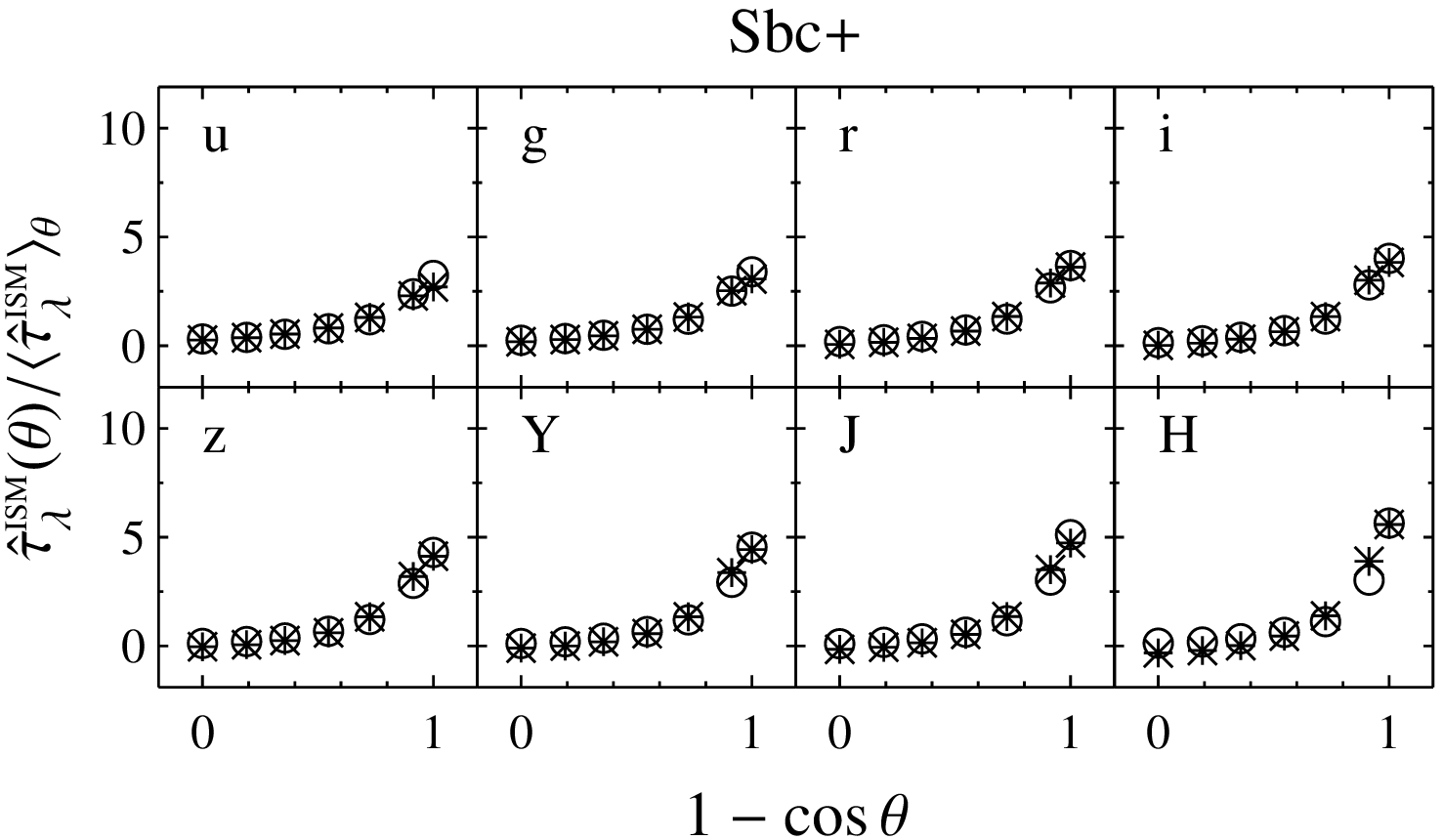}}
		\label{subfig:J10_a}}
	\subfigure{
		\resizebox{\hsize}{!}{\includegraphics{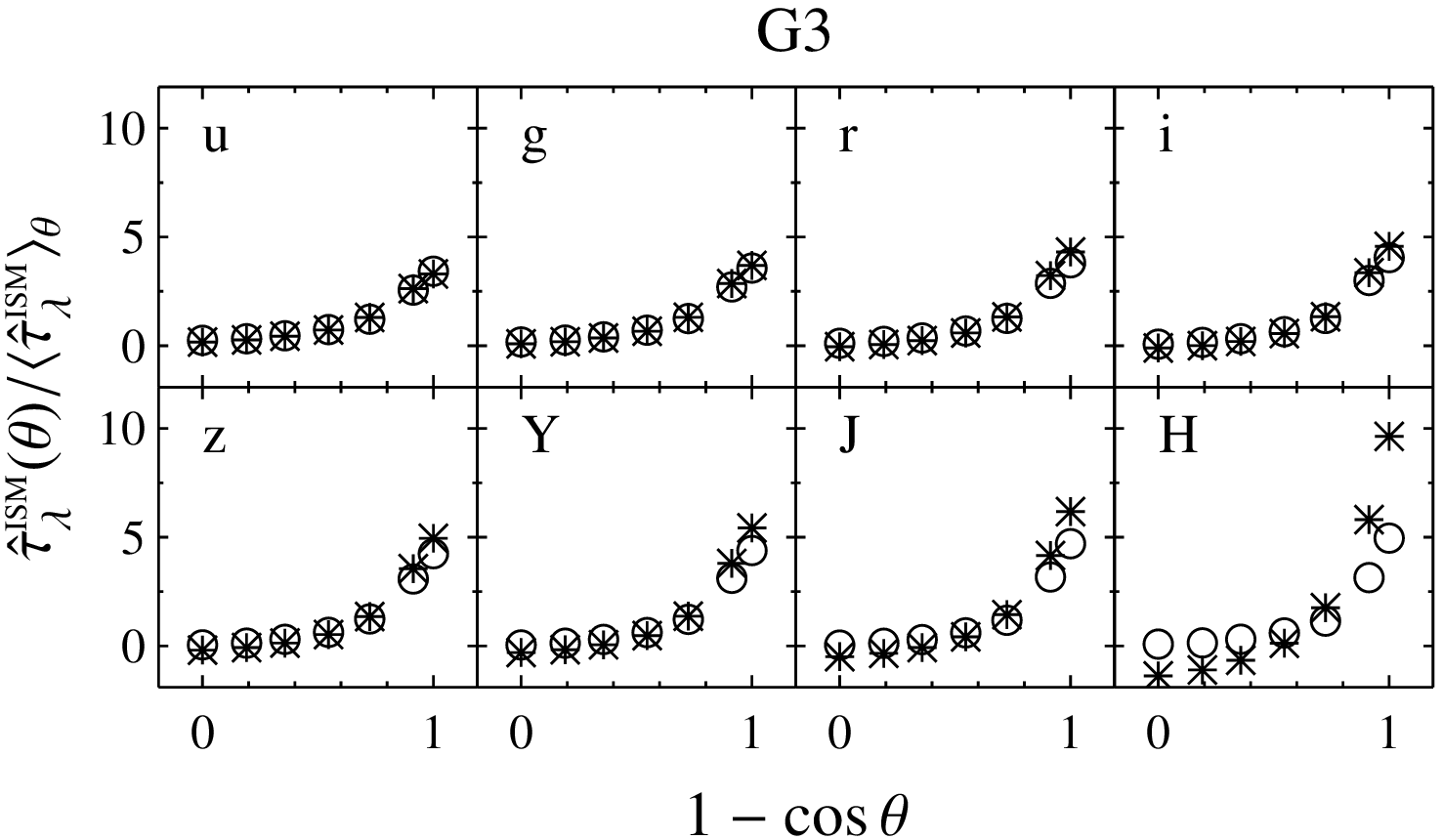}}
          	\label{subfig:J10_b}}
	\caption{Dust attenuation optical depth \tauLismTh\ at galaxy inclination $\theta$, in units of the angle-averaged value \meantaulISM, plotted against $1-\cos\theta$, in the $ugrizYJH$ photometric bands, as indicated. {\it Top}: reproduction of the Sbc+ model of J10 (stars) by a T04 model with $\taubP = 3.0$ and age parameters $\Tthin = 4.7$\,Gyr, $\Tthick = 10.8$\,Gyr and $\Tbulge=11.3$\,Gyr  (open circles). {\it Bottom}: reproduction of the G3 model of J10 (stars) by a T04 model with $\taubP = 1.8$ and age parameters $\Tthin = 0.5$\,Gyr, $\Tthick =11.3$\,Gyr and $ \Tbulge=11.5$\,Gyr (open circles). See Section~\ref{sec:tau_theta} for detail.}
	\label{fig:Fit_J10_disc}
\end{figure}

In Fig.~\ref{fig:Fit_J10_disc}, we present fits of the Sbc+ and G3 models of J10 obtained in a similar way by optimising the combination of T04 model components. We choose these galaxies because they are the largest ones simulated by J10 in their respective mass ranges (Section~\ref{sec:J10}), which should minimise the influence of numerical resolution on the results. To compute the relative intensities $\xi^{\,i}_{\lambda}$ of the different geometric components of the T04 model in this case (equation~\ref{eq:intensity}), we have adopted the original star formation histories of the J10 simulations. Fig.~\ref{fig:Fit_J10_disc}  shows that the fit of the Sbc+ model is overall excellent, despite some slight deviations in the near-infrared $Y$, $J$ and $H$ bands. The deviations in these bands are more pronounced for the fit of the G3 model. In fact, the G3 model reaches large negative attenuation optical depths at low inclination in the $Y$, $J$ and $H$ bands. This is again a signature of the isotropic scattering of infrared photons emitted along the equatorial plane of a galaxy, which can be scattered away from the plane and boost the emission in the face-on direction. The effect is more pronounced in the G3 model than in the Sbc+ model, possibly because of the lower dust attenuation optical depth of the former ($\meantauvISM = 0.24$) relative to the latter ($\meantauvISM=0.48$; see the discussion in Section~\ref{sec:S_A_rel} above). We note that the T04 model also reaches negative infrared attenuation optical depths at $\theta=0$, but only of the order of a few hundredths of magnitude. This discrepancy between the infrared predictions of the T04 and J10 models has a negligible effect on galaxy spectral analyses, since the infrared attenuation optical depths are generally very small (for reference, the angle-averaged $H$-band attenuation optical depths of the Sbc+ and G3 models are $\langle\tauHismTh\rangle_\theta=0.11$ and 0.03, respectively).

In summary, we find that the P04, T04 and J10 models all predict a roughly universal relation between attenuation optical depth of the dust and slope of the optical attenuation curve in the diffuse ISM of star-forming galaxies. We also find that, at fixed dust content, the three models predict a similar dependence of \tauLismTh\  on galaxy inclination $\theta$ in the $ugrizYJH$ photometric bands. We could demonstrate this because the versatility of the T04 model allows us to reproduce calculations of \tauLismTh\ by P04 and J10 at all these wavelengths for different types of galaxies.

\section{Insights into the content and spatial distribution of dust in observed galaxies}

In this section, we exploit the general properties of dust attenuation models outlined in the previous section to investigate the physical origin of observed systematic changes of the dust attenuation curve as a function of both inclination and star formation history in a sample of nearby star-forming galaxies. We start with a brief description of the observational sample. Then, we build a comprehensive library of model spectral energy distributions encompassing wide ranges of stellar and dust parameters to interpret the \HaHb\ ratio and $ugrizYJH$ photometry of the observed galaxies. This allows us to derive original constraints on the content and spatial distribution of dust from the integrated spectral properties of these galaxies.

\subsection{Observational sample}\label{sec:data}

\subsubsection{Sample selection and basic parameters}\label{sec:sample}

We consider the large sample of nearby star-forming galaxies assembled by \citet[hereafter W11]{Wild2011b} by combining optical spectroscopy and $ugriz$ photometry from the SDSS Data Release 7 \citep[DR7,][]{Abazajian2009} with near-infrared $YJHK$ photometry from the seventh data release of the UKIRT Infrared Deep Sky Survey-Large Area Survey \citep[UKIDSS-LAS,][]{Lawrence2007}. To build this sample, W11 first select SDSS galaxies with high-quality spectra ($\sn>6$ per pixel in the $g$ band) within the formal survey limits ($14.5<m_r<17.77$, where $m_r$ is the $r$-band Petrosian\footnote{\label{footnote:petrosian} This corresponds to the total flux within a circular aperture of twice the Petrosian radius, where the Petrosian radius is the largest radius at which the local $r$-band surface brightness is at least one-fifth the mean surface brightness interior to that radius.} magnitude corrected for Galactic extinction) and with physical-parameter entries in the SDSS-MPA value-added catalog.\footnote{\url{http://www.mpa-garching.mpg.de/SDSS/DR7/raw_data.html}} Potential hosts of active galactic nuclei are removed using the \citet{Kauffmann2003c} criterion in the \citet{BPT1981} diagram defined by the \Nii/\Ha\ and \Oiii/\Hb\ emission-line ratios (with the requirement that all four lines be measured with $\sn>3$, after rescaling of the noise as recommended on the SDSS-MPA website). W11 further exclude galaxies with post-starburst spectral features using the method of \citet{Wild2007}, since the rapidly changing spectral shapes of these galaxies can affect measurement of the attenuation curve through the pair-matching technique (see Section~\ref{sec:w11} below). W11 cross-match the SDSS and UKIDSS catalogs by locating all objects within 2\,arcsec of each other, selecting the nearest neighbour in case of multiple matches. They exclude galaxies for which UKIDSS photometry has been de-blended, as these suffer from flux calibration issues \citep{Hill2010}. The final sample consists of $\numprint{22902}$ galaxies at a median redshift of 0.07.

The median $r$-band Petrosian radius of these galaxies is 5.1\,arcsec, i.e. larger than the 3\,arcsec-diameter aperture of the SDSS spectroscopic fibres. Since we are interested in physical quantities derived from both spectroscopy and photometry, we do not use Petrosian magnitudes here but adopt instead fibre magnitudes measured in a 3\,arcsec-diameter aperture from the SDSS catalog. We combine this with infrared photometry measured in a matching aperture of 2.83\,arcsec diameter from the UKIDSS catalog (see W11 for more detail). 
 
\begin{figure}
	\centering
	\resizebox{\hsize}{!}{\includegraphics{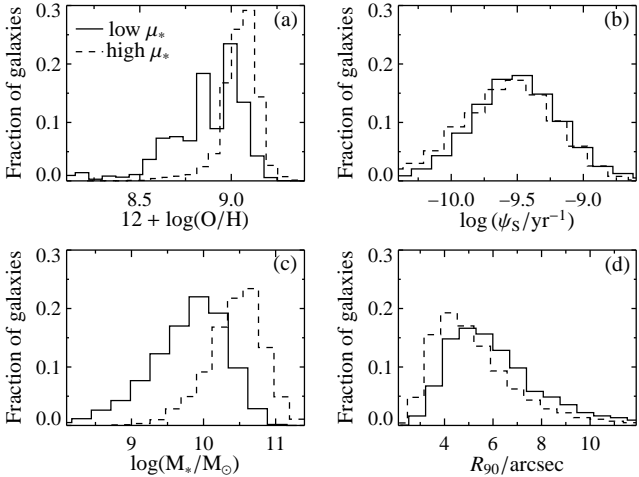}}
	\caption{Distributions of physical properties of the low-\muS\ (solid histograms) and high-\muS\ (dashed histograms) nearby star-forming galaxies in the W11 sample described in Section~\ref{sec:sample}. ({\it a})~Gas-phase oxygen abundance, \logOH. ({\it b})~Specific star formation rate, \psiS. ({\it c})~Stellar mass, $\log{\Mstar}$. ({\it d})~Radius containing 90 percent of the $r$-band Petrosian flux, $R_{90}$. The distributions of \logOH\ and \psiS\ are used as priors in the library of model spectral energy distributions described in Section~\ref{sec:model_lib}.}
	\label{fig:cur_params}
\end{figure}

Basic physical parameters are available for these galaxies from published SDSS analyses. This includes both total (\Mstar) and fibre (\Mfib) stellar masses, measured from Bayesian fits to the five-band SDSS photometry using a comprehensive library of model galaxy spectra similar to that used in \citet[][accounting for dust attenuation as prescribed by \citealt{CF00}]{Gallazzi2005};\footnote{\url{http://www.mpa-garching.mpg.de/SDSS/DR7/Data/stellarmass.html}} and gas-phase oxygen abundance [\logOH; \citealt{Tremonti2004}, corrected for dust attenuation using the model of \citealt{CF00}] and star formation rate ($\psi$; we adopt here the fibre estimates of W11, corrected for dust attenuation using the prescription of \citealt{Wild2011a}), measured from fibre spectroscopy. From these quantities, we compute the specific star formation rate within the fibre aperture, $\psiS=\psi/\Mfib$, and the stellar surface mass density, $\muS=\Mstar/(2\pi R_{50,z}^2)$, where $R_{50,z}$ is the radius encompassing 50 percent of the $z$-band Petrosian flux. W11 differentiate galaxies according to \muS. Following \citet{Kauffmann2003a}, they distinguish bulge-less galaxies, with $\muS< 3 \times 10^8\,\Msun\txn{pc}^{-2}$ (7451 galaxies), from galaxies with a bulge, with $\muS > 3 \times 10^8\,\Msun\txn{pc}^{-2}$ (15\,451 galaxies). We show in Fig.~\ref{fig:cur_params} the distributions of \logOH, \psiS\ and \Mstar\ for these two subsamples, along with the distribution of the radius enclosing 90 percent of the total Petrosian $r$-band flux, $R_{90}$. The low-\muS\ galaxies extend to significantly lower gas-phase oxygen abundances and stellar masses than their high-\muS\ counterparts, although in both cases, similar ranges in specific star formation rate and size are probed. Also available from the SDSS catalog for these galaxies is the ratio of minor- to major-axis length ($b/a$) measured from exponential fits to the SDSS $r$-band image. We convert this ratio into inclination angle $\theta$ using the standard formula for an oblate spheroid \citep[e.g.][]{Guthrie1992}
\begin{equation}
\cos{\theta} = \sqrt{\frac{(b/a)^2 - q_0^2}{1-q_0^2}} \, ,
\end{equation}
where $q_0$ is the intrinsic flattening of the ellipsoid representing the galaxy (i.e. the axis ratio when seen edge-on). We adopt $q_0=0.14$ and 0.20 for the low- and high-\muS\ samples, respectively \citep{Ryden2004, Ryden2006}. 

\subsubsection{Analysis of \citet{Wild2011b}}\label{sec:w11}

W11 use the above sample to explore the dependence of the shape of the optical and near-infrared attenuation curves on galaxy inclination and specific star formation rate. They achieve this by comparing the observed spectral energy distributions of pairs of galaxies with presumably similar star formation and chemical enrichment histories, but different dust content (as measured from the observed \HaHb\ ratio), at fixed inclination. This is similar to the approach introduced by \citet{Calzetti2001} to constrain the attenuation curve in nearby starburst galaxies, only generalised to galaxies with more complex star formation histories. In essence, W11 first identify pairs of galaxies with comparable gas-phase metallicity, specific star formation rate, axis ratio (inclination) and redshift but different dust content (replacing the match in gas-phase metallicity by one in stellar mass or surface mass density does not alter the conclusions). They normalise each spectral energy distribution to unit flux in the $K$ band ($\lambda_\txn{eff}=22\,010$\,\AA) to remove the dependence on galaxy mass. Then, the flux ratios of the two galaxies in the SDSS  $ugriz$ and UKIDSS  $YJH$ bandpasses provide an estimate of the dust attenuation curve. W11 compute average flux ratios in bins of fixed inclination $b/a$ and specific star formation rate \psiS\ to measure the dependence of the optical and near-infrared attenuation curves on these parameters.

The conclusions of Section~\ref{sec:S_A_rel} above allow us to reconsider the results of W11 from a new perspective. Mean attenuation curves derived from differential photometry, such as those of \citet{Calzetti1994} and W11, reflect the true attenuation curve of a class of galaxies only if the shape of this curve does not depend on the \HaHb\ ratio. However, as shown by Fig.~\ref{fig:V_slope}, current sophisticated models of radiative transfer predict that the shape of the dust attenuation curve should change with dust content in galaxies with similar relative distributions of stars and dust. Thus, differential photometry of pair-matched galaxies with similar intrinsic spectral energy distributions and different \HaHb\ ratio should trace not only the dust attenuation curve, but also changes in the shape of this curve with absolute dust content. The framework developed in Section~\ref{sec:models} allows us to account for these two contributions and better characterise the dependence of the attenuation curve on galaxy geometry and inclination. We note that W11 also analyse the dependence of the attenuation curve on specific star formation rate. We do not investigate this dependence here because the radiative transfer models described in Section~\ref{sec:RT} do not include any physical relation between attenuation by dust and star formation rate. 

\subsubsection{Correction of systematic biases in the dependence of dust attenuation signatures on inclination}\label{sec:bias_corr}

The first step toward characterising the dependence of the attenuation curve on galaxy geometry and inclination is to accurately measure the changes of observed  spectral properties with axis ratio. We consider here the broadband spectral energy distribution, which traces the dependence of attenuation on wavelength, and the \HaHb\ ratio, which quantifies the absolute attenuation optical depth. The \HaHb\ ratio reflects the attenuation of young stars by dust both in stellar birth clouds and in the ambient ISM (this ratio does not reveal the potential minor fraction of  dust in the ionised gas, which can absorb photons before they ionise hydrogen; e.g. \citealt{CF00}). We measure the attenuation relative to the dust-free case B recombination ratio $\HaHb\ \approx2.86$ (appropriate for electron densities $n_e \la 10^4 \: \txn{cm}^{-3}$ and temperatures $T_e\approx10\,000$\,K), the difference between the dust attenuation optical depths at H$\beta$ and H$\alpha$ being given by
\begin{equation}
\hat{\tau}{_{\txn{H}\beta}}(\theta)-\hat{\tau}{_{\txn{H}\alpha}}(\theta)=\ln\left({\HaHb\over{2.86}}\right)\,
\end{equation}
To proceed, we first divide the sample of Section~\ref{sec:sample} in 10 bins of equal axis ratio $b/a$. By analogy with W11, we normalise the $ugrizYJHK$ spectral energy distribution of each galaxy to unit flux in the $K$ band. Then, we compute the mean spectral energy distribution and mean \HaHb\ ratio in each bin of $b/a$. By comparing the mean spectral properties of galaxies in any bin to those in the face-one bin ($b/a \approx 1$; corresponding to the smallest attenuation), we can quantify the dependence of attenuation on galaxy orientation.

\begin{figure}
	\centering
	\subfigure{
		\resizebox{\hsize}{!}{\includegraphics{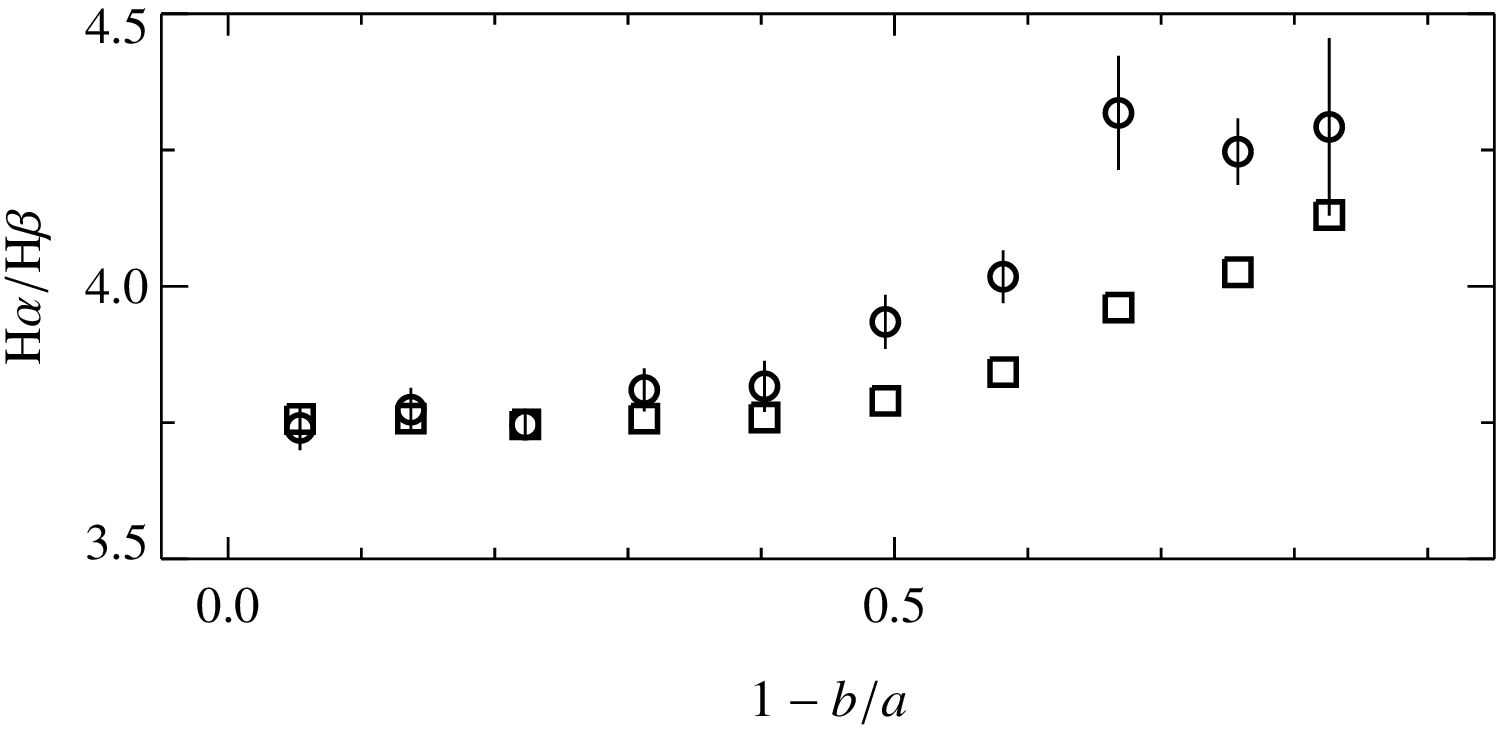}}}
	\subfigure{
		\resizebox{\hsize}{!}{\includegraphics{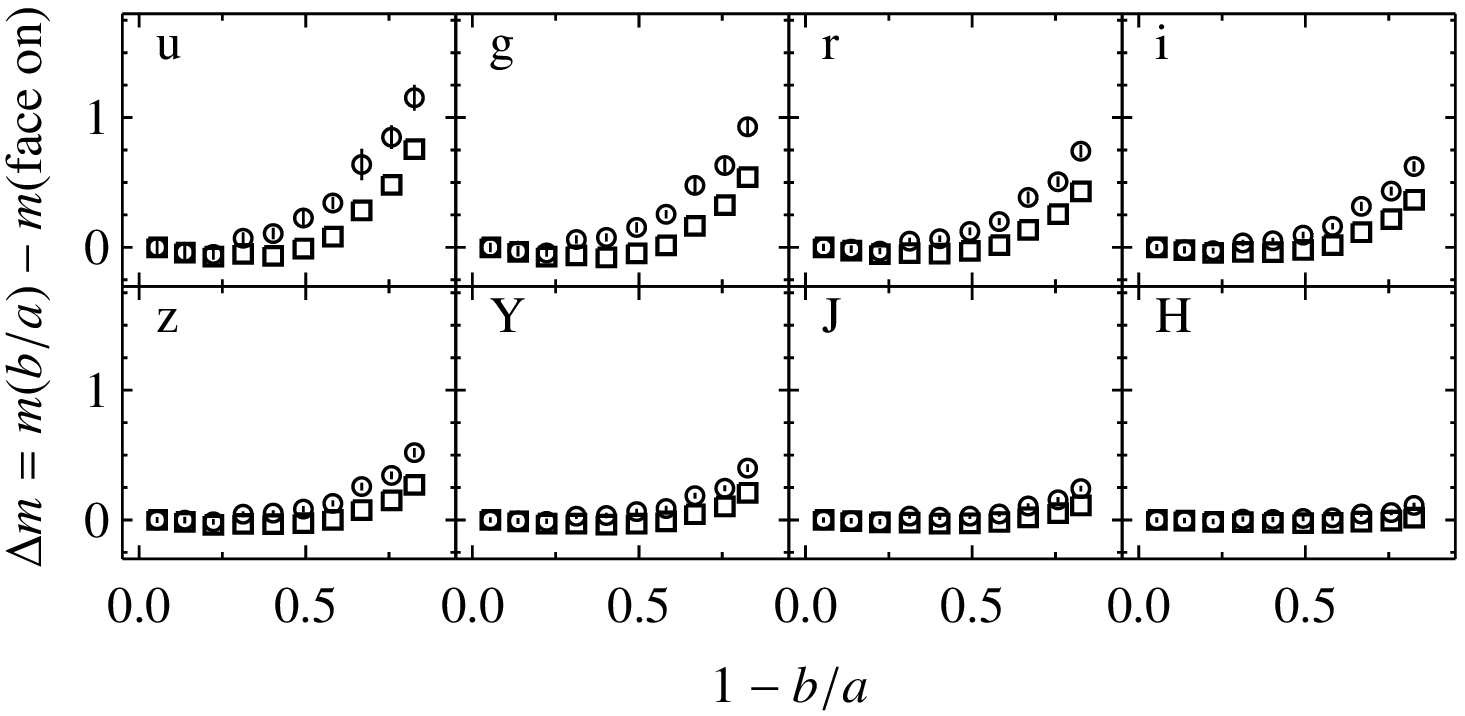}}}
	\caption{{\it Top}: Mean \HaHb\ ratio in bins of different galaxy inclination, from face-on ($1-b/a\approx0$) to edge-on ($1-b/a\approx1$), for the low-\muS\ galaxies in the W11 sample described in Section~\ref{sec:sample}. Square correspond to direct observations, while circles with error bars show the effect of correcting these observations for systematic biases using the method of importance sampling described in Section~\ref{sec:bias_corr} and Appendix~\ref{app:importance}. {\it Bottom}: difference between the mean magnitude of galaxies in bins of different inclination and that of face-on galaxies, at fixed $K$ band flux (i.e. fixed stellar mass), in the $ugrizYJH$ bandpasses, as indicated. The squares and circles have the same meaning as in {\it Top}.}
	\label{fig:data_low}
\end{figure}

\begin{figure}
	\centering
	\subfigure{
		\resizebox{\hsize}{!}{\includegraphics{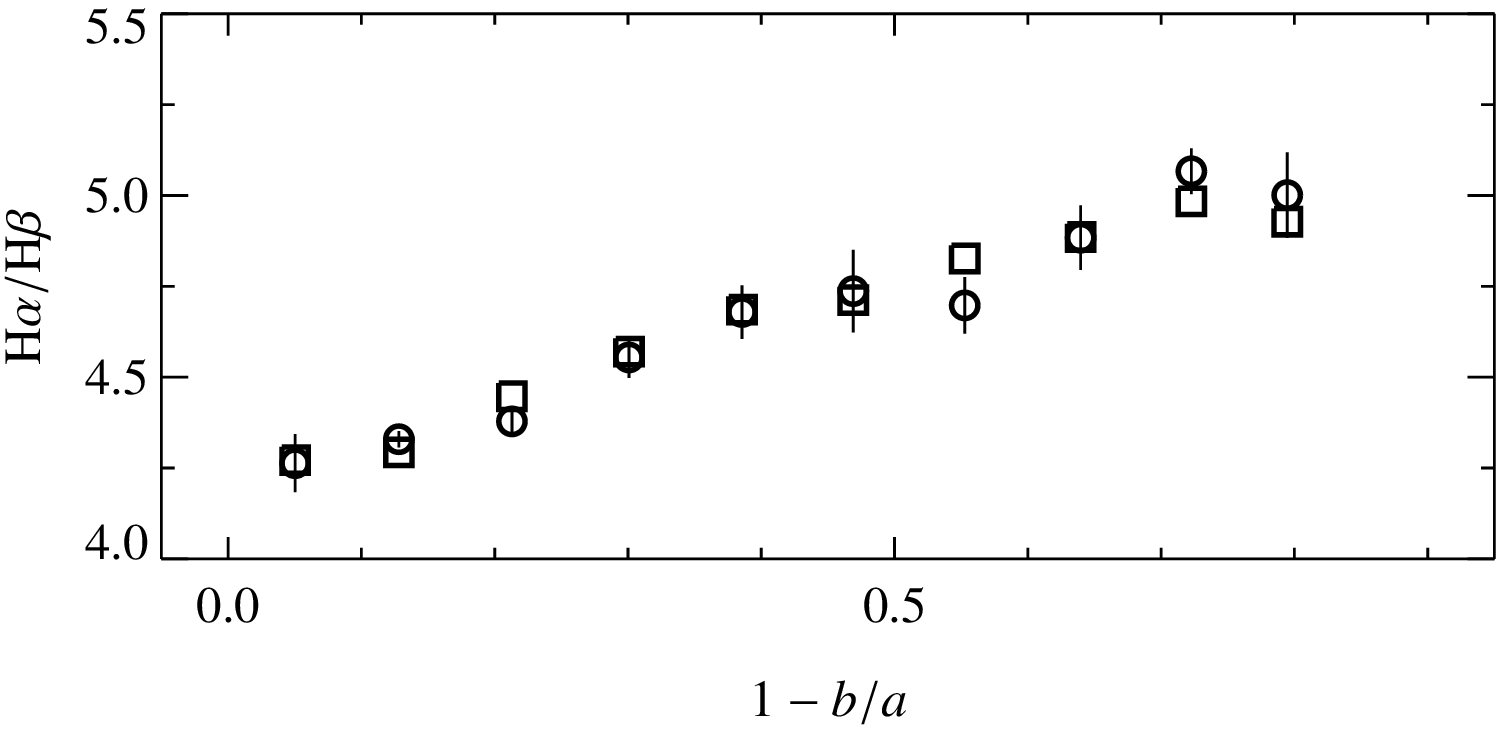}}}
	\subfigure{
		\resizebox{\hsize}{!}{\includegraphics{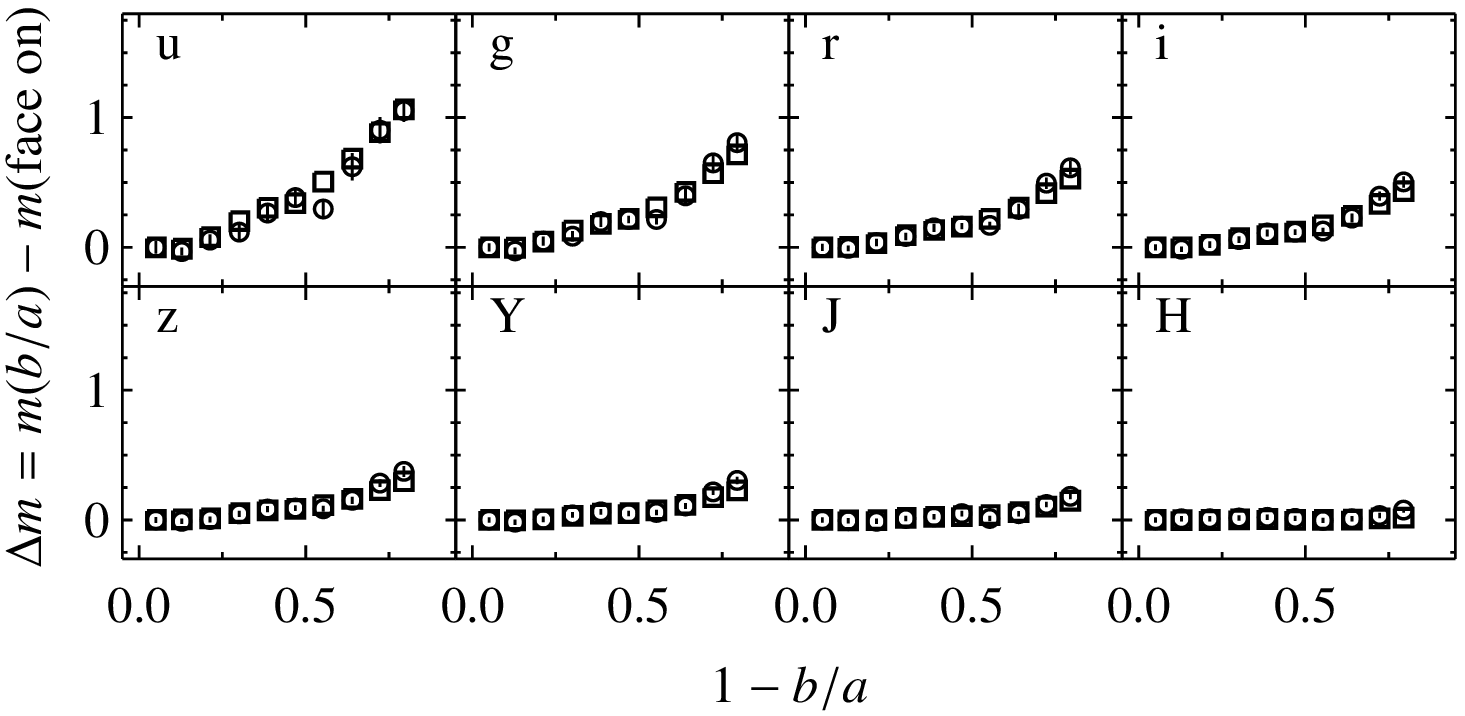}}}
	\caption{Same as Fig.~\ref{fig:data_low}, but for the high-\muS\ galaxies in the W11 sample described in Section~\ref{sec:sample}.}
	\label{fig:data_high}
\end{figure}

\begin{figure}
	\centering
	\resizebox{\hsize}{!}{\includegraphics{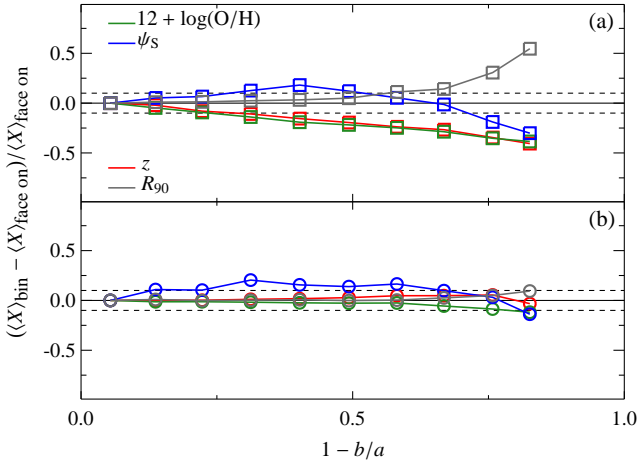}}
	\caption{Relative difference between the mean physical property of galaxies in bins of different inclination, from face-on ($1-b/a\approx0$) to edge-on ($1-b/a\approx1$), and those of face-on galaxies, for the low-\muS\ galaxies in the W11 sample described in Section~\ref{sec:sample} ({\it dark green}: \logOH; {\it blue}: \psiS; {\it red}: $z$; {\it dark grey}: $R_{90}$). ({\it a}) As measured in the original W11 sample. ({\it b}) After correcting these measurements for systematic biases using the method of importance sampling described in Section~\ref{sec:bias_corr} and Appendix~\ref{app:importance}. For reference, the dashed horizontal lines show the effect of a $\pm10$ percent bias.}
	\label{fig:bias_low}
\end{figure}

\begin{figure}
	\centering
	\resizebox{\hsize}{!}{\includegraphics{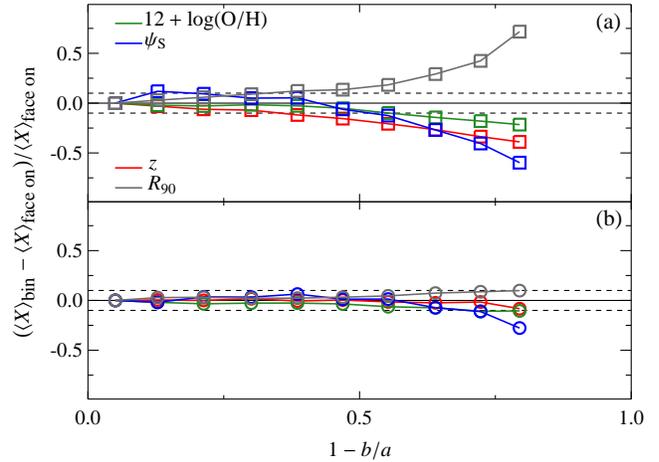}}
	\caption{Same as Fig.~\ref{fig:bias_low}, but for the high-\muS\ galaxies in the W11 sample described in Section~\ref{sec:sample}.}
	\label{fig:bias_high}
\end{figure}

In Figs~\ref{fig:data_low} and \ref{fig:data_high}, we show the \HaHb\ ratio and differential $ugrizYJH$ magnitudes, $\Delta m=m(b/a)-m(\txn{face-on})$, obtained in this way as a function of $b/a$, for the low- and high-\muS\ samples, respectively (open squares). As expected, when $b/a$ decreases, i.e. as galaxies are seen more edge-on, the \HaHb\ ratio rises, the ultraviolet and optical flux drop significantly ($\Delta m >0$) while the near-infrared flux remains roughly constant ($\Delta m\approx0$). To discriminate the true influence of inclination on these changes,  we must ascertain that other physical parameters, which correlate with dust content, do not also systematically vary with $b/a$. For this purpose, we show in Figs~\ref{fig:bias_low}a and \ref{fig:bias_high}a the difference between bin-averaged and sample-averaged values of several physical parameters as a function of $b/a$, for the low- and high-\muS\ samples, respectively. The parameters we consider are: gas-phase oxygen abundance, \logOH; specific star formation rate, \psiS; redshift, $z$; and radius enclosing 90 percent of the total Petrosian $r$-band flux, $R_{90}$. 

Figs~\ref{fig:bias_low}a and \ref{fig:bias_high}a reveal important biases in the physical properties of galaxies seen at different inclinations. In both the low- and high-\muS\ samples, galaxies seen more edge-on tend to have on average systematically lower \logOH, \psiS\  and $z$ and larger $R_{90}$ than galaxies seen more edge-on. This is likely a consequence of the high signal-to-noise ratio required to detect the \Nii, \Ha, \Oiii\ and \Hb\ emission lines in Section~\ref{sec:sample} (although part of the trends could also result from measurement biases). In particular, since the dust attenuation optical depth increases with inclination, and since the most metal-rich galaxies also contain the most dust (see fig.~6 of \citealt{Tremonti2004} and fig.~6 of \citealt{Brinchmann2004}), at high inclination, nebular emission lines of metal-rich galaxies are more difficult to detect than those of metal-poor galaxies. The fact that the specific star formation rate correlates with gas- and hence dust-mass fraction (see Fig.~6 of \citealt{daCunha2010}) produces a similar bias against highly-inclined, actively star-forming galaxies in Figs~\ref{fig:bias_low}a and \ref{fig:bias_high}. A bias against small $R_{90}$ at high inclination arises from a positive correlation between $R_{90}$ and surface brightness for the galaxies in our sample, and possibly from the blurring of the shapes of the smallest galaxies by the observational point spread function. This blurring may also be responsible for the bias against highly inclined galaxies at large $z$, together with cosmological dimming. These different biases are substantial, reaching roughly 40 (20) percent of the sample-averaged value in \logOH, 30 (60) percent in \psiS, 40 (40) percent in $z$ and 50 (70) in $R_{90}$ for low-\muS\  (high-\muS) galaxies. Such biases must be corrected to properly investigate the dependence of dust attenuation on galaxy inclination.

In the absence of a reliable model to describe complex correlations between \logOH , \psiS, $z$ and $R_{90}$, we can reduce the strength of the biases in Figs~\ref{fig:bias_low}a  and \ref{fig:bias_high}a using the data alone. Changes in the mean value of a given parameter from a bin to another in these figures reflect differences in the underlying distributions of that parameter at different $b/a$. The combined changes in mean \logOH , \psiS, $z$ and $R_{90}$ therefore reflect differences in the joint distributions of these parameters. To remove the influence of these differences on the relation between dust attenuation and galaxy inclination, we require a joint distribution of [\logOH , \psiS, $z$, $R_{90}$] common to all $b/a$ bins in Figs~\ref{fig:bias_low}a and \ref{fig:bias_high}a. This can be extracted from the original sample by appealing to the formalism of importance sampling, a well known Monte Carlo technique used to solve, for instance, multi-dimensional integrals \citep[e.g.][]{MonteCarloIntegration}. This method requires two conditions: that the joint probability distribution of [\logOH , \psiS, $z$, $R$] be evaluated in each bin (this implies computing probability distributions from a finite sampling of observables); and that the intersection of the joint probability distributions of different bins be non-zero. In Appendix~\ref{app:importance}, we describe in detail our implementation of this technique to derive a joint probability distribution of [\logOH , \psiS, $z$, $R_{90}$] common to all bins of $b/a$. The mean spectral properties of galaxies obeying this distribution are shown as a function of galaxy inclination by the open circles in Figs~\ref{fig:data_low} and \ref{fig:data_high}. The associated error bars reflect the standard error of the mean. Figs~\ref{fig:bias_low}b and \ref{fig:bias_high}b show that, after correction, any systematic change of a parameter with inclination amounts to at most about 10 percent of the face-on value of that parameter, for both the low- and high-\muS\ galaxy samples.

\begin{figure}
	\centering
		\resizebox{\hsize}{!}{\includegraphics{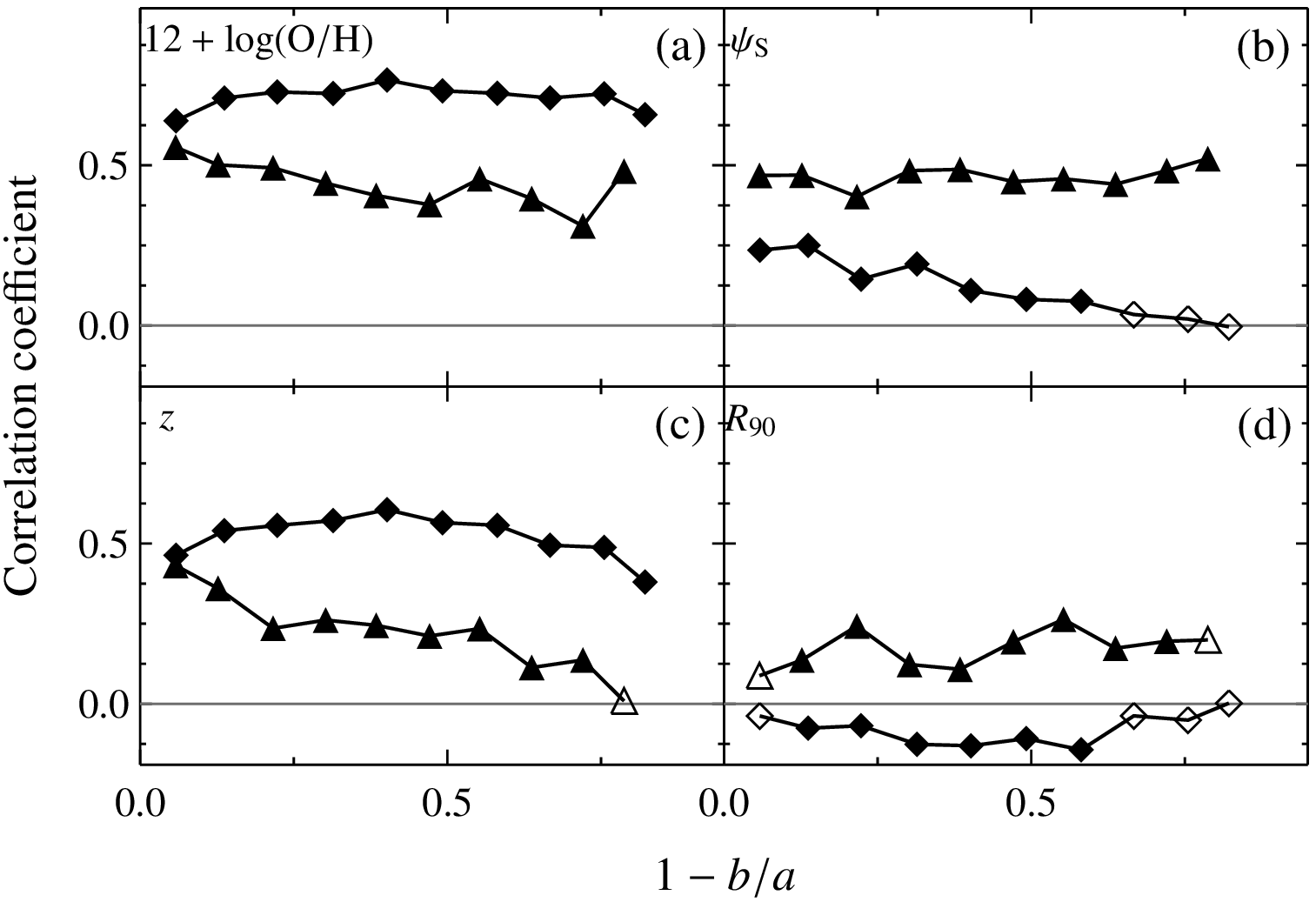}}
		\caption{\citet{Spearman1904} rank correlation coefficient for the relations between \HaHb\ ratio and different physical properties of low-\muS\ (diamonds) and high-\muS\ (triangles) galaxies in the W11 sample described in Section~\ref{sec:sample}, in bins of different inclination, from face-on ($1-b/a\approx0$) to edge-on ($1-b/a\approx1$). ({\it a}) \logOH; ({\it b}) \psiS; ({\it c}) $z$ and ({\it d}) $R_{90}$. In each panel, filled and open symbols indicate a significance level of the correlation larger and smaller than $99.99$ percent, respectively.}
	\label{fig:correl_coeff}
\end{figure}

Figs~\ref{fig:data_low} and \ref{fig:data_high} show that the corrections of systematic biases in the physical properties of galaxies as a function of inclination have a larger impact on the  \HaHb\ ratio and differential $ugrizYJH$ magnitudes  of low-\muS\ galaxies than on those of their high-\muS\ counterparts, especially at high inclination. To investigate the origin of this result, we plot in Fig.~\ref{fig:correl_coeff} the \citet{Spearman1904} rank correlation coefficient for the relations between \HaHb\ ratio and the four parameters \logOH , \psiS, $z$ and $R_{90}$, in each bin of $b/a$, for low- and high-\muS\ galaxies. Figs~\ref{fig:correl_coeff}a and c show that, for low-\muS\ galaxies (diamonds), \logOH\ and $z$ are strongly correlated with \HaHb\ ratio at all inclinations. These are also the quantities exhibiting the largest biases at all inclinations in Fig.~\ref{fig:bias_low}a. For high-\muS\ galaxies (triangles), Figs~\ref{fig:correl_coeff}a--c show that \logOH\ and $z$ are less strongly correlated with \HaHb\ ratio than for low-\muS\ galaxies, while \psiS\ is much more strongly correlated and $R_{90}$ equally weakly correlated but in the opposite sense. The complex interplay between these different trends is likely to be the reason for the different corrections of the relation between \HaHb\ ratio and inclination pertaining to low- and high-\muS\ galaxies in Figs~\ref{fig:data_low} and \ref{fig:data_high}. We cannot produce the equivalent of Fig.~\ref{fig:correl_coeff} for the differential $ugrizYJH$ magnitudes, because in any given bin of $b/a$, we compute only one measure of $\Delta m$ to average out the intrinsic variations in the spectral energy distributions of all galaxies in that bin (hence no correlation coefficient can be computed within a bin).

\subsection{Library of model spectral energy distributions}\label{sec:model_lib}

The dust attenuation models presented in Section~\ref{sec:models} provide an ideal framework to interpret the observed dependence of the \HaHb\ ratio and $ugrizYJH$ attenuation curve on galaxy geometry and inclination (Figs~\ref{fig:data_low} and \ref{fig:data_high}). The four  types of models we investigated in Section~\ref{sec:RT} all predict similar changes of the shape of the attenuation curve in the ambient ISM as a function of $V$-band attenuation optical depth \tauVismTh\ (Section~\ref{sec:S_A_rel}), and of the attenuation optical depth \tauLismTh\ in the $ugrizYJH$ bands as a function of galaxy inclination $\theta$ (Section~\ref{sec:tau_theta}). To explore the physical origin of the observed trends in Figs~\ref{fig:data_low} and \ref{fig:data_high}, we appeal to the most versatile of these models, by T04, in which galaxies are described as the sum of three stellar components (a thin and a thick disc, and a bulge) and two dust components (associated with the stellar discs; see Section~\ref{sec:T04}). In the next paragraphs, we combine the T04 model with a hierarchical galaxy formation model, a stellar population synthesis code and a prescription for dust attenuation in stellar birth clouds to build a  library of spectral energy distributions tailored to the analysis of the sample of Section~\ref{sec:data}.

\subsubsection{Spectral energy distributions of stellar populations}\label{sec:stelpops}

We follow the approach recently developed by \citet{Pacifici2012} and appeal to state-of-the-art models of star-formation and chemical-enrichment histories to build a broad library of as realistic as possible spectral energy distributions of galaxies. We start from a catalog of 40\,000 star formation and chemical enrichment histories produced by \citet{Pacifici2012} through the semi-analytic post-treatment (based on  recipes by \citealt{DeLucia2007}) of the Millennium cosmological simulation of \citet{Springel2005}. As discussed by \citet[see their section~2.1]{Pacifici2012}, star formation histories derived in this way are appropriate mainly to describe the properties of nearby SDSS galaxies more massive than $5 \times 10^9 h^{-1} \Msun$. The parameter space must be broadened to interpret observations of galaxies drawn from other samples and in other mass ranges, in particular by resampling the distributions of galaxy evolutionary stages (i.e. the age at which a galaxy is looked at) and current galaxy properties (i.e. the mean star formation rate and metallicity of new stars over a period of 10 Myr before the galaxy is looked at). Such resampling is compatible with the stochastic nature of star formation and chemical evolution in hierarchical scenarios of galaxy formation (see \citealt{Pacifici2012} for detail).

\begin{figure}
	\centering
	\resizebox{0.7\hsize}{!}{\includegraphics{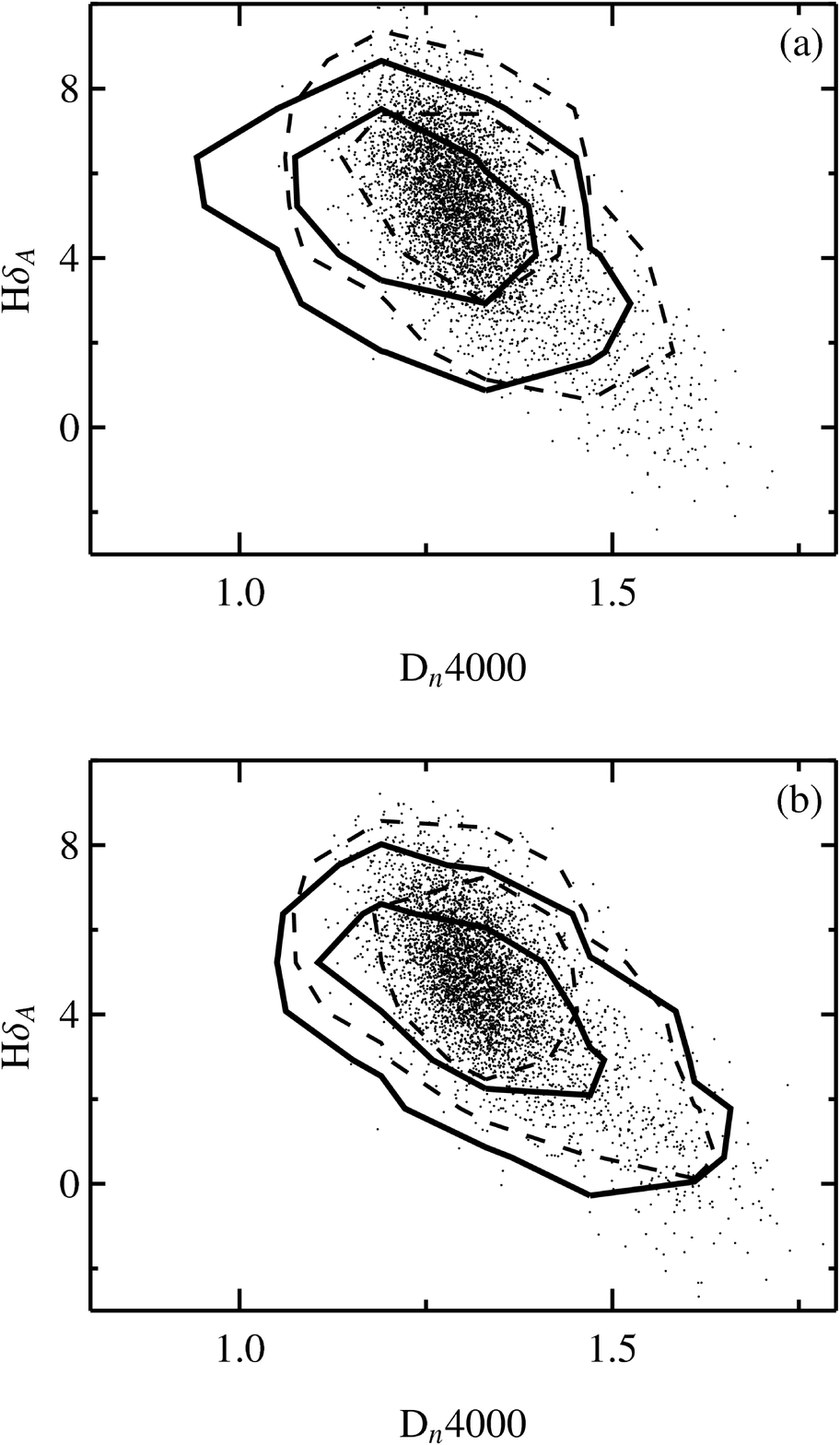}}
	\caption{({\it a}) Balmer stellar absorption-line index \Hd\ versus 4000-{\AA} discontinuity \Dn\ for the low-\muS\ galaxies in the W11 sample described in Section~\ref{sec:sample} (solid contours, encompassing areas of 1, 10 and 20 times the mean density) and the library of 5000 model spectral energy distributions generated in Section~\ref{sec:stelpops} to analyse this sample (dots and dashed contours). ({\it b})~Same as ({\it a}), but for the high-\muS\ galaxies.}
	\label{fig:D4000_Hdelta}
\end{figure}

From the 40\,000 star formation and chemical enrichment histories mentioned above, we therefore produce two smaller catalogs of 5000 galaxies each, appropriate to analyse the observed low- and high-\muS\ samples of Section~\ref{sec:data} (we have checked that generating twice-larger catalogs does not affect our conclusions). In practice, we extract models one by one from the original library and randomly redraw evolutionary stage, specific star formation rate and metallicity of newly-born stars from prior distributions appropriate for the low- and high-\muS\  samples. Specifically, we take the prior distributions of specific star formation rate and metallicity of newly-born stars to be the distributions of \psiS\ and \logOH\  of the observed galaxies in both samples (Fig.~\ref{fig:cur_params} of Section~\ref{sec:sample}). To redraw evolutionary stage, we appeal to the standard starburst-diagnostic diagram defined by the strengths of the 4000\,\AA\ break and the H$\delta_A$ stellar absorption-line indices \citep[e.g.][]{Kauffmann2003a}. This diagram reflects the balance between newly born stars with low \Dn\ and \Hd, intermediate-age stars with large \Hd\ and older stars with large \Dn. After some experimentation, we find that resampling the age at which the galaxy is looked at uniformly in redshift in the interval 0--1 (0--0.5) and rejecting galaxies that formed less than 15 (10) percent of their stars in the last 2.5\,Gyr (parameter \fSFH\ of \citealt{Pacifici2012}) provides model distributions in the \DnHd\ diagram in qualitative agreement with the observations of low-\muS\  (high-\muS) galaxies. This is illustrated by Figs~\ref{fig:D4000_Hdelta}a and \ref{fig:D4000_Hdelta}b for the low-\muS\  and high-\muS\ galaxy samples, respectively.

As in \citet{Pacifici2012}, we combine the two libraries of star formation and chemical enrichment histories generated in this way with the latest version of \citet{BC03} stellar population synthesis code (Charlot \& Bruzual, in preparation) to compute the spectral energy distribution of every galaxy at wavelengths between 91\,\AA\ and 160\,$\mu$m . This code incorporates a new library of observed stellar spectra at optical wavelengths \citep{MILES} and new prescriptions for the evolution of stars less massive than 20\,\Msun\ \citep{Bertelli2008,Bertelli2009} and of thermally pulsing asymptotic giant branch (TP-AGB) stars \citep{Marigo2008}. We adopt here the Galactic-disc stellar initial mass function of  \citet{Chabrier2003}. 

\subsubsection{Inclusion of dust modelling}\label{sec:dust}

We adopt the versatile radiative transfer code of T04 to describe the attenuation of starlight by dust in the ambient (i.e. diffuse) ISM (Section~\ref{sec:T04}). We must also account for the enhanced attenuation of newly born stars in their parent molecular clouds, which dissipate on a timescale typically of the order of 10\,Myr \citep[e.g.][]{Murray2010,Murray2011}. We achieve this by combining the T04 model with the simple two-component, angle-averaged dust model of \citet{CF00}, which accounts for the different attenuation affecting young and old stars in galaxies. We introduce a dependence on viewing angle in the prescription of \citet{CF00} and write the total attenuation optical depth at galaxy inclination $\theta$ affecting the radiation from stars of age $t$ as
\begin{equation}
\hat{\tau}^\txn{tot}_\lambda(\theta,t)=\left\{ \begin{array}{l l}
\tauLbc + \tauLismTt & \hspace{3mm} \txn{for} \hspace{3mm} t \leqslant 10\,\txn{Myr}\,,\\
\tauLismTt  & \hspace{3mm} \txn{for} \hspace{3mm}  t>10\,\txn{Myr}\,. \end{array}\right.
\label{eq:tau_cf00}
\end{equation}
Here \tauLbc\ is the attenuation optical depth, assumed isotropic, of the stellar birth clouds, and \tauLismTt\  that of the ambient ISM seen by stars of age $t$ at galaxy inclination $\theta$. Following \citet[see also \citealt{daCunha2008}]{Wild2007}, we take the attenuation curve in the birth clouds to be
\begin{equation}
\tauLbc = \tauVbc \, \left (\lambda/0.55\,\micron \right ) ^ {-1.3}\,
\end{equation}
where  \tauVbc\ is the angle-averaged $V$-band attenuation optical depth.

We describe \tauLismTt\ using the multi-component model of T04. By analogy with Section~\ref{sec:tau_theta}, we relate stars in different age ranges in a model galaxy to different geometric components of the T04 model. This is supported by observational evidence that stars in quiescent (i.e. non-starburst) star-forming galaxies tend to form along spiral arms in a thin disc close to the equatorial plane. The scale height of these stars can later increase as a consequence of dynamical heating and merging \citep{Yoachim2006,Roskar2012}, as suggested by the distinct stellar populations of the thin- and thick-disc components observed in external galaxies \citep{Yoachim2008A,Yoachim2012}. Finally, in normal star-forming galaxies, the oldest stellar populations are generally found in a bulge component \citep[e.g.][]{Kauffmann2003b}. We therefore consider the same three age parameters \Tthin, \Tthick\ and \Tbulge\  as in Section~\ref{sec:tau_theta}, such that $\Tthin \leq \Tthick \leq \Tbulge$, and associate stars younger than \Tthin\  with the thin-disc component, those in the age range $ \Tthin < t \leq \Tthick$ with the thick-disc component and those in the age range $ \Tthick < t \leq \Tbulge$ with the bulge component. We compute the attenuation optical depth in the ambient ISM seen by stars of age $t$ at galaxy inclination $\theta$ in equation~(\ref{eq:tau_cf00}) as
\begin{equation}
\tauLismTt=\left\{ \begin{array}{l l}
\tauLthinth & \hspace{3mm} \txn{for} \hspace{3mm} t \leqslant \Tthin \,,\\
\tauLthickth & \hspace{3mm} \txn{for} \hspace{3mm} \Tthin < t \leqslant \Tthick \,,\\
\tauLbulgeth  & \hspace{3mm} \txn{for} \hspace{3mm} \Tthick < t \leqslant \Tbulge \,, \\
\,\,0 & \hspace{3mm} \txn{for} \hspace{3mm}  t>\Tbulge\,,\end{array}\right.
\label{eq:tau_T04ism}
\end{equation}
where \tauLthinth, \tauLthickth\ and \tauLbulgeth\ are the attenuation optical depths of the thin-disc, thick-disc and bulge components in the T04 model. All three quantities are parametrized in terms of the central face-on $B$-band optical depth of the combined thin and thick dust discs, \taubP\ (Section~\ref{sec:T04}). We note that the age parameters  \Tthin, \Tthick\ and \Tbulge\ allow one to explore any combination of the T04 geometric components: a model can be described by a single component (i.e. a pure thin disc, if $\Tthin = \Tthick = \Tbulge$; a pure thick disc, if there are no stars younger than \Tthin\ and if $\Tthick = \Tbulge$; and a pure bulge, if there are no stars younger than $\Tthin = \Tthick$), two components (e.g. a thin and a thick stellar discs, if $\Tthin \ne \Tthick$ and $ \Tthick= \Tbulge$) or three components (if $\Tthin \ne \Tthick \ne \Tbulge$). The precise choice of  \Tthin, \Tthick\ and \Tbulge, combined with the star formation history,  determines the contributions by the different components to the total intensity produced by stars in the galaxy (weights $\xi_{\lambda}^i$; equation~\ref{eq:xi}). In summary, we model the attenuation of starlight by dust using five adjustable parameters,  \tauVbc,  \taubP, \Tthin, \Tthick\ and \Tbulge.

In Table~\ref{tab:prior}, we list the ranges in these parameters adopted in the next section to interpret observations of nearby star-forming galaxies. The largest observed \HaHb\ ratio in the sample of Section~\ref{sec:data} corresponds to a total $V$-band attenuation optical depth of roughly 1.3. This includes both attenuation in the birth clouds of ionising stars and through the intervening ambient ISM. We therefore sample \tauVbc\ uniformly in the interval from 0 to 1.5. Furthermore, we sample \taubP\ uniformly across the full range considered by T04, i.e. from 0 to 8. Finally, to satisfy the condition $\Tthin \leq \Tthick \leq \Tbulge$, we take \Tthin\ (in Gyr) to be distributed logarithmically between $-3$ and $1.13$,  \Tthick\ logarithmically between $\txn{max}[\lgTthin,-1.0]$ and 1.13, and \Tbulge\ logarithmically between $\txn{max}[\lgTthick,0.0]$ and 1.13.

\begin{table*}
	\centering
	\begin{tabular}{c c c c}

\toprule

{\bf Parameter}	         & {\bf Definition}	& {\bf Lower bound}  & {\bf Upper bound}    \\

\midrule

\vspace{\colhspace}
{\tauVbc}		                 &   \begin{minipage}{\LargecolWidth} $V$-band attenuation optical depth of dust in stellar birth clouds. \end{minipage} & 0.0 & 1.5   \\
\vspace{\colhspace}
{\taubP} 				&    \begin{minipage}{\LargecolWidth} Central face-on $B$-band attenuation optical depth of dust in the diffuse ISM. \end{minipage}  & 0 &  8   \\
\vspace{\colhspace}
$\lgTthin$ 		&   \begin{minipage}{\LargecolWidth}  Stars with ages $ t\leq \Tthin$ are associated with the T04 thin-disc component.  \end{minipage}  	& $-3.0$ &  1.13 \\
\vspace{\colhspace} 		 		
$\lgTthick$ 		&   \begin{minipage}{\LargecolWidth}  Stars with ages $ \Tthin< t \leq \Tthick$ are associated with the T04 thick-disc component. \end{minipage} 	& $\txn{max}[\lgTthin,-1.0]$ & 1.13 \\
$\lgTbulge$  	&   \begin{minipage}{\LargecolWidth} Stars with ages $ \Tthick < t \leq \Tbulge $ are associated with the T04 bulge component. \end{minipage}  		& $\txn{max}[\lgTthick,0.0]$ &  1.13  \\

\bottomrule

	\end{tabular}
	\caption{Prior distributions of the five main adjustable parameters of the dust model described in Section~\ref{sec:dust}.}
	\label{tab:prior}	
\end{table*}

\subsection{A Bayesian approach: Markov Chain Monte Carlo with the Metropolis-Hastings algorithm}\label{sec:bayes_approach}

We now use the library of model spectral energy distributions presented in Section \ref{sec:model_lib} to interpret the observed dependence of the \HaHb\ ratio and $ugrizYJH$ attenuation curve on geometry and inclination in the sample of nearby star-forming galaxies described in Section~\ref{sec:data}. To search for the combination of parameters  \tauVbc, \taubP,  \Tthin, \Tthick\  and \Tbulge\ that best reproduces the observed trends, we adopt a Bayesian approach. Technically, this amounts to computing posterior probability distributions of the adjustable parameters from the thorough exploration of the parameter space, given initial prior distributions and the data available to constraint the model. The Markov Chain Monte Carlo (MCMC) approach allows efficient exploration of complex, multi-dimensional parameter spaces. It differs from basic Monte Carlo methods through the inclusion of correlations between the random steps composing a chain, each step depending on the previous one (Markovian property). Such correlations allow an efficient exploration of the parameter space, making the algorithm spend most of the computation time in regions of highest probability (see Appendix~\ref{app:MCMC} for more detail). 

To sample in a Markovian way the posterior probability distributions of the five adjustable parameters of the dust model, we adopt the widely used random-walk Metropolis-Hastings algorithm \citep{Metropolis1953,Hastings1970}, as implemented in the publicly available code \CosmoMC\ \citep{CosmoMC}. At each draw (step $j$) of a new set of parameters from a `proposal' distribution (see below), we compute the posterior distribution of this set by comparing the predicted and observed mean spectral properties of galaxies as a function of axis ratio. In practice, the adoption of uniform priors makes the posterior distribution proportional to the likelihood distribution, which we compute assuming independent Gaussian errors on the observations, as
\begin{equation*}
-\ln{\mathcal{L}(\bmath{\Theta}^\prime_j)} \propto \sum_k\sum_i \frac{1}{2} \left ( \frac{\Phi_i^k(\bmath{\Theta}^\prime_j) - \tilde{\Phi}_i^k}{\sigma_i^k} \right ) ^2\,.
\end{equation*}
In this expression, the index $k$ runs across the 10 bins of $b/a$ and the index $i$ across all spectral properties, from \HaHb\ ratio to differential $ugrizYJH$ magnitudes. The quantity $\Phi_i^k(\bmath{\Theta}^\prime_j)$ is the value of the $i^\txn{th}$ observable obtained when using the set of parameters $\bmath{\Theta}^\prime_j$ in the dust model of Section~\ref{sec:dust}, while $\tilde{\Phi}_i^k$ is the observed (bias-corrected) value of that observable from Section~\ref{sec:bias_corr}. We estimate the observational error $\sigma_i^k$ from the variance of the importance weights used to correct the data for biases (Section~\ref{sec:bias_corr} and Appendix~\ref{app:importance}; we add an additional error of 0.02\,mag in quadrature to account for the residual biases for the differential $ugrizYJH$ magnitudes; such a correction is negligible for the \HaHb\ ratio).
The ratio of the posterior probabilities of two consecutive sets of parameters orients the move toward the high-probability region of the parameter space. The result of the algorithm is a sequence, or chain, of sets of parameters $\bmath{\Theta}_j$, which build up step by step the posterior probability density functions of \tauVbc, \taubP,  \Tthin, \Tthick\  and \Tbulge.

\begin{figure}
	\centering
	\resizebox{\hsize}{!}{\includegraphics{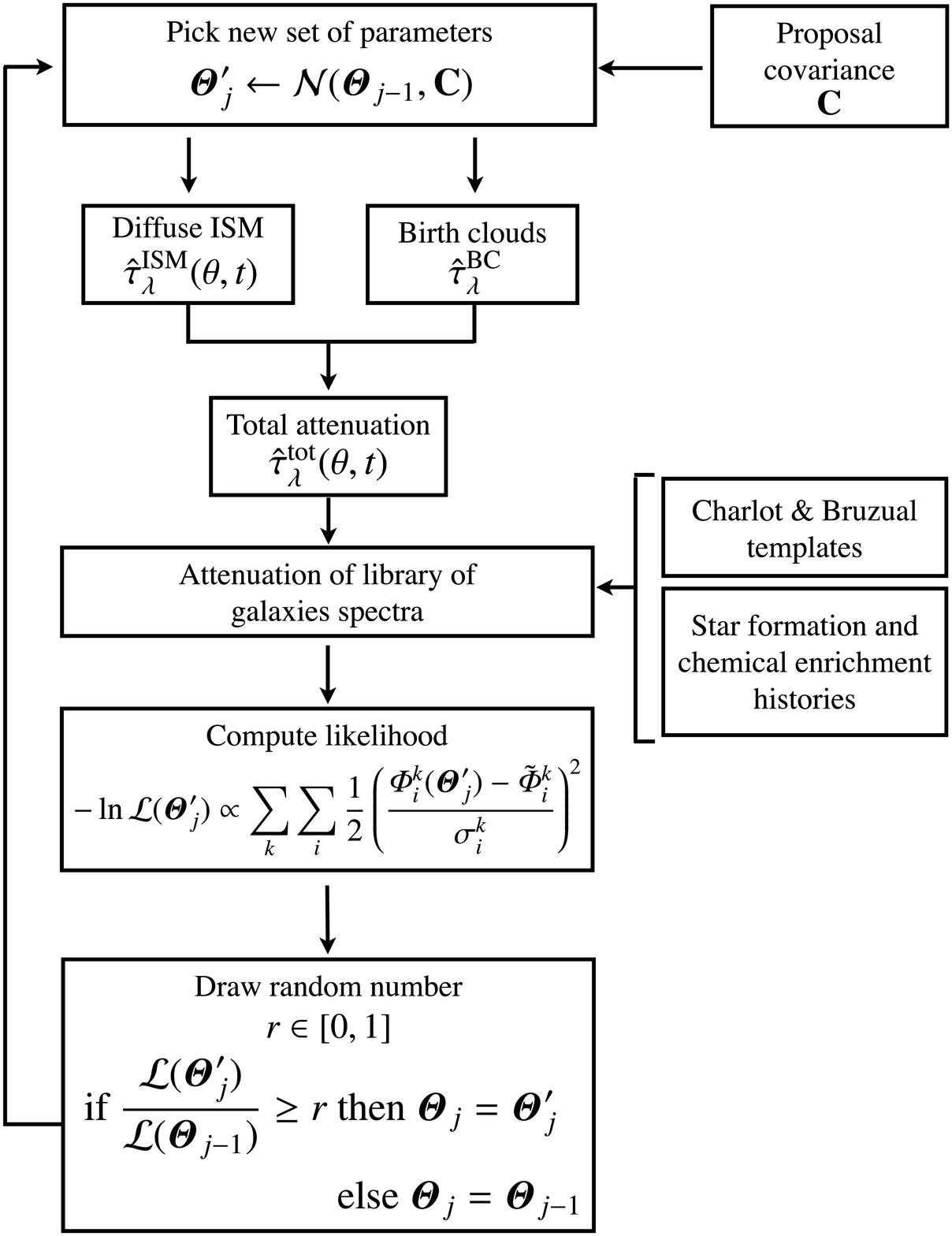}}
	\caption{Flow chart summarising the random-walk Metropolis-Hastings algorithm \citep{Metropolis1953,Hastings1970} adopted in Section~\ref{sec:dust_constraints} to sample the posterior probability distributions of the adjustable parameters \tauVbc , \taubP, \Tthin, \Tthick\ and \Tbulge, assuming flat, truncated priors in the ranges given in Table~\ref{tab:prior}.}
	\label{fig:flow_chart}
\end{figure}

A more detailed description of this procedure is presented in Fig.~\ref{fig:flow_chart}. The chain starts at step $j=0$ (not shown) with a random drawing of parameters within the ranges given in Table~\ref{tab:prior}. At each new step $j$, a candidate set of parameters $\bmath{\Theta}^\prime_j$ is drawn from the proposal distribution, which in \CosmoMC\ is a multi-variate normal distribution, noted $\mathcal{N}(\bmath{\Theta}_{j-1},\mathbfss{C})$, centred on the previous set $\bmath{\Theta}_{j-1}$ and with covariance matrix $\mathbfss{C}$.\footnote{
Theoretically, the optimum choice for $\mathbfss{C}$ is some scaled version of the covariance matrix of the posterior probability distributions \citep{Roberts2001}.  In practice, this quantity being unknown in our case, the possibility with \CosmoMC\ to build up $\mathbfss{C}$ through step-by-step adjustments is extremely time consuming. Following \citet{Trotta2011}, we therefore appeal to the freely available code \MultiNest\ \citep{Feroz2008,Feroz2009} to obtain a fast estimate of $\mathbfss{C}$ based on the nested sampling algorithm of \citet[see Appendix \ref{app:nested} for detail]{Skilling2004,Skilling2006}.} Then, we use the candidate set of parameters $\bmath{\Theta}^\prime_j$ to attenuate the spectral energy distributions of the 5000 stellar populations in the library assembled in Section~\ref{sec:stelpops}, as described in Section~\ref{sec:dust}. This allows us to compute the likelihood $\mathcal{L}(\bmath{\Theta}^\prime_j)$ of the candidate set by comparing the predicted and observed mean spectral properties of galaxies as a function of $b/a$ (we assume Gaussian observational errors). The candidate set is accepted, i.e. we adopt $\bmath{\Theta}_j=\bmath{\Theta}^\prime_j$, if the likelihood ratio $\mathcal{L}(\bmath{\Theta}^\prime_j)/\mathcal{L}(\bmath{\Theta}_{j-1})$ is larger than a critical value $r$ drawn uniformly at random from between 0 and 1. Otherwise, we draw a new candidate set of parameters $\bmath{\Theta}^\prime_j$. 

We perform these steps along six chains run in parallel, until stationary posterior probability density functions are reached (this is usually achieved for $j$ around a few thousand). We determine convergence through a comparison of the intra- and inter-chain variance for each parameter, setting the criterion of \citet{Gelman1992} to $1-\hat{R} = 0.01$. As is standard with \CosmoMC, we compute final posterior probability distributions by combining the samples produced in the last half of all chains. In Appendix \ref{sec:convergence}, we provide more detail on this procedure, as well as a visual representation of the build-up and adequate sampling of the posterior probability distribution of each parameter.

\subsection{Constraints on the content and spatial distribution of dust in nearby star-forming galaxies}\label{sec:dust_constraints}

\begin{figure*}
	\centering
	\resizebox{\hsize}{!}{\includegraphics{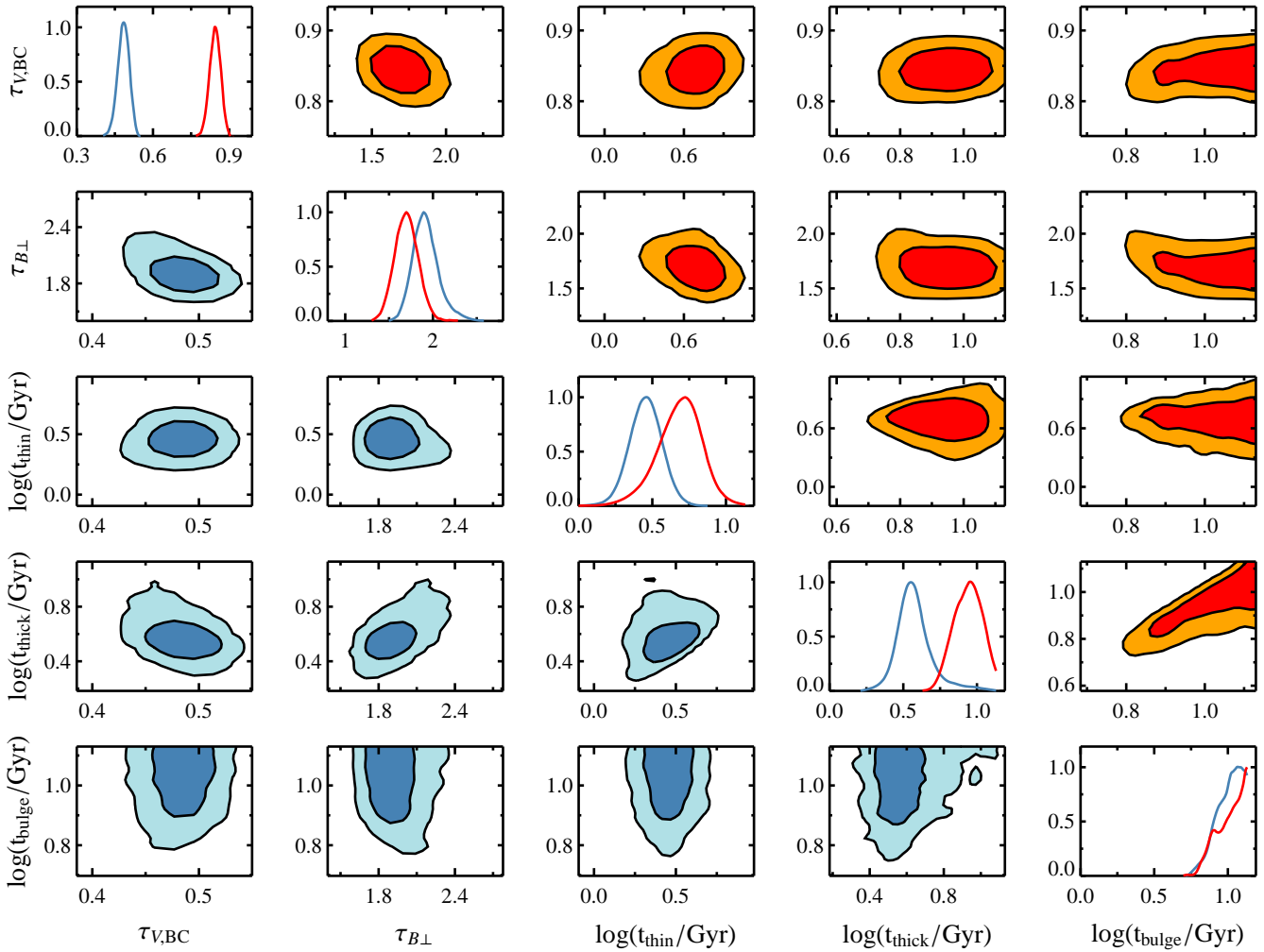}}
	\caption{Results of the Bayesian MCMC analysis of the dependence of the \HaHb\ ratio and $ugrizYJH$ attenuation curve on galaxy inclination in the W11 sample of nearby star-forming galaxies described in Section~\ref{sec:sample}, using the library of model spectral energy distributions assembled in Section~\ref{sec:model_lib}. The diagonal panels show the marginal posterior probability distributions for \tauVbc, \taubP, \Tthin, \Tthick\ and \Tbulge, for the low-\muS\ (blue lines) and high-\muS\ (red lines) galaxies. The off-diagonal panels show the joint probability distributions of the same parameters, for the low-\muS\ (blue contours) and high-\muS\ (red contours) galaxies. In each of these panels, the contours enclose 68 and 95 percent of the posterior probability.}
	\label{fig:joint}
\end{figure*} 

\begin{figure}
	\centering
	\subfigure
	{\resizebox{\hsize}{!}{\includegraphics{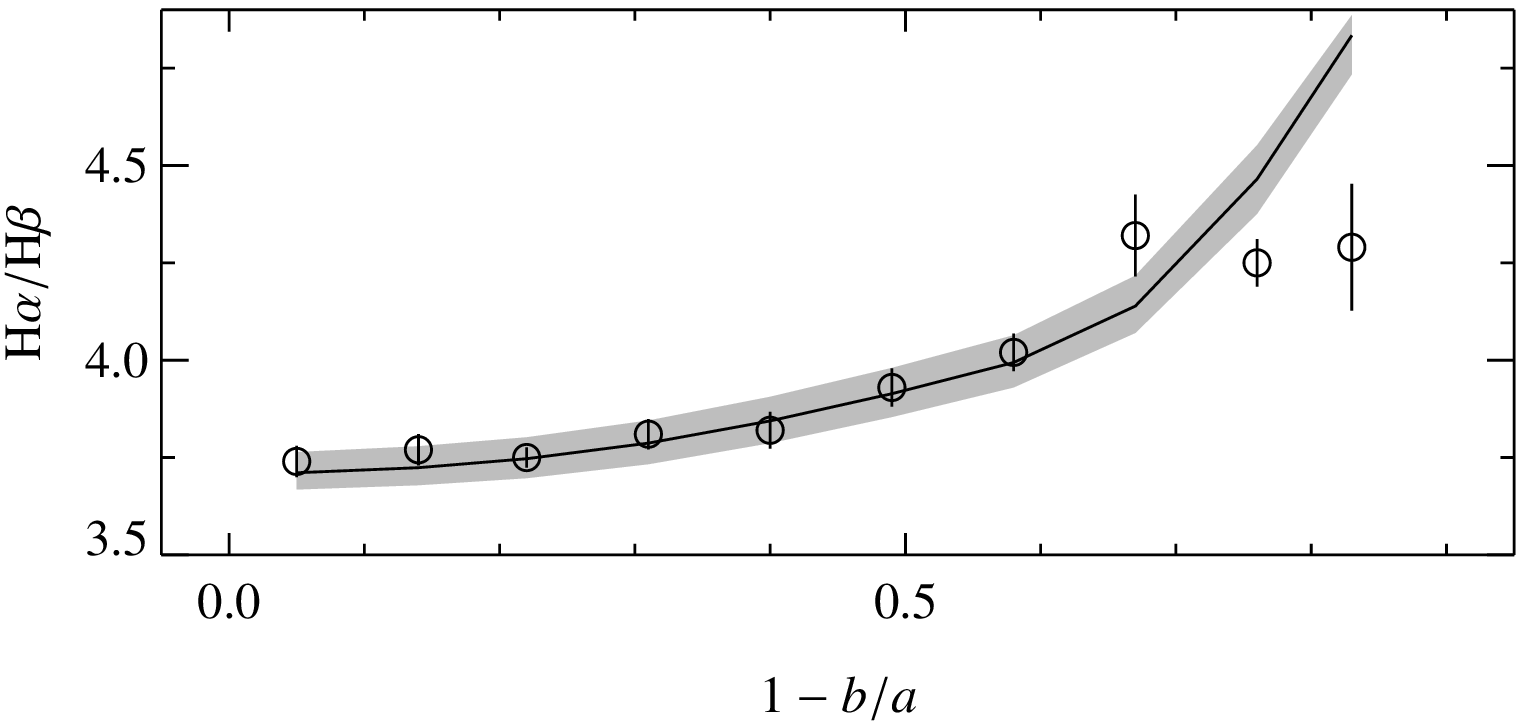}}}
	\subfigure
	{\resizebox{\hsize}{!}{\includegraphics{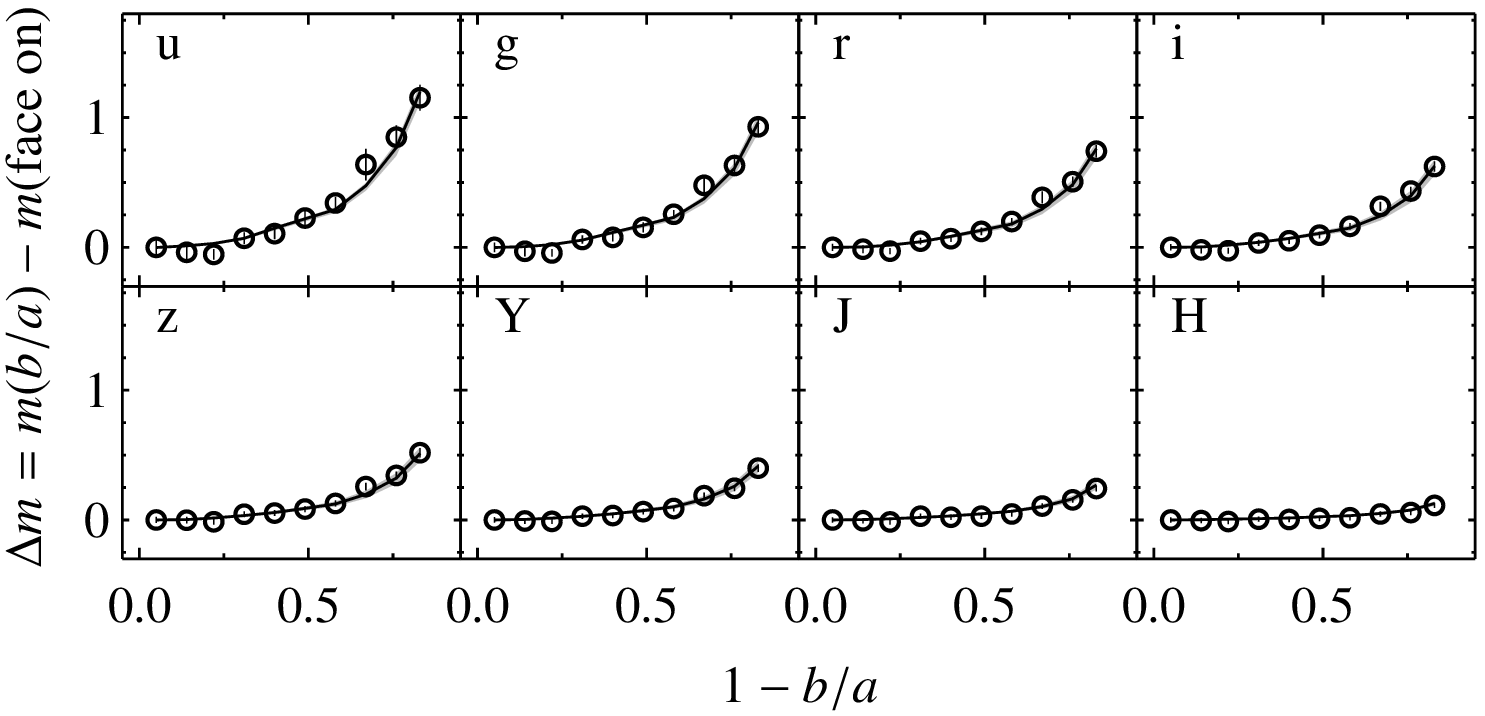}}}
	\caption{Dependence of integrated spectral properties on galaxy inclination, as predicted by models with \tauVbc,  \taubP, \Tthin, \Tthick\ and \Tbulge\ parameters corresponding to the median values reported in Table~\ref{tab:MCMC} for low-\muS\ galaxies (black lines), compared with the observed (bias-corrected) dependence in the W11 sample from Fig.~\ref{fig:data_low} (open circles). The grey shading locates all models with parameters spanning the 2.5th--97.5th percentile ranges of the posterior probability distributions of Table~\ref{tab:MCMC} for low-\muS\ galaxies.}
	\label{fig:fit_low_visual}
\end{figure}

Fig.~\ref{fig:joint} shows the posterior probability distributions of \tauVbc, \taubP, \Tthin, \Tthick\ and \Tbulge\ obtained from the MCMC analysis of the \HaHb\ ratios and $ugrizYJH$ attenuation curves of the sample of low-\muS, nearby star-forming galaxies of Section~\ref{sec:data}. Table~\ref{tab:MCMC} lists the corresponding median and 2.5th--97.5th percentile range of each parameter. To complement this information, we compare in Fig.~\ref{fig:fit_low_visual} the mean \HaHb\ ratio and differential $ugrizYJH$ magnitudes predicted using the best-estimate (i.e. median) parameters from Fig.~\ref{fig:joint}, with the (bias-corrected) observations of Fig.~\ref{fig:data_low},  as function of galaxy inclination. The agreement between model and data is remarkable at all wavelengths in this figure.

We now describe these results for the low-\muS\ galaxies. The blue lines on the diagonal panels of Fig.~\ref{fig:joint} show the marginal posterior probability distributions of the five dust parameters. For \tauVbc, \taubP,  \Tthin\ and \Tthick, the distributions are nearly Gaussian with well-defined peaks and tails. For \Tbulge, the distribution  peaks at $\lgTbulge = 1.1$ and is truncated at the maximum allowed age of $\lgTbulge=1.13$ ($\Tbulge=13.5$\,Gyr). To interpret the marginal distributions of Fig.~\ref{fig:joint} (Table~\ref{tab:MCMC}), it is helpful to return to the definition of the age parameters in Table~\ref{tab:prior}. In this context, we infer from the median values in Table~\ref{tab:MCMC} that stars younger than $t\approx2.6$ Gyr [$\log(t/\txn{Gyr}) \approx 0.42$] in nearby low-\muS\ galaxies are attenuated in the same way as T04 thin-disc stars, those in the age range $ 2.6 \leq t < 3.9 $\,Gyr [$0.42 \leq \log(t/\txn{Gyr}) < 0.59 $] in the same way as T04 thick-disc stars and those in the age range $ 3.9 \leq t < 12.6 $ Gyr [$0.59 \leq\log(t/\txn{Gyr})< 1.1$] in the same was as T04 bulge stars. It is also useful to remember that \Tthin, \Tthick\  and \Tbulge, when combined with the star formation histories of model galaxies (Section~\ref{sec:model_lib}), determine the intensity weights $\xi_{\lambda}^i$ of the thin-disc, thick-disc and bulge components in the attenuation arising form the diffuse ISM (see equation~\ref{eq:xi}). Hence, we can relate \Tthin, \Tthick\ and \Tbulge\ to $V$-band intensity weights  \xiVthin , \xiVthick and \xiVbulge, which we report in Table~\ref{tab:MCMC}. The results suggest that 80 percent of the $V$-band attenuation in the diffuse ISM of low-\muS\ galaxies in our sample is characteristic of that affecting thin-disc stars in the T04 model, 9 percent to that affecting thick-disc stars, and the remaining 10 percent to that affecting bulge stars. We note that, according to the results of Section~\ref{sec:impact_bias} below, the finding of a non-zero contribution by bulge stars to the integrated light of low-\muS\ galaxies, which are expected to be bulge-less (Section~\ref{sec:sample}), could arise from an imperfect correction of systematic biases in the dependence of dust attenuation signatures on inclination in the W11 sample (Section~\ref{sec:bias_corr}). The median $\taubP\approx1.9$ favoured by our analysis of low-\muS\ galaxies corresponds to an angle-averaged $V$-band attenuation optical depth $\meantauvISM\approx 0.26$ in the ambient ISM, and the median attenuation optical depth of stellar birth clouds is $\tauVbc\approx0.48$.

The bottom-left off-diagonal panels of Fig.~\ref{fig:joint} show the joint posterior probability distributions of all pair of parameters. The joint probability distribution of \tauVbc\  and \taubP\  does not show any degeneracy, indicating that the combined analysis of the \HaHb\ ratio and $ugrizYJH$ attenuation curve allows us to efficiently separate the attenuation arising from the diffuse ISM from that pertaining to stellar birth clouds. The slight apparent degeneracies affecting the age parameters \Tthick\ and \Tthin\, and the tails of \Tbulge\ and \Tthick , arise from the presence of likely models at the limit of the $\Tthin \leq \Tthick \leq \Tbulge$ criterion (i.e. around  $\Tthin=\Tthick$ and $\Tthick =\Tbulge$; see prior definitions in Table~\ref{tab:prior}).

\begin{figure}
	\centering
	\subfigure
	{\resizebox{\hsize}{!}{\includegraphics{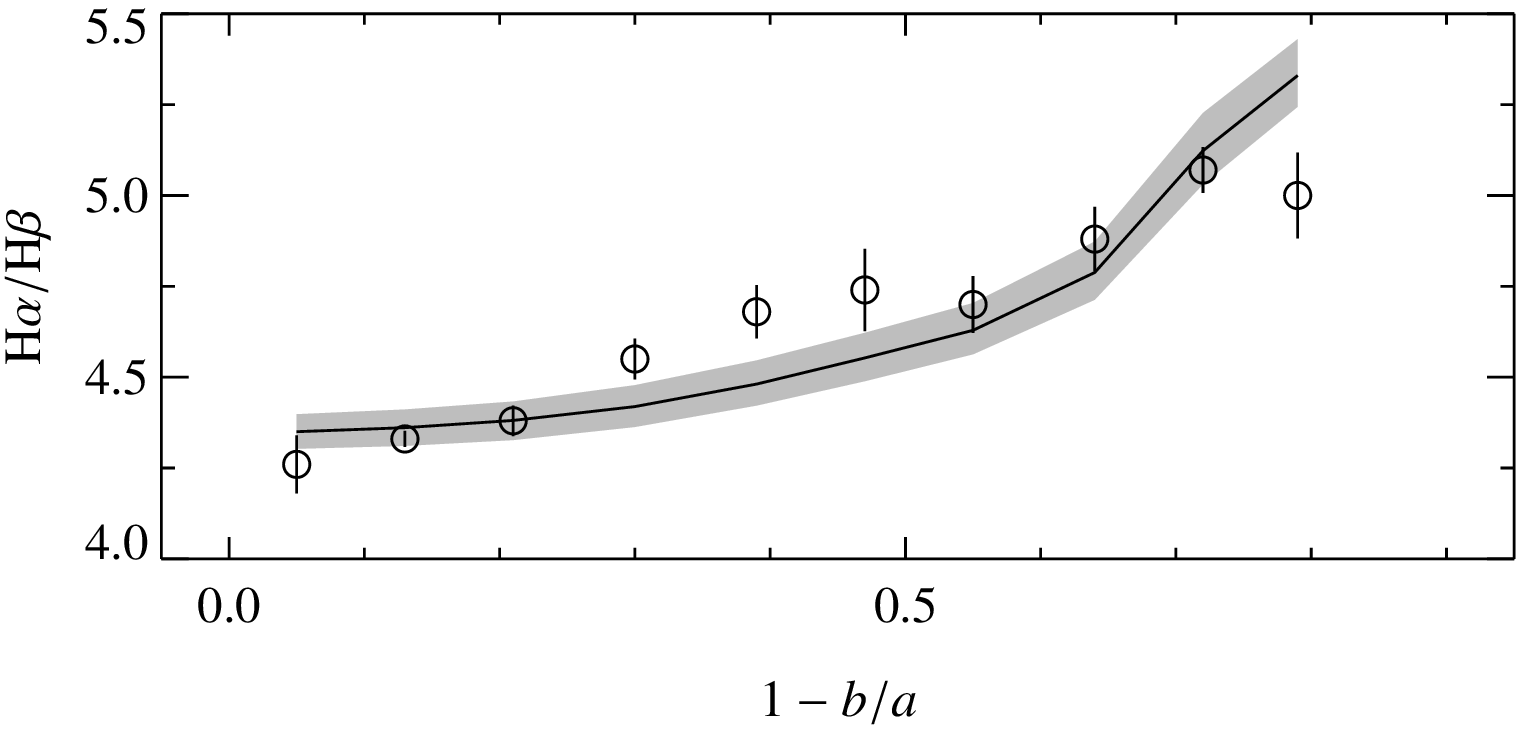}}}
	\subfigure
	{\resizebox{\hsize}{!}{\includegraphics{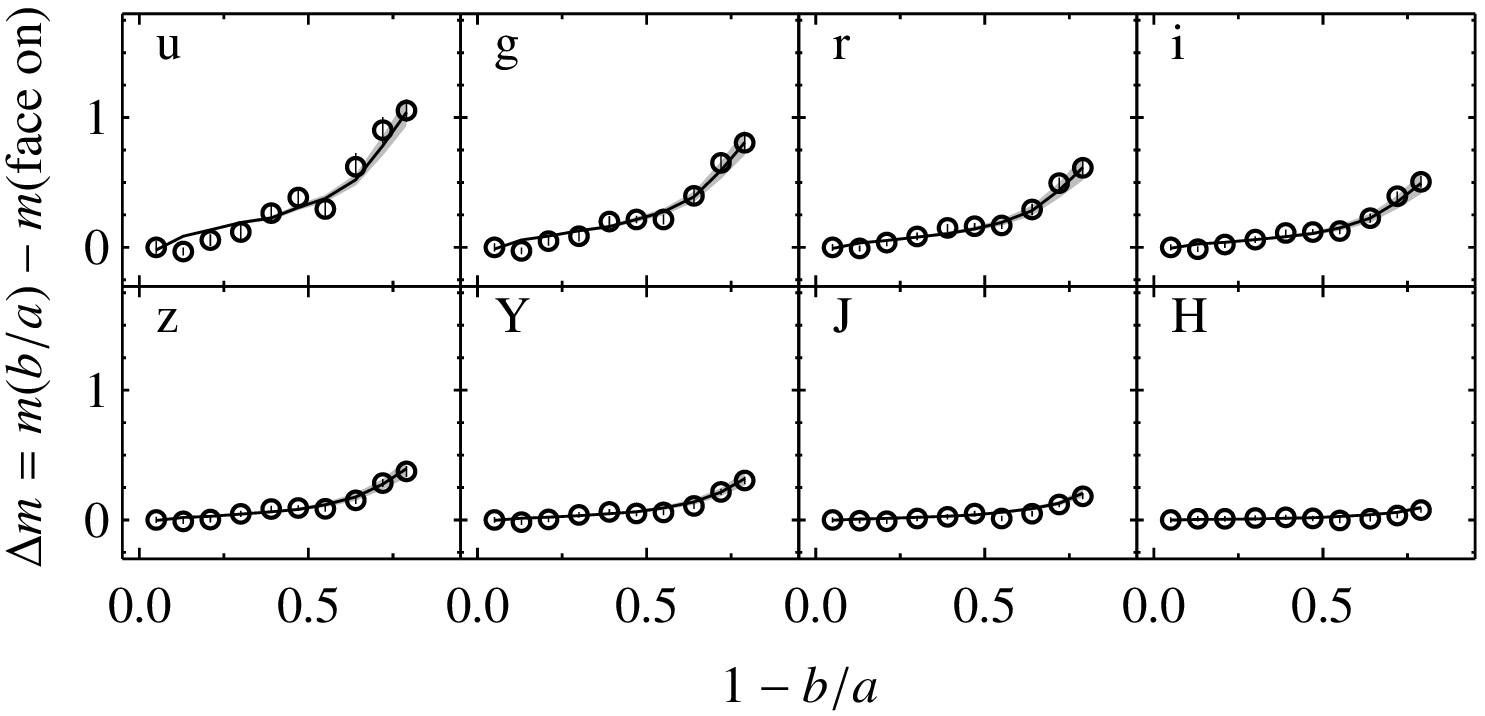}}}
	\caption{Same as Fig.~\ref{fig:fit_low_visual}, but for the high-\muS\ galaxies. The (bias-corrected) observations are from Fig.~\ref{fig:data_high}.}
	\label{fig:fit_high_visual}
\end{figure}

Figs~\ref{fig:joint} and \ref{fig:fit_high_visual} show the analog of Figs~\ref{fig:joint} and \ref{fig:fit_low_visual} for the high-\muS\ galaxy sample. As before, we report in Table \ref{tab:MCMC} the median and 2.5th--97.5th percentile range of the marginal posterior probability distribution of each parameter. In Fig.~\ref{fig:fit_high_visual}, the agreement  between predicted and observed mean spectral properties as a function of galaxy inclination is again remarkable.

As in the case of low-\muS\ galaxies, the marginal distributions of \tauVbc , \taubP, \Tthin, \Tthick\ and \Tbulge\ shown as red lines in Fig.~\ref{fig:joint} are nearly Gaussian with well-defined peaks and tails and with a cutoff for \Tbulge\ at the maximum age $\lgTbulge=1.13$. The median parameter values favoured by observations of high-\muS\ galaxies  in Table~\ref{tab:MCMC} differ slightly from those found above for low-\muS\ galaxies. The slightly larger \Tthin\ (4.7 versus 2.6\,Gyr), \Tthick\ (8.2 versus 3.9\,Gyr) and smaller \Tbulge\ (8.7 versus 12.6\,Gyr) imply small changes in the $V$-band intensity weights  \xiVthin , \xiVthick and \xiVbulge\ reported in Table~\ref{tab:MCMC}. The results are similar to those for low-\muS\ galaxies in that the $V$-band attenuation of the diffuse ISM of high-\muS\ galaxies is primarily characteristic of that affecting thin-disc stars in the T04 model, while 16 percent is more similar to that affecting thick-disc stars, and 3 percent bulge stars. Also, we find a median $\taubP\approx1.7$, corresponding to an angle-averaged $V$-band attenuation optical depth $\meantauvISM\approx0.31$ in the ambient ISM, very similar to the value found for low-\muS\ galaxies. In contrast, the favoured median attenuation optical depth of stellar birth clouds is roughly twice larger for high- than for low-\muS\ galaxies ($\tauVbc\approx0.85$ versus 0.48).

The top-right off-diagonal panels of Fig.~\ref{fig:joint} show the joint posterior probability distributions of all parameters, revealing a weak degeneracy between \tauVbc\ and \taubP. As for the low-\muS\ galaxies, the joint posterior probability distributions of the age parameters \Tthin, \Tthick\ and \Tbulge\ appear degenerate because of the presence of likely models close to the $\Tthin \leq \Tthick \leq \Tbulge$ limit.

We have now established that the model presented in Section \ref{sec:model_lib} can account remarkably well for the observed dependence of the \HaHb\ ratio and $ugrizYJH$ attenuation curve of both low- and high-\muS\ galaxies on inclination (Figs~ \ref{fig:fit_low_visual} and \ref{fig:fit_high_visual}). It is worth recalling that the predictions of the T04 model adopted in our MCMC analysis are similar to those of other types of state-of-the-art radiative transfer models (Section~\ref{sec:S_A_rel} and \ref{sec:tau_theta}). Therefore, we do not expect the results presented in this section to depend critically on the details of the radiative transfer model. These results have several important implications for our ability to constrain the content and spatial distribution of dust in nearby star-forming galaxies. Firstly, the ability with a single best estimate of the central face-on $B$-band optical depth \taubP\ to reproduce the dependence of the \HaHb\ ratio and $ugrizYJH$ attenuation curve on inclination across the full observed range in $b/a$ implies that geometric effects alone can account for these observational trends. Also, the similarity of the best estimates of \taubP\ for low- and high-\muS\ galaxies (1.9 versus 1.7) implies that the surface mass density (and hence the mass, since both samples have similar size distributions; see Fig.~\ref{fig:cur_params}d) of diffuse-ISM dust should be roughly similar in both samples. 

Another important implication of the agreement between model and data in Figs~ \ref{fig:fit_low_visual} and \ref{fig:fit_high_visual} is that it validates, a posteriori, our combination of the T04 model with a galaxy spectral evolution code by associating stars in different age ranges with the thin-disc, thick-disc and bulge components of T04. This demonstrates the possibility to relate integrated spectral properties of structurally unresolved galaxies, such as the \HaHb\ ratio and differential $ugrizYJH$ magnitudes, to different geometric components and the spatial distribution of dust in these galaxies. For example, we find here that over 80 percent of the $V$-band attenuation in the diffuse ISM of both low- and high-\muS\ star-forming galaxies is characteristic of that affecting thin-disc stars in the T04 model. This by itself means that the luminosity of these galaxies is dominated by a component in which the spatial distribution of stars and dust have the same scale heights, as in the thin-disc model of T04. Finally, the fact that the best estimate of \tauVbc\ is about twice as large for high- than low-\muS\ galaxies, while that of \taubP\ is similar, sets interesting constraints on the distribution of dust in the ISM. This could suggest that, for example, low-\muS\ galaxies, which are typically more metal-poor and hence have presumably lower dust-to-gas ratio than their high-\muS\ counterparts (Fig.~\ref{fig:cur_params}), contain proportionally more gas than high-\muS\ galaxies, while the masses of individual stellar birth clouds are more  similar. Since both types of galaxies have similar specific star formation rates (Fig.~\ref{fig:cur_params}), this would also require that low-\muS\ galaxies form stars less efficiently than high-\muS\ galaxies out of the available gas.

\begin{table*}
\centering
	\begin{tabular}{c c c c c c c}
\toprule

\multicolumn{3}{c}{\bf Original model parameter}	& \multicolumn{3}{c}{\bf Associated physical parameter}        &  {\bf Definition}	 \\
\midrule

\multirow{2}[4]{*}{\centering \tauVbc }  & low-\muS\  	& $0.48^{+0.02}_{-0.02}$ \bigstrut	&   \multirow{2}[4]{*}{\centering \tauVbc}   & low-\muS\  & $0.48^{+0.02}_{-0.02}$  & \multirow{2}[4]{*}{\centering  \begin{minipage}{\colWidth}  $V$-band attenuation optical depth of dust in stellar birth clouds. \end{minipage}}  \\

\vspace{\colhspace}
& high-\muS\ & $0.85^{+0.02}_{-0.01}$ \bigstrut &  &  high-\muS\ & $0.85^{+0.02}_{-0.01}$ & \\

\multirow{2}[4]{*}{\centering \taubP }  & low-\muS\  	& $1.90^{+0.23}_{-0.10}$ \bigstrut	&  \multirow{2}[4]{*}{\centering \meantauvISM}   & low-\muS\  & $0.26^{+0.06}_{-0.06}$  & \multirow{2}[4]{*}{\centering \begin{minipage}{\colWidth}  Angle-averaged  $V$-band attenuation optical depth of dust in the diffuse ISM. \end{minipage}}  \\

\vspace{\colhspace}

& high-\muS\ & $1.68^{+0.19}_{-0.10}$ \bigstrut &  &  high-\muS\ & $0.31^{+0.04}_{-0.03}$ & \\

\multirow{2}[4]{*}{\centering \lgTthin }  & low-\muS\  	& $0.42^{+0.17}_{-0.11} $ \bigstrut	&  \multirow{2}[4]{*}{\centering \xiVthin}   & low-\muS\  & $0.80^{+0.08}_{-0.08}$  &   \multirow{2}[4]{*}{\centering \begin{minipage}{\colWidth}  $V$-band intensity weight of thin-disc stars in the T04 model.  \end{minipage}}  \\

\vspace{\colhspace}
& high-\muS\ & $ 0.67^{+0.15}_{-0.27} $ \bigstrut &  &  high-\muS\ & $0.80^{+0.16}_{-0.13}$ & \\

\multirow{2}[4]{*}{\centering \lgTthick }  & low-\muS\  	& $0.59^{+0.18}_{-0.15}$ \bigstrut	&  \multirow{2}[4]{*}{\centering \xiVthick}   & low-\muS\  & $0.09^{+0.13}_{-0.09}$   &   \multirow{2}[4]{*}{\centering \begin{minipage}{\colWidth} $V$-band intensity weight of thick-disc stars in the T04 model.  \end{minipage}}  \\

\vspace{\colhspace}
& high-\muS\ & $ 0.91^{+0.11}_{-0.13} $ \bigstrut &  & high-\muS\ & $0.16^{+0.15}_{-0.16}$ & \\

\multirow{2}[4]{*}{\centering \lgTbulge }  & low-\muS\  	& $1.08^{+0.05}_{-0.20} $ \bigstrut &  \multirow{2}[4]{*}{\centering \xiVbulge}   & low-\muS\  & $0.10^{+0.12}_{-0.10}$   &   \multirow{2}[4]{*}{\centering \begin{minipage}{\colWidth}  $V$-band intensity weight of bulge stars in the T04 model.  \end{minipage}} \\

\vspace{\colhspace}
& high-\muS\ & $ 0.94^{+0.19}_{-0.10}  $ \bigstrut &  &  high-\muS\ & $0.03^{+0.08}_{-0.03}$ & \\

\bottomrule

	\end{tabular}
	\caption{Median and 2.5th--97.5th percentile range of the posterior probability distributions of the five adjustable model parameters of Table~\ref{tab:prior} derived from the MCMC analysis of the low- and high-\muS\  galaxies in the W11 sample described in Section~\ref{sec:sample}, using the library of model spectral energy distributions assembled in Section~\ref{sec:model_lib}.}
	\label{tab:MCMC}	
\end{table*}

\subsection{Integrated versus fibre properties}

\begin{figure}
	\centering
	\subfigure
	{\resizebox{\hsize}{!}{\includegraphics{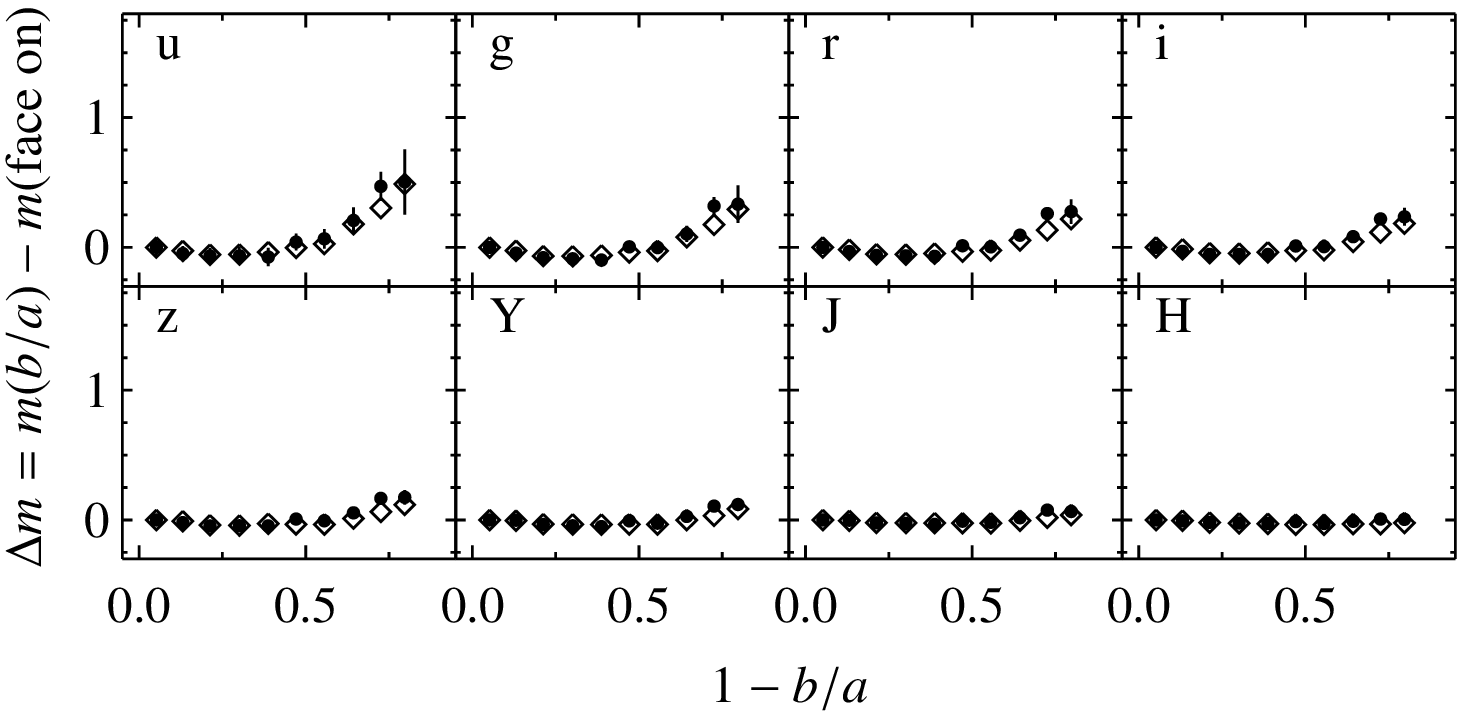}}}
	\subfigure
	{\resizebox{\hsize}{!}{\includegraphics{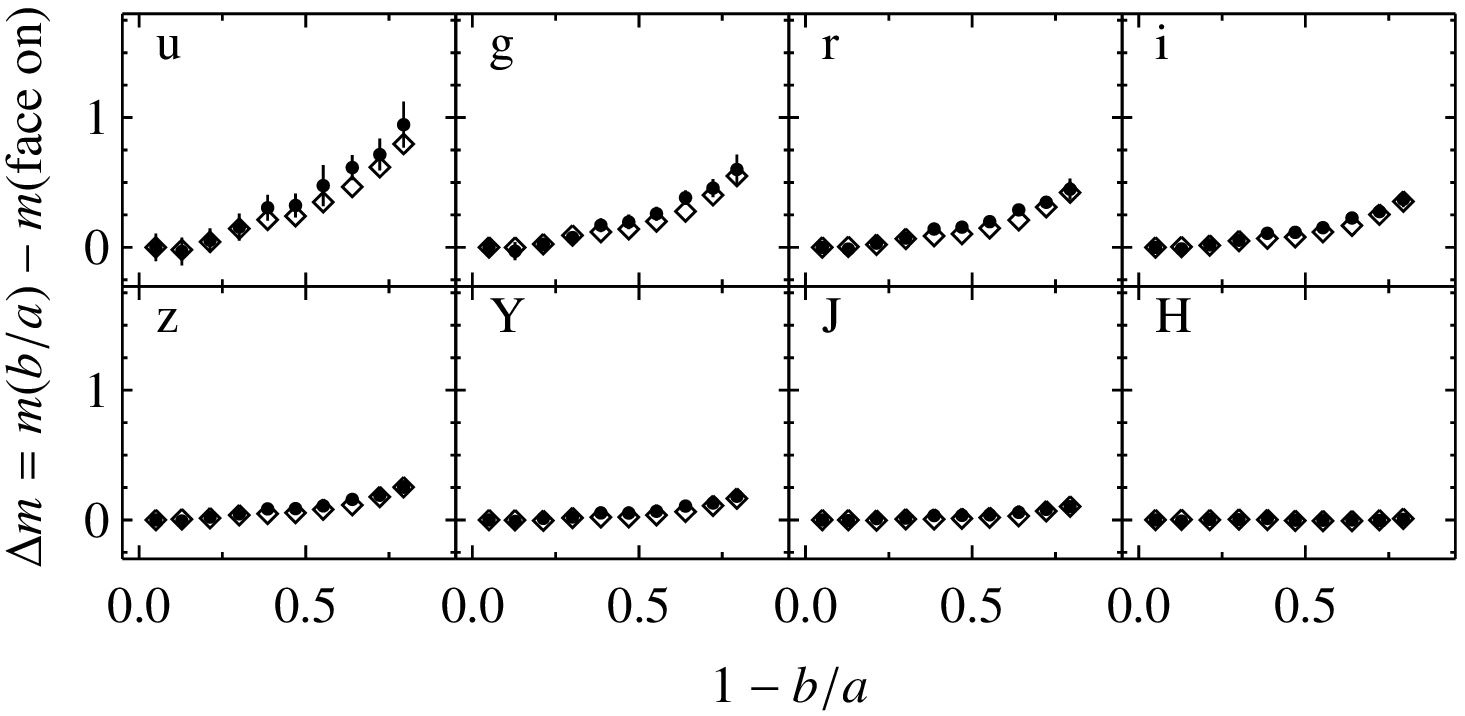}}}
	\caption{Difference between the mean magnitude of galaxies in bins of different inclination, from face-on ($1-b/a\approx0$) to edge-on ($1-b/a\approx1$), and that of face-on galaxies, at fixed $K$ band flux (i.e. fixed stellar mass), for the low-\muS\ ({\it top}) and high-\muS\  ({\it bottom}) galaxies of the W11 sample described in Section~\ref{sec:sample}, in the $ugrizYJH$ bandpasses, as indicated. Different symbols refer to different aperture sizes. Open diamonds refer to differential magnitudes measured in a large circular aperture of size the $r$-band Petrosian radius. Filled circles with error bars refer to differential magnitudes measured in a 3-arcsec-diameter aperture and corrected for the bias in galaxy size $R_{90}$ using the procedure described in Section~\ref{sec:bias_corr} and Appendix~\ref{app:importance}.}
	\label{fig:aperture}
\end{figure} 

The constraints derived above on the content and spatial distribution of dust in nearby star-forming galaxies pertain to an inner region of 3-arcsec diameter  probed by the SDSS fibre. To check the applicability of these results to entire galaxies, we first build integrated $ugrizYJH$ spectral energy energy distributions of the galaxies in the W11 sample by computing, for each waveband, the flux comprised in a circular aperture of size the $r$-band Petrosian radius (footnote~\ref{footnote:petrosian}). In practice, we interpolate the flux within this radius from observations in different apertures (available at 8 radii between 0.23 and 11.42~arcsec for SDSS data, and 12 radii between 0.5 and 10.0~arcsec for UKIDSS data; we exclude $\sim5$~percent of galaxies with $r$-band Petrosian radius larger than 10~arcsec in the original sample; see Fig.~\ref{fig:cur_params}d). Then, we follow the procedure outlined in Section~\ref{sec:sample} and compute the dependence of the attenuation curve on galaxy orientation, by comparing the mean total magnitudes of galaxies in different bins of $b/a$ to those in the face-one bin. 

We do not have access to spectroscopic measurements outside the SDSS-fibre aperture. Thus, unlike in Section~\ref{sec:sample}, we can neither measure the dependence of the integrated \HaHb\ ratio on inclination, nor correct the dependence on inclination of the total $ugrizYJH$ attenuation curve for systematic biases in gas-phase oxygen abundance and specific star formation rate. Yet, we can compare in a meaningful way fibre versus total dust properties by correcting the dependence on inclination of the fibre-derived attenuation curves purely for the bias in galaxy size $R_{90}$, and not applying any correction to the total attenuation curves. This is because the fraction of total light sampled by the SDSS fibre depends on galaxy size, which in the W11 sample correlates with axis ratio (Figs~\ref{fig:bias_low} and \ref{fig:bias_high}), while magnitudes measured within the $r$-band Petrosian radius should include the majority of the light of a galaxy, at any inclination.
 
Fig.~\ref{fig:aperture} shows the differential $ugrizYJH$ magnitudes, $\Delta m=m(b/a)-m(\txn{face-on})$, obtained in this way as a function of $b/a$, for the low- and high-\muS\ samples, respectively. Open diamonds refer to the total dust attenuation curves, computed directly from observations. Filled circles refer to the fibre-derived attenuation curves, corrected for the bias in galaxy size $R_{90}$ using the procedure described in Section~\ref{sec:bias_corr} and Appendix~\ref{app:importance}. The remarkable agreement, within the errors, between the total and fibre-derived attenuation curves in Fig.~\ref{fig:aperture} suggests that the results derived in Section~\ref{sec:dust_constraints} (Table~\ref{tab:MCMC}) from fibre observations are applicable to entire galaxies. We note that this agreement in differential $ugrizYJH$ magnitudes cannot reveal potential offsets in the absolute normalisations of the attenuation curves. In fact, W11 find that the $V$-band attenuation optical depth of galaxies in their sample is typically $\sim15$~percent larger in the central regions than in an aperture of size $R_{90}$, independent of axis ratio and stellar-mass density (see their fig.~6). Hence, the constraints on \meantauvISM\ derived from our analysis of fibre observations in Section~\ref{sec:dust_constraints} may slightly overestimate the galaxy-wide attenuation of W11 galaxies. 

\subsection{Impact of the correction of systematic biases in the dependence of dust attenuation signatures on inclination}\label{sec:impact_bias}

The results reported in Figs~\ref{fig:joint}--\ref{fig:fit_high_visual} were obtained after correcting the observed dependence of the \HaHb\ ratio and $ugrizYJH$ attenuation curve of nearby star-forming galaxies on inclination for various systematic biases, as described in Section~\ref{sec:bias_corr}. To quantify the impact of these corrections on our results, we have repeated the same analysis as in Section~\ref{sec:dust_constraints}, but fitting this time the uncorrected data (open circles in Figs~\ref{fig:data_low} and \ref{fig:data_high}) in place of the corrected ones (filled circles).

We find that, in this case, the model presented in Section~\ref{sec:model_lib} can still account reasonably well the observations of low-\muS\ galaxies in Fig.~\ref{fig:data_low}, although the best-estimate (i.e. median) parameters differ from those derived in Section~\ref{sec:dust_constraints}. The favoured optical depths are $\taubP\approx0.41$ and $\tauVbc\approx0.59$, and the age parameters  $\Tthin\approx0.02$\,Gyr, $\Tthick\approx$0.32\,Gyr and $\Tbulge\approx8.0$\,Gyr indicate that the T04 bulge accounts for 60 percent of the $V$-band attenuation in the diffuse ISM. The reason for this is that the $ugrizYJH$ attenuation curve of uncorrected data depends only weakly on galaxy orientation in Fig.~\ref{fig:data_low}, which is characteristic of the T04 bulge model.

Similarly, the model can still account well for the observations of high-\muS\ galaxies if no bias correction is applied to the data. The best-estimate  optical depths,  $\taubP\approx2.3$ and $\tauVbc\approx0.78$, are close to the values found above when accounting for bias corrections. The constraints on the age parameters are slightly different, corresponding to $\Tthin\approx3.0$\,Gyr, $\Tthick\approx4.4$\,Gyr and $\Tbulge\approx4.8$\,Gyr. The fact that all three ages are close together indicates that the $V$-band attenuation in the diffuse ISM is almost exclusively characteristic of that affecting thin-disc stars in the T04 model. This makes the fraction of $V$-band luminosity produced by unattenuated stars, with ages greater than \Tbulge, increase from roughly 0 to 20 percent.

Hence, as expected from the stronger bias corrections derived in Section~\ref{sec:bias_corr} for low- than high-\muS\ galaxies, the changes in parameter estimates induced by these corrections are significantly larger for low- than for high-\muS\ galaxies.

\section{Discussion and conclusion}

\subsection{General results on dust attenuation curves from models of radiative transfer in dusty media}

We have presented a new approach to investigate the content and spatial distribution of dust in structurally unresolved star-forming galaxies from the dependence of integrated spectral properties on inclination. To achieve this, we first established that four different types of existing models of radiative transfer in disc galaxies all predict a similar dependence of the dust attenuation optical depth in the ambient ISM, \tauLismTh, on viewing angle $\theta$ in the $ugrizYJH$ bandpasses. We find that, in fact, these models predict a quasi-universal relation between slope of the attenuation curve at any wavelength from the ultraviolet to the near infrared, \nLismTh, and $V$-band attenuation optical depth of the dust, \tauVismTh, which we parametrize by means of a simple analytic expression (equation~\ref{eq:n_lambda} of Section~\ref{sec:S_A_rel}). As shown by Figs~\ref{fig:V_slope}a and \ref{fig:V_slope}c, the range in \nLismTh\  spanned by this relation at $\lambda=0.55$ and $1.6\,\mu$m across the full range in \tauVismTh\ is much wider than that sampled by the \citet{Calzetti2001} attenuation curve and the Milky-Way, LMC and SMC extinction curves, even though all the models considered in the present study rely on optical properties of Milky Way-type dust. This confirms that, in normal star-forming (disc) galaxies, changes in the shape of the optical attenuation curve induced by geometry and orientation effects are likely to dominate over those induced by differences in the optical properties of dust grains \citep[e.g.,][]{Granato2000,Fontanot2009}. 

The quasi-universal relation between \nLismTh\ and \tauVismTh\ can be easily implemented in any spectral analysis of structurally unresolved disc galaxies to account in a simple yet physically consistent way for the effect of galaxy orientation on dust attenuation \citep[see e.g.][]{Pacifici2012}. This relation may also be relevant to the interpretation of recent observational results by \citet{Wuyts2011} and Arnouts et. al (in preparation). \citet{Wuyts2011} find that, at redshifts $z \la 3$, the rest-frame ultraviolet, optical and near-infrared colours of actively star-forming galaxies can be best modelled using a \citet{Calzetti2001} attenuation curve, while those of more quiescent star-forming galaxies are better reproduced with a steeper, SMC-like curve. \citet{Wuyts2011}  suggest that systematic changes in the grain size distribution could account for this effect. A similar correlation between shape of the optical attenuation curve and star formation activity has been found by Arnouts et. al (in preparation) in a sample of $z \la 1.5$ star-forming galaxies. We argue here that behaviours of this type could result naturally from the relation between \nLismTh\ and \tauVismTh, if star formation activity correlates with dust attenuation optical depth in the ambient ISM of galaxies. Such a correlation is expected from the combination of the Schmidt-Kennicutt relation \citep{Schmidt1959,Kennicutt1998} with the relation between metallicity and star formation rate at fixed stellar mass \citep{Mannucci2010}. According to these trends, actively star-forming galaxies should contain more gas and metals, and hence presumably more dust than more quiescent star-forming galaxies \citep[see also][]{Zahid2013}. It is possible that, therefore, the observed systematic changes of galaxy attenuation curves with star formation activity might arise purely from geometric effect, rather than from changes in the physical properties of dust grains.

\subsection{Beyond angle-averaged dust prescriptions}

These findings demonstrate the importance of accounting for the dependence of integrated spectral properties on inclination when constraining galaxy physical parameters. In this paper, we have proposed an original, computationally inexpensive approach to accomplish this goal by extending simple angle-averaged models of dust attenuation using the generic output of sophisticated radiative transfer calculations. Specifically, we have extended the simple angle-averaged  model of \citet{CF00}, which accounts for the different attenuation affecting young stars in their birth clouds and older stars in the ambient ISM, by associating stars in different age ranges in the ambient ISM with different geometric components of galaxies (Section~\ref{sec:dust}). This is motivated by the observation that different stellar populations reside in different spatial components of nearby  star-forming galaxies. Stars in these galaxies tend to form along spiral arms in a thin disc close to the equatorial plane (as a consequence of gas dissipation). The scale height of these stars can later increase as a result of dynamical heating and merging \citep{Yoachim2006,Roskar2012}, as suggested by the presence of different stellar populations in the thin and thick discs of external galaxies \citep{Yoachim2008A,Yoachim2012}, while the oldest stars reside typically in a bulge \citep{Kauffmann2003b}. The association of different stellar-age ranges with different geometric components of galaxies presents several notable advantages with respect to traditional investigations of spatially resolved nearby galaxies using detailed radiative transfer modelling \citep[e.g.,][]{Baes2010,MacLachlan2011,DeLooze2012,DeGeyter2013}. Firstly, it can be easily combined with sophisticated models of spectral evolution. Also, it is applicable to the study of structurally unresolved galaxies. And finally, it can be efficiently applied to the statistical analysis of large samples of galaxies.

We have implemented this approach by combining the multi-component radiative transfer model of \citet{Tuffs2004} with a comprehensive library of star formation and chemical enrichment histories from \citet{Pacifici2012} and a stellar population synthesis code (\citealt{BC03}; Charlot \& Bruzual, in preparation). The resulting library of galaxy spectral energy distributions allowed us to investigate the dependence of dust attenuation on inclination in the sample of nearby star-forming galaxies assembled by \citet{Wild2011b}. To avoid unwanted biases in the observational measurements, we also developed an original method, based on importance sampling, to measure the dependence of the \HaHb\ ratio and $ugrizYJH$ attenuation curve as a function of galaxy axis ratio in this sample (Section~\ref{sec:bayes_approach}). From the Bayesian MCMC analysis of two subsamples of galaxies with low (bulge-less) and high (with a bulge) stellar-mass density \muS, we can draw valuable conclusions on the content and spatial distribution of dust in nearby star-forming galaxies, which are encoded in the posterior probability distributions of the five main adjustable model parameters: the $V$-band attenuation optical depth of stellar birth clouds,  \tauVbc\ (assumed independent of angle); the central face-on $B$-band optical depth of dust in the ambient ISM, \taubP, and the age parameters \Tthin, \Tthick\ and \Tbulge\ separating stars in the thin-disc, thick-disc and bulge components (Tables~\ref{tab:prior} and \ref{tab:MCMC}). Our results suggest that \tauVbc\ is significantly larger in high-\muS\ than in low-\muS\ galaxies, while \taubP\ is roughly similar in both cases. As mentioned in Section~\ref{sec:dust_constraints}, this can arise if, for example, high-\muS\ galaxies have higher star formation efficiency than their low-\muS\ counterparts at fixed specific star formation rate, and dustier stellar birth clouds because of the higher metallicity (Fig.~\ref{fig:cur_params}). 

\subsection{Content and spatial distribution of dust in the W11 sample of star forming galaxies}

The central face-on $B$-band optical depth of dust in the ambient ISM favoured by our analysis of nearby star-forming galaxies is $\taubP\approx1.8^{+0.2}_{-0.1}$ (Table~\ref{tab:MCMC}). This is slightly larger than the range $0.5\la \taubP\la1.1$ derived from detailed modelling of the $BVRIK$ vertical surface brightness profiles of seven edge-on late-type galaxies by \citet{Xilouris1999}. At the same time, the value favoured by our analysis is smaller than the constraint  $\taubP\approx3.7\pm0.7$ inferred by \citet{Driver2007} from an analysis of the local $B$-band luminosity function of galactic discs. These authors measure the dependence of the turnover of the galaxy luminosity function on inclination and interpret the fading of the turnover luminosity from low to high inclination as a signature of attenuation by dust. \citet{Driver2007} find that the observations can be well reproduced by a T04 thin-disc model with $\taubP\approx3.7$ (we note that, according to their fig.~10, a value $\taubP\approx2$ appears to also provide excellent agreement with their data). An advantage of the approach developed in the present paper over this study is that it makes use of the entire spectral information in the $ugrizYJH$ bandpasses.
 
Our results on dust attenuation in nearby star-forming galaxies are also consistent with the conclusions from recent studies of low-redshift galaxies observed at infrared wavelengths with the {\it Herschel Space Observatory}. In particular, based on the spectral analysis of ultraviolet, optical and infrared spectral energy distributions of a sample of about 300 nearby galaxies, \citet{Cortese2012} find a strong anti-correlation between dust-to-stellar mass ratio, \MdMs, and stellar surface mass density, \muS. This relation supports our finding that low-\muS\ galaxies, which have lower stellar masses than high-\muS\ galaxies (Fig.~\ref{fig:cur_params}c), also have similar dust masses (as probed by the central face-on $B$-band optical depth \taubP; Section~\ref{sec:dust_constraints}). We note that the similar distributions of specific star formation rates of low- and high-\muS\ galaxies in the W11 sample (Fig.~\ref{fig:cur_params}b) prevent us from testing the correlation between \MdMs\ and \psiS\ identified by \citet{daCunha2010} and \citet{Smith2012} in two independent samples of  about 1500 nearby star-forming galaxies. Our results are consistent with the large scatter about this mean relation.

We also note that, during the revision of this manuscript, \citet{Grootes2013} proposed a relation between  \taubP\ and stellar surface mass density \muS, of the form $\log \taubP = 1.12(\pm0.11) \, \log (\muS/\Msun\txn{pc}^{-2}) -8.6(\pm0.8)$, based on the analysis of the ultraviolet, optical and infrared spectral energy distributions of a small sample of 74 nearby disc galaxies. This implies $\taubP=3.0^{+1.3}_{-0.9}$ for low-\muS\ galaxies in the W11 sample [median $\log (\muS/\Msun\txn{pc}^{-2})  = 8.1$] and $\taubP=13.9^{+6.2}_{-4.3}$ for high-\muS\ galaxies [median $\log(\muS/\Msun\txn{pc}^{-2})  = 8.7$]. The value for high-\muS\ galaxies is much larger than that inferred in our analysis (and in any of the previously quoted studies of nearby disc galaxies). The origin of this discrepancy is unclear. It may arise, at least in part, from the different range in \psiS\ spanned at fixed \muS\ by galaxies in the \citet{Grootes2013} sample compared to the W11 sample. We note that the dispersion in \taubP\ ($\tau_{\rm B}^{\rm f}$ in their notation)  at fixed \muS\  exceeds an order of magnitude in the small sample of nearby galaxies studied by \citet{Grootes2013}. 
   
In our approach, the age parameters \Tthin, \Tthick\  and \Tbulge, along with the adopted library of star formation histories, determine the relative weights of the different geometric components of galaxies in the overall dust attenuation curve. We find that, for both low- and high-\muS\ galaxies in the \citet{Wild2011b} sample, over 80 percent of the attenuation is characteristic of that affecting thin-disc stars in the T04 model, while the remaining 20 percent is more similar to that affecting thick-disc and bulge stars (Table~\ref{tab:MCMC}; we note that \citealt{Wild2011b} also suggest that their observations might be explained by a scale height for the dust that is smaller than that of old stars). It is interesting to examine the ratio of thick-disc to thin-disc stellar luminosity predicted by our analysis, since this quantity has been estimated observationally. Our results suggest that the median unattenuated $V$-band luminosity ratio of thick-disc to thin-disc stars is $\xiVthick / \xiVthin\approx 0.1$ and 0.2 for low- and high-\muS\ galaxies, respectively. Observationally, \citet{Yoachim2006} find that the $R$-band luminosity ratio of thick-disc to thin-disc stars (corrected for internal extinction) in a sample of 34 nearby edge-on disc galaxies decreases as a function of circular (i.e. disc rotation) velocity roughly as $\xi^\txn{thick}_R/\xi^\txn{thin}_R=0.25(V_c/100\,\txn{km}\,\txn{s}^{-1})^{-2.1}$ (their fig.~18; see also \citealt{Comeron2012}). We can estimate the circular velocities of galaxies in the \citet{Wild2011b}  sample from the available SDSS-MPA stellar masses  using the recipes of \citet{Reyes2011}. We find median circular velocities  $V_c \approx 150 \, \txn{km}\,\txn{s}^{-1}$ and $190 \,\txn{km}\,\txn{s}^{-1}$  for low- and high-\muS\ galaxies, respectively (corresponding to median masses $6.3\times10^9\,\Msun$ and $3.2\times10^{10}\,\Msun$), which implies $\xi^\txn{thick}_R/\xi^\txn{thin}_R\approx0.1$ for both types of galaxies. This is in remarkable agreement with the results of our analysis, given the negligible differences expected between $V$- and $R$-band luminosity ratios. 

\subsection{Effect of galaxy inclination and spatial distribution of dust on broadband colours}

\begin{figure}
	\centering
	\resizebox{\hsize}{!}{\includegraphics{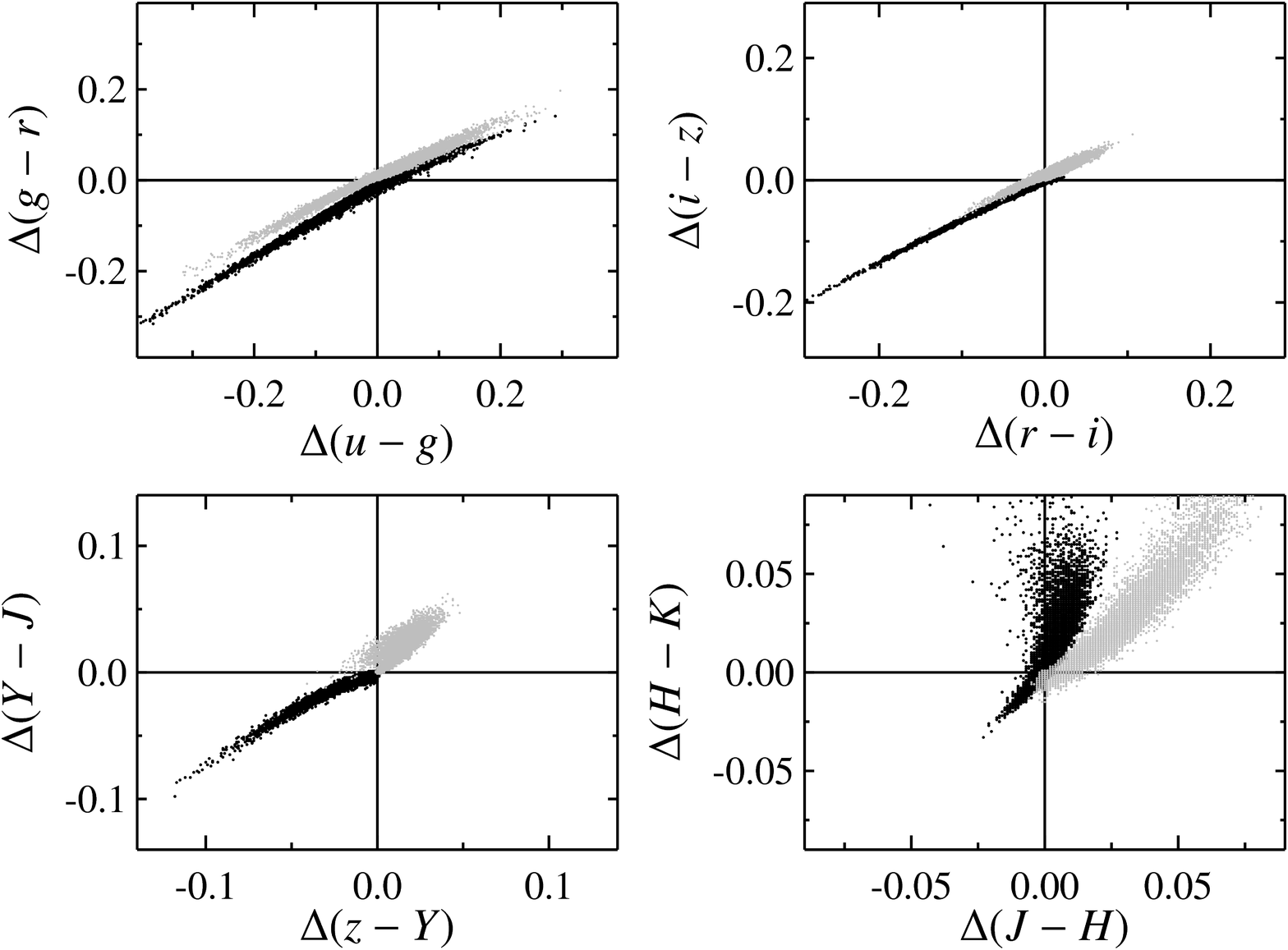}}
	\caption{Difference in the broadband colours of 5000 model galaxies computed using the same star formation and chemical enrichment histories as in Fig.~\ref{fig:D4000_Hdelta}a, but different prescriptions for attenuation by dust. {\it Black points}: colours computed using the approach presented in Section~\ref{sec:dust} above, which account for the effects of galaxy inclination on the dust attenuation curve, minus colours computed with the prescriptions of \citet{Calzetti2001}.  {\it Grey points}: colours computed using the approach presented in Section~{sec:dust} above, minus colours computed using the two-component dust model of \citet{CF00}. In both cases, the substantial differences illustrate the importance of accounting for the influence of geometry and orientation effects on galaxy spectral energy distributions.}
	\label{fig:color_th}
\end{figure}

The much wider range of slopes of attenuation curves probed by the models in Fig.~\ref{fig:V_slope} relative to commonly adopted dust prescriptions (e.g. Milky-Way and SMC extinction curves; \citealt{Calzetti2001} attenuation curve; \citealt{CF00} two-component dust model) has important implications for galaxy spectral analyses. We can illustrate this by comparing the spectral energy distributions of galaxies computed using standard dust prescriptions with those derived using equations~(\ref{eq:pow_law})--(\ref{eq:b_tau_fit}) above, which account for the effects of galaxy inclination on the dust attenuation curve. For simplicity, we appeal to the same library of 5000 star formation and chemical enrichment histories as used in Section~\ref{sec:stelpops} above to describe the properties of low-\muS\ galaxies in the \citet{Wild2011b} sample. We compute attenuation by dust in these galaxies using different prescriptions. For each galaxy in the library, we first draw the $V$-band attenuation optical depth in the ambient ISM from a Gaussian distribution centred on $\tauVism=0.65$ with a dispersion of 0.40. Then, we draw the parameters $\nVism$ and $b$ defining the attenuation curve (see equation~\ref{eq:n_lambda}) from Gaussian distributions centred on the values predicted by equations~(\ref{eq:n_tau_fit}) and (\ref{eq:b_tau_fit}), with dispersions of 25 and 10 percent, respectively. Fig.~\ref{fig:color_th} shows the difference between the $ugrizYJHK$ colours of galaxies computed using this prescription and those obtained using the \citet{Calzetti2001}  attenuation curve corresponding to the same $V$-band attenuation optical depth, for each galaxy (black points). The differences are substantial, reaching 0.3--0.4\,mag in optical colours and 0.1\,mag in near-infrared ones. To compare our prescription with the two-component dust model of \citet{CF00}, we also include attenuation by dust in stellar birth clouds by drawing the ratio $\tauVism/(\tauVism+\tauVbc)$ from a Gaussian distribution centred on 0.4 with a dispersion of 0.2. The colour differences in Fig.~\ref{fig:color_th} in this case are again substantial (grey points). The results of these tests indicate that neglecting the effect of geometry and orientation on attenuation can severely bias the interpretation of galaxy spectral energy distributions (for similar conclusions, from a modelling perspective, see \citealt{Ferrara1999, Granato2000, Panuzzo2007, Rocha2008}; from an observational perspective, see \citealt{Boquien2009, Boquien2012, Mao2012}).

\subsection{Summary}

In conclusion, therefore, the approach presented in this paper should help us gain valuable insight into the content and spatial distribution of stars and dust from spectral analyses of structurally unresolved star-forming galaxies. This is enabled by the possibility to associate stars in different age ranges with different geometric components of sophisticated models of radiative transfer. It is worth emphasising that the results obtained in this way should not depend critically on the details of the adopted radiative transfer model, as the predictions of the versatile T04 model adopted in our MCMC analysis are similar to those of other types of state-of-the-art radiative transfer models (Section~\ref{sec:S_A_rel} and \ref{sec:tau_theta}). We have illustrated the usefulness of this approach to interpret the dependence of integrated spectral properties on galaxy inclination in a sample of nearby star-forming galaxies. Our approach also opens new possibilities to straightforwardly implement the effects of different spatial components and galaxy inclination in the predictions of semi-analytic models of galaxy evolution. This will be particularly useful to gauge the importance of selection effects and systematic biases in the derivation of physical parameters from large samples of galaxies using such models.

We can summarise our main results as follows: 
\begin{enumerate}
	\item Four different types of radiative transfer models all predict a quasi-universal relation between slope of the attenuation curve at any wavelength from the ultraviolet to the near infrared,  \nLismTh, and $V$-band attenuation optical depth of the dust in the diffuse ISM, \tauVismTh , at all galaxy inclinations $\theta$ (Section~\ref{sec:S_A_rel}).
	\item We parametrize the quasi-universal relation between \nLismTh\ and \tauVismTh\ by mean of the simple analytic expression given in equation~(\ref{eq:n_lambda}) of Section~\ref{sec:S_A_rel}. This expression can be easily implemented in any spectral analysis of structurally unresolved disc galaxies to account in a simple yet physically consistent way for the effect of galaxy orientation on dust attenuation.
	\item The relation between \nLismTh\ and \tauVismTh\ given in equation~(\ref{eq:n_lambda}) predicts steeper curves than the \citet{Calzetti2001} and Milky-Way curves at small \tauVismTh, and shallower curves at larger \tauVismTh . This may be relevant to the interpretation of recent observational results by \citet{Wuyts2011} and Arnouts et. al (in preparation), who find a systematic flattening of the optical attenuation curve with increased star formation activity in distant galaxies.
	\item We propose an original, computationally inexpensive approach to combine the output of sophisticated radiative transfer models with models of spectral evolution, by associating stars in different age ranges in the ambient ISM with different geometric components of galaxies (Section~\ref{sec:dust}).
	\item We find that the combination of the radiative transfer model of \citet{Tuffs2004} with the galaxy spectral evolution code of Charlot \& Bruzual (in preparation) accounts remarkably well for the dependence of the \HaHb\ ratio and $ugrizYJH$ attenuation curve on axis ratio (i.e. inclination) in the W11 sample of nearby star-forming galaxies (Section~\ref{sec:dust_constraints}). 
	\item The central face-on $B$-band optical depth favoured by this analysis is $\taubP\approx1.8\pm0.2$ (corresponding to an angle-average $\meantauvISM \approx 0.3$), slightly larger than the range $0.5\la \taubP\la1.1$ derived by \citet{Xilouris1999} from the study of nearby spatially resolved galaxies, and smaller than the constraint $\taubP\approx3.7\pm0.7$ inferred by \citet{Driver2007} from the analysis of a sample of spatially unresolved nearby galaxies. 
	\item The roughly equal central face-on $B$-band optical depths of low- and high-\muS\ galaxies, which have similar size distributions in the W11 sample, are consistent with the anticorrelation between dust-to-stellar mass ratio and stellar surface mass density inferred from the ultraviolet, optical and infrared spectral analysis of low-redshift star-forming galaxies \citep{Cortese2012}.
	\item For the galaxies in the W11 sample, over 80 percent of the attenuation is characteristic of that affecting thin-disc stars in the T04 model, while the remaining 20 percent is more similar to that affecting thick-disc and bulge stars (Section~\ref{sec:dust_constraints}).
	\item The median unattenuated $V$-band luminosity ratio of thick-disc to thin-disc stars inferred from our analysis of structurally unresolved, randomly oriented galaxies in the W11 sample is in remarkable agreement with that derived from the analysis of a sample of spatially resolved, nearly edge-on galaxies by \citet{Yoachim2006}.
	\item The influence of galaxy orientation and of the spatial distribution of dust relative to stars on the shape of the attenuation curve can have a large impact on broadband colours of galaxies, reaching up to 0.3--0.4\,mag at optical wavelengths and 0.1\,mag at near-infrared ones. 
\end{enumerate}

\section*{Acknowledgements}

We thank the anonymous referee for insightful comments, which helped improve the quality of this paper. We also thank Patrick Boiss\'e for useful discussion on the impact of galaxy geometry and orientation on dust attenuation curves. We are grateful to Kenny Wood and John MacLachlan for kindly providing us with the radiative transfer calculations used to produce Fig.~\ref{fig:scattering}. We also thank Patrik Jonsson and Brent Groves for useful discussions and for providing us with their radiative transfer calculations in electronic format, separating the contributions by young and old stars to the overall spectral energy distribution of a galaxy. We thank Richard Tuffs and Daniele Pierini for providing us with their radiative transfer calculations in electronic format and Camilla Pacifici for sharing her library of star formation and chemical enrichment histories. JC and SC acknowledge the support of the European Commission through the Marie Curie Initial Training Network ELIXIR under contract PITN-GA-2008-214227. BW acknowledges the support of the NSF via the grant AST 07-08849 and of the Agence Nationale de la Recherche via the Chaire d'Excellence ANR-10-CEXC-004-01. VW acknowledges the support of the European Research Council via a Starting Grant (SEDMorph, P.I. Wild) and an Advanced Grant (P.I. James Dunlop).

\bibliographystyle{mn2e} % style aa.bst
\bibliography{BibDust_2010} % your references Yourfile.bib

\newcommand{\noop}[1]{}
\begin{thebibliography}{128}
\expandafter\ifx\csname natexlab\endcsname\relax\def\natexlab#1{#1}\fi

\bibitem[{{Abazajian} {et~al}\mbox{.}(2009){Abazajian}, {Adelman-McCarthy},
  {Ag{\"u}eros}, {Allam}, {Allende Prieto}, {An}, {Anderson}, {Anderson},
  {Annis}, {Bahcall}, \& et~al.}]{Abazajian2009}
{Abazajian} K.~N. {et~al.}, 2009, \apjs, 182, 543

\bibitem[{{Arzoumanian} {et~al}\mbox{.}(2011){Arzoumanian}, {Andr{\'e}},
  {Didelon}, {K{\"o}nyves}, {Schneider}, {Men'shchikov}, {Sousbie}, {Zavagno},
  {Bontemps}, {di Francesco}, {Griffin}, {Hennemann}, {Hill}, {Kirk}, {Martin},
  {Minier}, {Molinari}, {Motte}, {Peretto}, {Pezzuto}, {Spinoglio},
  {Ward-Thompson}, {White}, \& {Wilson}}]{Arzoumanian2011}
{Arzoumanian} D. {et~al.}, 2011, \aap, 529, L6+

\bibitem[{{Baes} {et~al}\mbox{.}(2010){Baes}, {Fritz}, {Gadotti}, {Smith},
  {Dunne}, {da Cunha}, {Amblard}, {Auld}, {Bendo}, {Bonfield}, {Burgarella},
  {Buttiglione}, {Cava}, {Clements}, {Cooray}, {Dariush}, {de Zotti}, {Dye},
  {Eales}, {Frayer}, {Gonzalez-Nuevo}, {Herranz}, {Ibar}, {Ivison}, {Lagache},
  {Leeuw}, {Lopez-Caniego}, {Jarvis}, {Maddox}, {Negrello}, {Micha{\l}owski},
  {Pascale}, {Pohlen}, {Rigby}, {Rodighiero}, {Samui}, {Serjeant}, {Temi},
  {Thompson}, {van der Werf}, {Verma}, \& {Vlahakis}}]{Baes2010}
{Baes} M. {et~al.}, 2010, \aap, 518, L39

\bibitem[{{Baldwin} {et~al}\mbox{.}(1981){Baldwin}, {Phillips}, \&
  {Terlevich}}]{BPT1981}
{Baldwin} J.~A., {Phillips} M.~M., {Terlevich} R., 1981, \pasp, 93, 5

\bibitem[{{Bertelli} {et~al}\mbox{.}(2008){Bertelli}, {Girardi}, {Marigo}, \&
  {Nasi}}]{Bertelli2008}
{Bertelli} G., {Girardi} L., {Marigo} P., {Nasi} E., 2008, \aap, 484, 815

\bibitem[{{Bertelli} {et~al}\mbox{.}(2009){Bertelli}, {Nasi}, {Girardi}, \&
  {Marigo}}]{Bertelli2009}
{Bertelli} G., {Nasi} E., {Girardi} L., {Marigo} P., 2009, \aap, 508, 355

\bibitem[{{Boquien} {et~al}\mbox{.}(2012){Boquien}, {Buat}, {Boselli}, {Baes},
  {Bendo}, {Ciesla}, {Cooray}, {Cortese}, {Eales}, {Gavazzi}, {Gomez},
  {Lebouteiller}, {Pappalardo}, {Pohlen}, {Smith}, \&
  {Spinoglio}}]{Boquien2012}
{Boquien} M. {et~al.}, 2012, \aap, 539, A145

\bibitem[{{Boquien} {et~al}\mbox{.}(2009){Boquien}, {Calzetti}, {Kennicutt},
  {Dale}, {Engelbracht}, {Gordon}, {Hong}, {Lee}, \& {Portouw}}]{Boquien2009}
{Boquien} M. {et~al.}, 2009, \apj, 706, 553

\bibitem[{{Bouchet} {et~al}\mbox{.}(1985){Bouchet}, {Lequeux}, {Maurice},
  {Prevot}, \& {Prevot-Burnichon}}]{Bouchet1985}
{Bouchet} P., {Lequeux} J., {Maurice} E., {Prevot} L., {Prevot-Burnichon}
  M.~L., 1985, \aap, 149, 330

\bibitem[{{Box} \& {Muller}(1958)}]{Box_Muller}
{Box} G.~E.~P., {Muller} M.~E., 1958, Annals of Mathematical Statistics, 29,
  610

\bibitem[{{Bressan} {et~al}\mbox{.}(2002){Bressan}, {Silva}, \&
  {Granato}}]{Bressan2002}
{Bressan} A., {Silva} L., {Granato} G.~L., 2002, \aap, 392, 377

\bibitem[{{Brinchmann} {et~al}\mbox{.}(2004){Brinchmann}, {Charlot}, {White},
  {Tremonti}, {Kauffmann}, {Heckman}, \& {Brinkmann}}]{Brinchmann2004}
{Brinchmann} J., {Charlot} S., {White} S.~D.~M., {Tremonti} C., {Kauffmann} G.,
  {Heckman} T., {Brinkmann} J., 2004, \mnras, 351, 1151

\bibitem[{{Brooks} {et~al}\mbox{.}(2011){Brooks}, {Gelman}, {Galin}, \&
  {Xiao-Li}}]{Brooks2011}
{Brooks} S., {Gelman} A., {Galin} L.~J., {Xiao-Li} M., 2011, Handbook of Markov
  Chain Monte Carlo, Handbook of Modern Statistical Methods. Chapman \&
  Hall/CRC

\bibitem[{{Bruzual} \& {Charlot}(2003)}]{BC03}
{Bruzual} G., {Charlot} S., 2003, \mnras, 344, 1000

\bibitem[{{Calzetti}(2001)}]{Calzetti2001}
{Calzetti} D., 2001, \pasp, 113, 1449

\bibitem[{{Calzetti} {et~al}\mbox{.}(1994){Calzetti}, {Kinney}, \&
  {Storchi-Bergmann}}]{Calzetti1994}
{Calzetti} D., {Kinney} A.~L., {Storchi-Bergmann} T., 1994, \apj, 429, 582

\bibitem[{{Cardelli} {et~al}\mbox{.}(1988){Cardelli}, {Clayton}, \&
  {Mathis}}]{Cardelli1988}
{Cardelli} J.~A., {Clayton} G.~C., {Mathis} J.~S., 1988, \apjl, 329, L33

\bibitem[{{Cardelli} {et~al}\mbox{.}(1989){Cardelli}, {Clayton}, \&
  {Mathis}}]{Cardelli1989}
{Cardelli} J.~A., {Clayton} G.~C., {Mathis} J.~S., 1989, \apj, 345, 245

\bibitem[{{Chabrier}(2003)}]{Chabrier2003}
{Chabrier} G., 2003, \pasp, 115, 763

\bibitem[{{Charlot} \& {Fall}(2000)}]{CF00}
{Charlot} S., {Fall} S.~M., 2000, \apj, 539, 718

\bibitem[{{Cherchneff}(2010)}]{Cherchneff2010}
{Cherchneff} I., 2010, in Astronomical Society of the Pacific Conference
  Series, Vol. 425, Hot and Cool: Bridging Gaps in Massive Star Evolution,
  {C.~Leitherer, P.~Bennett, P.~Morris, \& J.~van Loon}, ed., pp. 237--+

\bibitem[{{Clayton} \& {Martin}(1985)}]{Clayton1985}
{Clayton} G.~C., {Martin} P.~G., 1985, \apj, 288, 558

\bibitem[{{Comer{\'o}n} {et~al}\mbox{.}(2012){Comer{\'o}n}, {Elmegreen},
  {Salo}, {Laurikainen}, {Athanassoula}, {Bosma}, {Knapen}, {Gadotti}, {Sheth},
  {Hinz}, {Regan}, {Gil de Paz}, {Mu{\~n}oz-Mateos}, {Men{\'e}ndez-Delmestre},
  {Seibert}, {Kim}, {Mizusawa}, {Laine}, {Ho}, \& {Holwerda}}]{Comeron2012}
{Comer{\'o}n} S. {et~al.}, 2012, \apj, 759, 98

\bibitem[{{Cortese} {et~al}\mbox{.}(2012){Cortese}, {Ciesla}, {Boselli},
  {Bianchi}, {Gomez}, {Smith}, {Bendo}, {Eales}, {Pohlen}, {Baes}, {Corbelli},
  {Davies}, {Hughes}, {Hunt}, {Madden}, {Pierini}, {di Serego Alighieri},
  {Zibetti}, {Boquien}, {Clements}, {Cooray}, {Galametz}, {Magrini},
  {Pappalardo}, {Spinoglio}, \& {Vlahakis}}]{Cortese2012}
{Cortese} L. {et~al.}, 2012, \aap, 540, A52

\bibitem[{{Cowles} \& {Carlin}(1996)}]{Cowles1996}
{Cowles} M.~K., {Carlin} B.~P., 1996, Journal of the American Statistical
  Association

\bibitem[{{Cox} {et~al}\mbox{.}(2006){Cox}, {Jonsson}, {Primack}, \&
  {Somerville}}]{Cox2006}
{Cox} T.~J., {Jonsson} P., {Primack} J.~R., {Somerville} R.~S., 2006, \mnras,
  373, 1013

\bibitem[{{Cox} {et~al}\mbox{.}(2008){Cox}, {Jonsson}, {Somerville}, {Primack},
  \& {Dekel}}]{Cox2008}
{Cox} T.~J., {Jonsson} P., {Somerville} R.~S., {Primack} J.~R., {Dekel} A.,
  2008, \mnras, 384, 386

\bibitem[{{da Cunha} {et~al}\mbox{.}(2008){da Cunha}, {Charlot}, \&
  {Elbaz}}]{daCunha2008}
{da Cunha} E., {Charlot} S., {Elbaz} D., 2008, \mnras, 388, 1595

\bibitem[{{da Cunha} {et~al}\mbox{.}(2010){da Cunha}, {Eminian}, {Charlot}, \&
  {Blaizot}}]{daCunha2010}
{da Cunha} E., {Eminian} C., {Charlot} S., {Blaizot} J., 2010, \mnras, 403,
  1894

\bibitem[{{De Geyter} {et~al}\mbox{.}(2013){De Geyter}, {Baes}, {Fritz}, \&
  {Camps}}]{DeGeyter2013}
{De Geyter} G., {Baes} M., {Fritz} J., {Camps} P., 2013, \aap, 550, A74

\bibitem[{{de Looze} {et~al}\mbox{.}(2012){de Looze}, {Baes}, {Bendo},
  {Ciesla}, {Cortese}, {de Geyter}, {Groves}, {Boquien}, {Boselli}, {Brondeel},
  {Cooray}, {Eales}, {Fritz}, {Galliano}, {Gentile}, {Gordon}, {Hony}, {Law},
  {Madden}, {Sauvage}, {Smith}, {Spinoglio}, \& {Verstappen}}]{DeLooze2012}
{de Looze} I. {et~al.}, 2012, \mnras, 427, 2797

\bibitem[{{De Lucia} \& {Blaizot}(2007)}]{DeLucia2007}
{De Lucia} G., {Blaizot} J., 2007, \mnras, 375, 2

\bibitem[{{Draine}(2003)}]{Draine2003}
{Draine} B.~T., 2003, \araa, 41, 241

\bibitem[{{Draine} \& {Lee}(1984)}]{Draine1984}
{Draine} B.~T., {Lee} H.~M., 1984, \apj, 285, 89

\bibitem[{{Draine} \& {Li}(2007)}]{Draine2007}
{Draine} B.~T., {Li} A., 2007, \apj, 657, 810

\bibitem[{{Driver} {et~al}\mbox{.}(2007){Driver}, {Popescu}, {Tuffs}, {Liske},
  {Graham}, {Allen}, \& {de Propris}}]{Driver2007}
{Driver} S.~P., {Popescu} C.~C., {Tuffs} R.~J., {Liske} J., {Graham} A.~W.,
  {Allen} P.~D., {de Propris} R., 2007, \mnras, 379, 1022

\bibitem[{{Dwek}(1998)}]{Dwek1998}
{Dwek} E., 1998, \apj, 501, 643

\bibitem[{{Feroz} \& {Hobson}(2008)}]{Feroz2008}
{Feroz} F., {Hobson} M.~P., 2008, \mnras, 384, 449

\bibitem[{{Feroz} {et~al}\mbox{.}(2009){Feroz}, {Hobson}, \&
  {Bridges}}]{Feroz2009}
{Feroz} F., {Hobson} M.~P., {Bridges} M., 2009, \mnras, 398, 1601

\bibitem[{{Ferrara} {et~al}\mbox{.}(1999){Ferrara}, {Bianchi}, {Cimatti}, \&
  {Giovanardi}}]{Ferrara1999}
{Ferrara} A., {Bianchi} S., {Cimatti} A., {Giovanardi} C., 1999, \apjs, 123,
  437

\bibitem[{{Fontanot} {et~al}\mbox{.}(2009){Fontanot}, {Somerville}, {Silva},
  {Monaco}, \& {Skibba}}]{Fontanot2009}
{Fontanot} F., {Somerville} R.~S., {Silva} L., {Monaco} P., {Skibba} R., 2009,
  \mnras, 392, 553

\bibitem[{{Gallazzi} {et~al}\mbox{.}(2005){Gallazzi}, {Charlot}, {Brinchmann},
  {White}, \& {Tremonti}}]{Gallazzi2005}
{Gallazzi} A., {Charlot} S., {Brinchmann} J., {White} S.~D.~M., {Tremonti}
  C.~A., 2005, \mnras, 362, 41

\bibitem[{Gelman(2004)}]{Gelman2004}
Gelman A., 2004, Bayesian data analysis, Texts in statistical science. Chapman
  \& Hall/CRC

\bibitem[{Gelman {et~al}\mbox{.}(2010)Gelman, Brooks, Jones, \&
  Meng}]{MCMC_handbook}
Gelman A., Brooks S., Jones G., Meng X., 2010, Handbook of Markov Chain Monte
  Carlo: Methods and Applications, Chapman \& Hall/CRC Handbooks of Modern
  Statistical Methods. Taylor and Francis

\bibitem[{Gelman \& Rubin(1992)}]{Gelman1992}
Gelman A., Rubin D.~B., 1992, Statistical Science, 7, 457

\bibitem[{Geweke(1989)}]{Geweke1989}
Geweke J., 1989, Econometrica, 57, 1317

\bibitem[{{Gordon} {et~al}\mbox{.}(2003){Gordon}, {Clayton}, {Misselt},
  {Landolt}, \& {Wolff}}]{Gordon2003}
{Gordon} K.~D., {Clayton} G.~C., {Misselt} K.~A., {Landolt} A.~U., {Wolff}
  M.~J., 2003, \apj, 594, 279

\bibitem[{{Gordon} {et~al}\mbox{.}(2001){Gordon}, {Misselt}, {Witt}, \&
  {Clayton}}]{Gordon2001}
{Gordon} K.~D., {Misselt} K.~A., {Witt} A.~N., {Clayton} G.~C., 2001, \apj,
  551, 269

\bibitem[{{Granato} \& {Danese}(1994)}]{Granato1994}
{Granato} G.~L., {Danese} L., 1994, \mnras, 268, 235

\bibitem[{{Granato} {et~al}\mbox{.}(2000){Granato}, {Lacey}, {Silva},
  {Bressan}, {Baugh}, {Cole}, \& {Frenk}}]{Granato2000}
{Granato} G.~L., {Lacey} C.~G., {Silva} L., {Bressan} A., {Baugh} C.~M., {Cole}
  S., {Frenk} C.~S., 2000, \apj, 542, 710

\bibitem[{{Grootes} {et~al}\mbox{.}(2013){Grootes}, {Tuffs}, {Popescu},
  {Pastrav}, {Andrae}, {Gunawardhana}, {Kelvin}, {Liske}, {Seibert}, {Taylor},
  {Graham}, {Baes}, {Baldry}, {Bourne}, {Brough}, {Cooray}, {Dariush}, {De
  Zotti}, {Driver}, {Dunne}, {Gomez}, {Hopkins}, {Hopwood}, {Jarvis},
  {Loveday}, {Maddox}, {Madore}, {Micha{\l}owski}, {Norberg}, {Parkinson},
  {Prescott}, {Robotham}, {Smith}, {Thomas}, \& {Valiante}}]{Grootes2013}
{Grootes} M.~W. {et~al.}, 2013, ArXiv e-prints

\bibitem[{{Groves} {et~al}\mbox{.}(2004){Groves}, {Dopita}, \&
  {Sutherland}}]{Groves2004}
{Groves} B., {Dopita} M., {Sutherland} R., 2004, in IAU Symposium, Vol. 222,
  The Interplay Among Black Holes, Stars and ISM in Galactic Nuclei,
  {T.~Storchi-Bergmann, L.~C.~Ho, \& H.~R.~Schmitt}, ed., pp. 263--266

\bibitem[{{Groves} {et~al}\mbox{.}(2008){Groves}, {Dopita}, {Sutherland},
  {Kewley}, {Fischera}, {Leitherer}, {Brandl}, \& {van Breugel}}]{Groves2008}
{Groves} B., {Dopita} M.~A., {Sutherland} R.~S., {Kewley} L.~J., {Fischera} J.,
  {Leitherer} C., {Brandl} B., {van Breugel} W., 2008, \apjs, 176, 438

\bibitem[{{Guthrie}(1992)}]{Guthrie1992}
{Guthrie} B.~N.~G., 1992, \aaps, 93, 255

\bibitem[{Haario {et~al}\mbox{.}(2001)Haario, Saksman, \&
  Tamminen}]{Haario2001}
Haario H., Saksman E., Tamminen J., 2001, Bernoulli, 7, 223

\bibitem[{Hastings(1970)}]{Hastings1970}
Hastings W.~K., 1970, Biometrika, 57, 97

\bibitem[{{Hill} {et~al}\mbox{.}(2010){Hill}, {Driver}, {Cameron}, {Cross},
  {Liske}, \& {Robotham}}]{Hill2010}
{Hill} D.~T., {Driver} S.~P., {Cameron} E., {Cross} N., {Liske} J., {Robotham}
  A., 2010, \mnras, 404, 1215

\bibitem[{{H{\"o}fner}(2009)}]{Hofner2009}
{H{\"o}fner} S., 2009, in Astronomical Society of the Pacific Conference
  Series, Vol. 414, Cosmic Dust - Near and Far, {T.~Henning, E.~Gr{\"u}n, \&
  J.~Steinacker}, ed., pp. 3--+

\bibitem[{{Jones}(2004)}]{Jones2004}
{Jones} A.~P., 2004, in Astronomical Society of the Pacific Conference Series,
  Vol. 309, Astrophysics of Dust, {A.~N.~Witt, G.~C.~Clayton, \& B.~T.~Draine},
  ed., pp. 347--+

\bibitem[{{Jones} \& {Nuth}(2011)}]{Jones2011}
{Jones} A.~P., {Nuth} J.~A., 2011, \aap, 530, A44+

\bibitem[{{Jonsson}(2006)}]{SUN2006}
{Jonsson} P., 2006, \mnras, 372, 2

\bibitem[{{Jonsson} {et~al}\mbox{.}(2006){Jonsson}, {Cox}, {Primack}, \&
  {Somerville}}]{Jonsson2006}
{Jonsson} P., {Cox} T.~J., {Primack} J.~R., {Somerville} R.~S., 2006, \apj,
  637, 255

\bibitem[{{Jonsson} {et~al}\mbox{.}(2010){Jonsson}, {Groves}, \&
  {Cox}}]{Jonsson2010}
{Jonsson} P., {Groves} B.~A., {Cox} T.~J., 2010, \mnras, 403, 17

\bibitem[{{Kass} {et~al}\mbox{.}(1998){Kass}, {Carlin}, {Gelman}, \&
  {Radford}}]{Kass1997}
{Kass} R.~E., {Carlin} B.~P., {Gelman} A., {Radford} M.~N., 1998, The American
  Statistician, 52, 93

\bibitem[{{Kauffmann} {et~al}\mbox{.}(2003c){Kauffmann}, {et al.}, \&
  {}}]{Kauffmann2003c}
{Kauffmann} G., {et al.}, {}, 2003c, \mnras, 346, 1055

\bibitem[{{Kauffmann} {et~al}\mbox{.}(2003a){Kauffmann}, {Heckman}, {White},
  {Charlot}, {Tremonti}, {Brinchmann}, {Bruzual}, {Peng}, {Seibert},
  {Bernardi}, \& et~al.}]{Kauffmann2003a}
{Kauffmann} G. {et~al.}, 2003a, \mnras, 341, 33

\bibitem[{{Kauffmann} {et~al}\mbox{.}(2003b){Kauffmann}, {Heckman}, {White},
  {Charlot}, {Tremonti}, {Peng}, {Seibert}, {Brinkmann}, {Nichol}, {SubbaRao},
  \& {York}}]{Kauffmann2003b}
{Kauffmann} G. {et~al.}, 2003b, \mnras, 341, 54

\bibitem[{{Kennicutt}(1998)}]{Kennicutt1998}
{Kennicutt}, Jr. R.~C., 1998, \apj, 498, 541

\bibitem[{{Kylafis} \& {Bahcall}(1987)}]{Kylafis1987}
{Kylafis} N.~D., {Bahcall} J.~N., 1987, \apj, 317, 637

\bibitem[{{Laor} \& {Draine}(1993)}]{Laor1993}
{Laor} A., {Draine} B.~T., 1993, \apj, 402, 441

\bibitem[{{Lawrence} {et~al}\mbox{.}(2007){Lawrence}, {Warren}, {Almaini},
  {Edge}, {Hambly}, {Jameson}, {Lucas}, {Casali}, {Adamson}, {Dye}, {Emerson},
  {Foucaud}, {Hewett}, {Hirst}, {Hodgkin}, {Irwin}, {Lodieu}, {McMahon},
  {Simpson}, {Smail}, {Mortlock}, \& {Folger}}]{Lawrence2007}
{Lawrence} A. {et~al.}, 2007, \mnras, 379, 1599

\bibitem[{{Lewis} \& {Bridle}(2002)}]{CosmoMC}
{Lewis} A., {Bridle} S., 2002, \prd, 66, 103511

\bibitem[{{MacLachlan} {et~al}\mbox{.}(2011){MacLachlan}, {Matthews}, {Wood},
  \& {Gallagher}}]{MacLachlan2011}
{MacLachlan} J.~M., {Matthews} L.~D., {Wood} K., {Gallagher} J.~S., 2011, \apj,
  741, 6

\bibitem[{{Mannucci} {et~al}\mbox{.}(2010){Mannucci}, {Cresci}, {Maiolino},
  {Marconi}, \& {Gnerucci}}]{Mannucci2010}
{Mannucci} F., {Cresci} G., {Maiolino} R., {Marconi} A., {Gnerucci} A., 2010,
  \mnras, 408, 2115

\bibitem[{{Mao} {et~al}\mbox{.}(2012){Mao}, {Kennicutt}, {Hao}, {Kong}, \&
  {Zhou}}]{Mao2012}
{Mao} Y.-W., {Kennicutt}, Jr. R.~C., {Hao} C.-N., {Kong} X., {Zhou} X., 2012,
  \apj, 757, 52

\bibitem[{{Marigo} {et~al}\mbox{.}(2008){Marigo}, {Girardi}, {Bressan},
  {Groenewegen}, {Silva}, \& {Granato}}]{Marigo2008}
{Marigo} P., {Girardi} L., {Bressan} A., {Groenewegen} M.~A.~T., {Silva} L.,
  {Granato} G.~L., 2008, \aap, 482, 883

\bibitem[{{Mathis} {et~al}\mbox{.}(1977){Mathis}, {Rumpl}, \&
  {Nordsieck}}]{Mathis1977}
{Mathis} J.~S., {Rumpl} W., {Nordsieck} K.~H., 1977, \apj, 217, 425

\bibitem[{Metropolis {et~al}\mbox{.}(1953)Metropolis, Rosenbluth, Rosenbluth,
  Teller, \& Teller}]{Metropolis1953}
Metropolis N., Rosenbluth A.~W., Rosenbluth M.~N., Teller A.~H., Teller E.,
  1953, The Journal of Chemical Physics, 21, 1087

\bibitem[{Mills(2011)}]{Mills2011}
Mills P., 2011, International Journal of Remote Sensing, 32, 6109

\bibitem[{{Mookerjea} {et~al}\mbox{.}(2011){Mookerjea}, {Kramer}, {Buchbender},
  {Boquien}, {Verley}, {Rela{\~n}o}, {Quintana-Lacaci}, {Aalto}, {Braine},
  {Calzetti}, {Combes}, {Garcia-Burillo}, {Gratier}, {Henkel}, {Israel},
  {Lord}, {Nikola}, {R{\"o}llig}, {Stacey}, {Tabatabaei}, {van der Tak}, \&
  {van der Werf}}]{Mookerjea2011}
{Mookerjea} B. {et~al.}, 2011, \aap, 532, A152

\bibitem[{{Murray}(2011)}]{Murray2011}
{Murray} N., 2011, \apj, 729, 133

\bibitem[{{Murray} {et~al}\mbox{.}(2010){Murray}, {Quataert}, \&
  {Thompson}}]{Murray2010}
{Murray} N., {Quataert} E., {Thompson} T.~A., 2010, \apj, 709, 191

\bibitem[{{O'Donnell}(1994)}]{ODonnell1994}
{O'Donnell} J.~E., 1994, \apj, 422, 158

\bibitem[{{Pacifici} {et~al}\mbox{.}(2012){Pacifici}, {Charlot}, {Blaizot}, \&
  {Brinchmann}}]{Pacifici2012}
{Pacifici} C., {Charlot} S., {Blaizot} J., {Brinchmann} J., 2012, \mnras, 421,
  2002

\bibitem[{{Panuzzo} {et~al}\mbox{.}(2003){Panuzzo}, {Bressan}, {Granato},
  {Silva}, \& {Danese}}]{Panuzzo2003}
{Panuzzo} P., {Bressan} A., {Granato} G.~L., {Silva} L., {Danese} L., 2003,
  \aap, 409, 99

\bibitem[{{Panuzzo} {et~al}\mbox{.}(2007){Panuzzo}, {Granato}, {Buat}, {Inoue},
  {Silva}, {Iglesias-P{\'a}ramo}, \& {Bressan}}]{Panuzzo2007}
{Panuzzo} P., {Granato} G.~L., {Buat} V., {Inoue} A.~K., {Silva} L.,
  {Iglesias-P{\'a}ramo} J., {Bressan} A., 2007, \mnras, 375, 640

\bibitem[{{Pei}(1992)}]{Pei1992}
{Pei} Y.~C., 1992, \apj, 395, 130

\bibitem[{{Pierini} {et~al}\mbox{.}(2004){Pierini}, {Gordon}, {Witt}, \&
  {Madsen}}]{Pierini2004}
{Pierini} D., {Gordon} K.~D., {Witt} A.~N., {Madsen} G.~J., 2004, \apj, 617,
  1022

\bibitem[{{Prevot} {et~al}\mbox{.}(1984){Prevot}, {Lequeux}, {Prevot},
  {Maurice}, \& {Rocca-Volmerange}}]{Prevot1984}
{Prevot} M.~L., {Lequeux} J., {Prevot} L., {Maurice} E., {Rocca-Volmerange} B.,
  1984, \aap, 132, 389

\bibitem[{Rasmussen \& Williams(2006)}]{Rasmussen2006}
Rasmussen C., Williams C., 2006, Gaussian processes for machine learning,
  Vol.~1. MIT press Cambridge, MA

\bibitem[{{Reyes} {et~al}\mbox{.}(2011){Reyes}, {Mandelbaum}, {Gunn},
  {Pizagno}, \& {Lackner}}]{Reyes2011}
{Reyes} R., {Mandelbaum} R., {Gunn} J.~E., {Pizagno} J., {Lackner} C.~N., 2011,
  \mnras, 417, 2347

\bibitem[{Robert \& Casella(2004)}]{MonteCarloIntegration}
Robert C., Casella G., 2004, Monte Carlo Statistical Methods, Springer Texts in
  Statistics. Springer

\bibitem[{Roberts \& Rosenthal(2001)}]{Roberts2001}
Roberts G., Rosenthal J., 2001, Statistical Science, 16, 351

\bibitem[{{Rocha} {et~al}\mbox{.}(2008){Rocha}, {Jonsson}, {Primack}, \&
  {Cox}}]{Rocha2008}
{Rocha} M., {Jonsson} P., {Primack} J.~R., {Cox} T.~J., 2008, \mnras, 383, 1281

\bibitem[{Roskar {et~al}\mbox{.}(2012)Roskar, Debattista, \&
  Loebman}]{Roskar2012}
Roskar R., Debattista V.~P., Loebman S.~R., 2012, arXiv.org, astro-ph.GA

\bibitem[{{Ryden}(2004)}]{Ryden2004}
{Ryden} B.~S., 2004, \apj, 601, 214

\bibitem[{{Ryden}(2006)}]{Ryden2006}
{Ryden} B.~S., 2006, \apj, 641, 773

\bibitem[{{S{\'a}nchez-Bl{\'a}zquez}
  {et~al}\mbox{.}(2006){S{\'a}nchez-Bl{\'a}zquez}, {Peletier},
  {Jim{\'e}nez-Vicente}, {Cardiel}, {Cenarro}, {Falc{\'o}n-Barroso}, {Gorgas},
  {Selam}, \& {Vazdekis}}]{MILES}
{S{\'a}nchez-Bl{\'a}zquez} P. {et~al.}, 2006, \mnras, 371, 703

\bibitem[{{Savage} \& {Mathis}(1979)}]{Savage1979}
{Savage} B.~D., {Mathis} J.~S., 1979, \araa, 17, 73

\bibitem[{{Schmidt}(1959)}]{Schmidt1959}
{Schmidt} M., 1959, \apj, 129, 243

\bibitem[{{Silva} {et~al}\mbox{.}(1998){Silva}, {Granato}, {Bressan}, \&
  {Danese}}]{Silva1998}
{Silva} L., {Granato} G.~L., {Bressan} A., {Danese} L., 1998, \apj, 509, 103

\bibitem[{{Silverman}(1986)}]{Silverman1986}
{Silverman} B.~W., 1986, {Density Estimation for Statistics and Data Analysis
  (Chapman \& Hall/CRC Monographs on Statistics \& Applied Probability)}.
  Chapman and Hall/CRC

\bibitem[{{Skilling}(2004)}]{Skilling2004}
{Skilling} J., 2004, in AIP Conference Proceedings, Vol. 735, p. 395

\bibitem[{{Skilling}(2006)}]{Skilling2006}
{Skilling} J., 2006, Bayesian Analysis, 1, 833

\bibitem[{{Smith} {et~al}\mbox{.}(2012){Smith}, {Dunne}, {da Cunha},
  {Rowlands}, {Maddox}, {Gomez}, {Bonfield}, {Charlot}, {Driver}, {Popescu},
  {Tuffs}, {Dunlop}, {Jarvis}, {Seymour}, {Symeonidis}, {Baes}, {Bourne},
  {Clements}, {Cooray}, {De Zotti}, {Dye}, {Eales}, {Scott}, {Verma}, {van der
  Werf}, {Andrae}, {Auld}, {Buttiglione}, {Cava}, {Dariush}, {Fritz},
  {Hopwood}, {Ibar}, {Ivison}, {Kelvin}, {Madore}, {Pohlen}, {Rigby},
  {Robotham}, {Seibert}, \& {Temi}}]{Smith2012}
{Smith} D.~J.~B. {et~al.}, 2012, \mnras, 427, 703

\bibitem[{Spearman(1904)}]{Spearman1904}
Spearman C., 1904, The American journal of psychology, 15, 72

\bibitem[{{Spitzer}(1978)}]{Spitzer1978}
{Spitzer} L., 1978, {Physical processes in the interstellar medium}

\bibitem[{{Springel}(2005)}]{GADGET2}
{Springel} V., 2005, \mnras, 364, 1105

\bibitem[{{Springel} {et~al}\mbox{.}(2005){Springel}, {White}, {Jenkins},
  {Frenk}, {Yoshida}, {Gao}, {Navarro}, {Thacker}, {Croton}, {Helly},
  {Peacock}, {Cole}, {Thomas}, {Couchman}, {Evrard}, {Colberg}, \&
  {Pearce}}]{Springel2005}
{Springel} V. {et~al.}, 2005, \nat, 435, 629

\bibitem[{{Springel} {et~al}\mbox{.}(2001){Springel}, {Yoshida}, \&
  {White}}]{Springel2001}
{Springel} V., {Yoshida} N., {White} S.~D.~M., 2001, \na, 6, 79

\bibitem[{{Tielens}(2005)}]{Tielens2005}
{Tielens} A.~G.~G.~M., 2005, {The Physics and Chemistry of the Interstellar
  Medium}

\bibitem[{{Tremonti} {et~al}\mbox{.}(2004){Tremonti}, {Heckman}, {Kauffmann},
  {Brinchmann}, {Charlot}, {White}, {Seibert}, {Peng}, {Schlegel}, {Uomoto},
  {Fukugita}, \& {Brinkmann}}]{Tremonti2004}
{Tremonti} C.~A. {et~al.}, 2004, \apj, 613, 898

\bibitem[{{Trotta} {et~al}\mbox{.}(2011){Trotta}, {J{\'o}hannesson},
  {Moskalenko}, {Porter}, {Ruiz de Austri}, \& {Strong}}]{Trotta2011}
{Trotta} R., {J{\'o}hannesson} G., {Moskalenko} I.~V., {Porter} T.~A., {Ruiz de
  Austri} R., {Strong} A.~W., 2011, \apj, 729, 106

\bibitem[{{Tuffs} {et~al}\mbox{.}(2004){Tuffs}, {Popescu}, {V{\"o}lk},
  {Kylafis}, \& {Dopita}}]{Tuffs2004}
{Tuffs} R.~J., {Popescu} C.~C., {V{\"o}lk} H.~J., {Kylafis} N.~D., {Dopita}
  M.~A., 2004, \aap, 419, 821

\bibitem[{{Wandelt}(2012)}]{AstrostatsWandelt2012}
{Wandelt} B.~D., 2012, in Astrostatistical Challenges for the New Astronomy,
  Hilbe J.~M., ed., Springer Series in Astrostatistics, Springer

\bibitem[{{Weingartner} \& {Draine}(2001)}]{Weing2001}
{Weingartner} J.~C., {Draine} B.~T., 2001, \apj, 548, 296

\bibitem[{{Wild} {et~al}\mbox{.}(2011b){Wild}, {Charlot}, {Brinchmann},
  {Heckman}, {Vince}, {Pacifici}, \& {Chevallard}}]{Wild2011b}
{Wild} V., {Charlot} S., {Brinchmann} J., {Heckman} T., {Vince} O., {Pacifici}
  C., {Chevallard} J., 2011b, \mnras, 417, 1760

\bibitem[{{Wild} {et~al}\mbox{.}(2011a){Wild}, {Groves}, {Heckman},
  {Sonnentrucker}, {Armus}, {Schiminovich}, {Johnson}, {Martins}, \&
  {Lamassa}}]{Wild2011a}
{Wild} V. {et~al.}, 2011a, \mnras, 410, 1593

\bibitem[{{Wild} {et~al}\mbox{.}(2007){Wild}, {Kauffmann}, {Heckman},
  {Charlot}, {Lemson}, {Brinchmann}, {Reichard}, \& {Pasquali}}]{Wild2007}
{Wild} V., {Kauffmann} G., {Heckman} T., {Charlot} S., {Lemson} G.,
  {Brinchmann} J., {Reichard} T., {Pasquali} A., 2007, \mnras, 381, 543

\bibitem[{{Witt} \& {Gordon}(2000)}]{Witt2000}
{Witt} A.~N., {Gordon} K.~D., 2000, \apj, 528, 799

\bibitem[{{Wuyts} {et~al}\mbox{.}(2011){Wuyts}, {F{\"o}rster Schreiber},
  {Lutz}, {Nordon}, {Berta}, {Altieri}, {Andreani}, {Aussel}, {Bongiovanni},
  {Cepa}, {Cimatti}, {Daddi}, {Elbaz}, {Genzel}, {Koekemoer}, {Magnelli},
  {Maiolino}, {McGrath}, {P{\'e}rez Garc{\'{\i}}a}, {Poglitsch}, {Popesso},
  {Pozzi}, {Sanchez-Portal}, {Sturm}, {Tacconi}, \& {Valtchanov}}]{Wuyts2011}
{Wuyts} S. {et~al.}, 2011, \apj, 738, 106

\bibitem[{{Xilouris} {et~al}\mbox{.}(1999){Xilouris}, {Byun}, {Kylafis},
  {Paleologou}, \& {Papamastorakis}}]{Xilouris1999}
{Xilouris} E.~M., {Byun} Y.~I., {Kylafis} N.~D., {Paleologou} E.~V.,
  {Papamastorakis} J., 1999, \aap, 344, 868

\bibitem[{{Yip} {et~al}\mbox{.}(2010){Yip}, {Szalay}, {Wyse}, {Dobos},
  {Budav{\'a}ri}, \& {Csabai}}]{Yip2010}
{Yip} C.-W., {Szalay} A.~S., {Wyse} R.~F.~G., {Dobos} L., {Budav{\'a}ri} T.,
  {Csabai} I., 2010, \apj, 709, 780

\bibitem[{{Yoachim} \& {Dalcanton}(2006)}]{Yoachim2006}
{Yoachim} P., {Dalcanton} J.~J., 2006, \aj, 131, 226

\bibitem[{{Yoachim} \& {Dalcanton}(2008)}]{Yoachim2008A}
{Yoachim} P., {Dalcanton} J.~J., 2008, \apj, 683, 707

\bibitem[{{Yoachim} {et~al}\mbox{.}(2012){Yoachim}, {Ro{\v s}kar}, \&
  {Debattista}}]{Yoachim2012}
{Yoachim} P., {Ro{\v s}kar} R., {Debattista} V.~P., 2012, \apj, 752, 97

\bibitem[{{York} {et~al}\mbox{.}(2000){York}, {Adelman}, {Anderson},
  {Anderson}, {Annis}, {Bahcall}, {Bakken}, {Barkhouser}, {Bastian}, {Berman},
  {Boroski}, {Bracker}, {Briegel}, {Briggs}, {Brinkmann}, {Brunner}, {Burles},
  {Carey}, {Carr}, {Castander}, {Chen}, {Colestock}, {Connolly}, {Crocker},
  {Csabai}, {Czarapata}, {Davis}, {Doi}, {Dombeck}, {Eisenstein}, {Ellman},
  {Elms}, {Evans}, {Fan}, {Federwitz}, {Fiscelli}, {Friedman}, {Frieman},
  {Fukugita}, {Gillespie}, {Gunn}, {Gurbani}, {de Haas}, {Haldeman}, {Harris},
  {Hayes}, {Heckman}, {Hennessy}, {Hindsley}, {Holm}, {Holmgren}, {Huang},
  {Hull}, {Husby}, {Ichikawa}, {Ichikawa}, {Ivezi{\'c}}, {Kent}, {Kim},
  {Kinney}, {Klaene}, {Kleinman}, {Kleinman}, {Knapp}, {Korienek}, {Kron},
  {Kunszt}, {Lamb}, {Lee}, {Leger}, {Limmongkol}, {Lindenmeyer}, {Long},
  {Loomis}, {Loveday}, {Lucinio}, {Lupton}, {MacKinnon}, {Mannery}, {Mantsch},
  {Margon}, {McGehee}, {McKay}, {Meiksin}, {Merelli}, {Monet}, {Munn},
  {Narayanan}, {Nash}, {Neilsen}, {Neswold}, {Newberg}, {Nichol}, {Nicinski},
  {Nonino}, {Okada}, {Okamura}, {Ostriker}, {Owen}, {Pauls}, {Peoples},
  {Peterson}, {Petravick}, {Pier}, {Pope}, {Pordes}, {Prosapio},
  {Rechenmacher}, {Quinn}, {Richards}, {Richmond}, {Rivetta}, {Rockosi},
  {Ruthmansdorfer}, {Sandford}, {Schlegel}, {Schneider}, {Sekiguchi}, {Sergey},
  {Shimasaku}, {Siegmund}, {Smee}, {Smith}, {Snedden}, {Stone}, {Stoughton},
  {Strauss}, {Stubbs}, {SubbaRao}, {Szalay}, {Szapudi}, {Szokoly}, {Thakar},
  {Tremonti}, {Tucker}, {Uomoto}, {Vanden Berk}, {Vogeley}, {Waddell}, {Wang},
  {Watanabe}, {Weinberg}, {Yanny}, \& {Yasuda}}]{SDSS}
{York} D.~G. {et~al.}, 2000, \aj, 120, 1579

\bibitem[{{Zahid} {et~al}\mbox{.}(2013){Zahid}, {Yates}, {Kewley}, \&
  {Kudritzki}}]{Zahid2013}
{Zahid} H.~J., {Yates} R.~M., {Kewley} L.~J., {Kudritzki} R.~P., 2013, \apj,
  763, 92

\end{thebibliography}

\appendix

%\include{App1}

%\clearpage

\section{Gaussian Random Process extrapolation}\label{app:GRP}

\subsection{Method}\label{sec:GRP_method}

We describe here the algorithm adopted to infer the attenuation curves of the T04 thick disc and bulge models in the range $0.1 < \lambda < 0.45$ \micron\ from the region $0.45 < \lambda < 2.1$ \micron , employing the T04 thin disc model, which is computed in the whole range $0.1 < \lambda < 2.1$ \micron . To make such an educated guess, we appeal to a Gaussian Random Proces \citep[e.g.][]{Rasmussen2006,AstrostatsWandelt2012}\footnote{The \citet{Rasmussen2006} book is freely available online at \url{http://www.Gaussianprocess.org/gpml/chapters/RW.pdf}.}. This is a generalisation of the Gaussian probability distribution, which determines the relative likelihood of random variables, to functions. A Gaussian Random Process hence measures the relative likelihood of different functions, that is of different relationships among random variables. To ease the understanding of a Gaussian Random Process, we recall here the definition of function: a function is a relation between a set of inputs and a set of admissible outputs with the property that each input is related to exactly a single output. Analytic functions (e.g. polynomials, power laws, splines) are a particular class of functions commonly used to solve interpolation and extrapolation problems, while a Gaussian Random Process is a powerful non-parametric method able to learn the (non-linear) relationships among a finite set of inputs and outputs. In particular, by appealing to a Gaussian Random Process we can solve any problem of interpolation and extrapolation (i.e. of inference) at arbitrary points from a finite set of known points (e.g. observations, simulations), within a fully Bayesian framework. A Gaussian Random Process is fully determined by a prior (over functions), and by a covariance matrix, which encodes the relationships among the known points. The result of such an approach is a posterior distribution over functions, which naturally accounts for the uncertainties due to a finite sample and to different possible relationships among the known variables.

To apply the algorithm, we compute a library of $n=2500$ attenuation curves from the T04 thin disc model (i.e. the `training library'), shown in Fig.~\ref{fig:GRF_T04}a. We consider 50 values of \taubP\ and $\theta$ uniformly distributed in the ranges $0.0 \leq \taubP \leq 8.0$ and $0 \leq \cos{\theta} \leq 1$, respectively, and 100 wavelength bins uniformly distributed in $\log{\lambda}$ in the range $0.1 < \lambda < 2.1$ \micron . We refer to any thin disc attenuation curve in the training library in the range $0.1 < \lambda < 0.45$ and $0.45 < \lambda < 2.1$ \micron\ as $X_i$ and $Y_i$ respectively. Any $X_i$ curve contains $n_X=49$ wavelength bins, and any $Y_i$ curve $n_Y=51$ bins. We then calculate the mean attenuation curves $\mu_X = 1/n \sum_i X_i$ and $\mu_Y = 1/n \sum_i Y_i$. Next, we compute the $n_Y \times n_Y$ sample covariance matrix $\mathbfss{C}_{YY}$\footnote{We adopt here the maximum likelihood estimate of the sample covariance matrix, so the denominator is $n$ and not $n-1$.}
\begin{equation}\label{eq:cyy}
{C}^{jk}_{YY} = \frac{1}{n} \sum_{i=1}^n \left ( Y^{j}_i - \mu^j_Y \right ) \left ( Y^{k}_i - \mu^k_Y \right ) \, ,
\end{equation} 
where the indices $k,j=1,n_Y$ span the wavelength bins in the range $0.45 < \lambda < 2.1$ \micron .
Then, we calculate the $n_X \times n_Y$ sample covariance matrix $\mathbfss{C}_{XY}$
\begin{equation}\label{eq:cxy}
{C}^{lk}_{XY} = \frac{1}{n} \sum_{i=1}^n \left ( X^{l}_i - \mu^l_X \right ) \left ( Y^{k}_i - \mu^k_Y \right ) \, ,
\end{equation} 
where the indices $l=i,n_X$ and $j=1,n_Y$ span the wavelength bins in the range $0.1 < \lambda < 0.45$ and $0.45 < \lambda < 2.1$ \micron\ respectively.
Finally, to infer the attenuation curve for any T04 thick disc and bulge model in the range $0.1 < \lambda < 0.45$ \micron , we calculate the attenuation curve $Y^{\ast}$ in the range $0.45 < \lambda < 2.1$ \micron, and obtain the posterior mean over functions $\mu_{X^{\ast}|Y}$, which we call $X^{\ast}$, in the range  $0.1 < \lambda < 0.45$ \micron
\begin{equation}\label{eq:GRF_retrieve}
X^{\ast}  \equiv \mu_{X^{\ast}|Y}  = \mu_X + \mathbfss{C}_{XY} \mathbfss{C}^{-1}_{YY} \left ( Y^{\ast} - \mu_Y \right ) \, .
\end{equation}
The curve $X^{\ast}$ is the most probable estimate of any thin disc and bulge curve in the range  $0.1 < \lambda < 0.45$ \micron , given the curve in the range  $0.45 < \lambda < 2.1$ \micron\ and the T04 thin disc model. 
The posterior covariance, which is a detailed quantification of the uncertainty in this prediction method, can be computed as
\begin{equation}\label{eq:GRF_covar}
 \mathbfss{C}_{X^{\ast}|Y} = \mathbfss{C}_{XX} - \mathbfss{C}_{XY} \mathbfss{C}^{-1}_{YY} \mathbfss{C}_{YX} \, .
\end{equation}

In the next two sections we test the accuracy of this approach by computing a set of attenuation curves in the range  $0.1 < \lambda < 2.1$ \micron, then comparing the curves predicted in the range  $0.1 < \lambda < 0.45$ \micron\ with the `true' ones.

\subsection{Tests of the accuracy of the method}

\subsubsection{Test on the T04 thin disc model}

\begin{figure*}
	\centering
	\subfigure
	{\resizebox{.45\hsize}{!}{\includegraphics{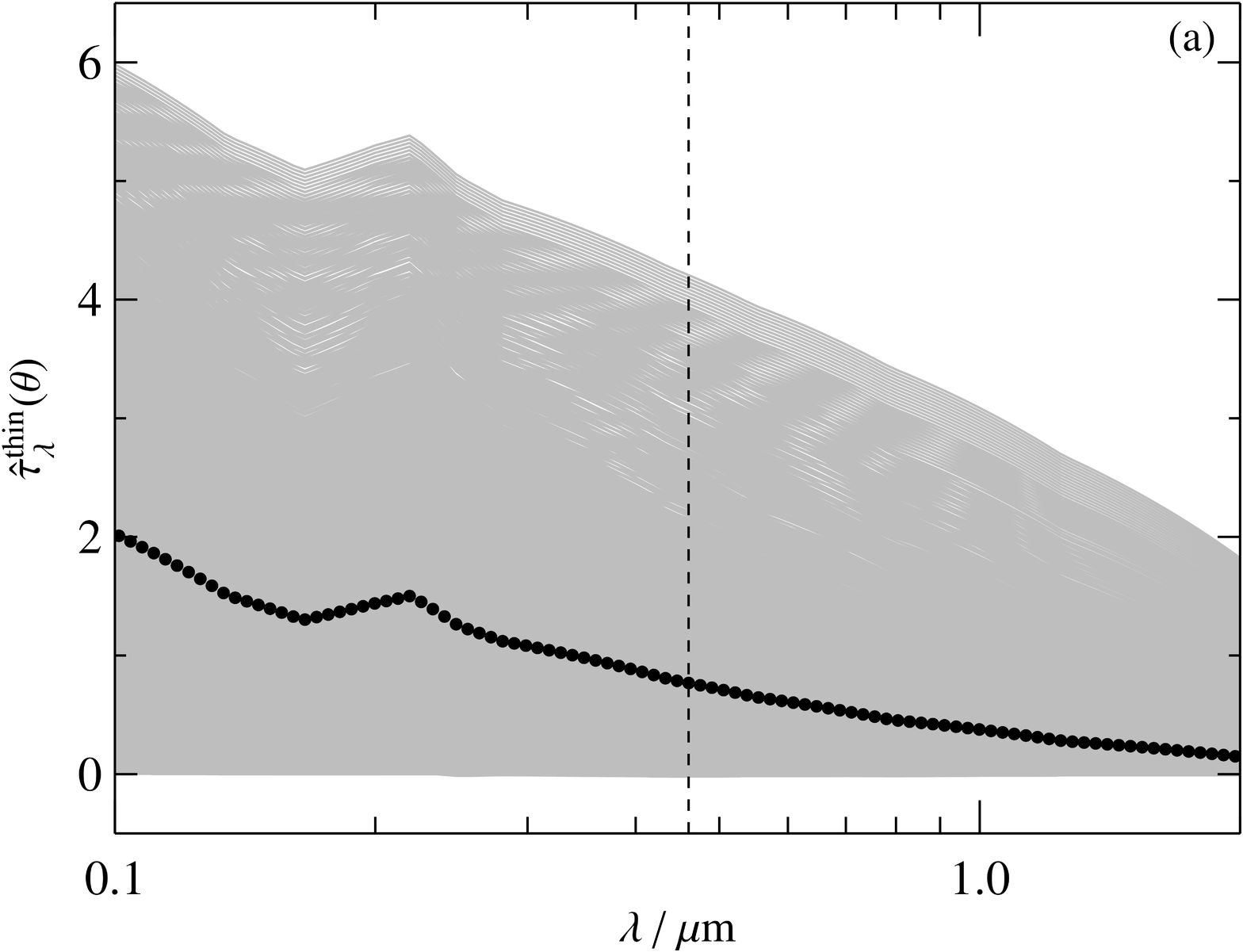}}}
	\hspace{.05\hsize}	
	\subfigure
	{\resizebox{.45\hsize}{!}{\includegraphics{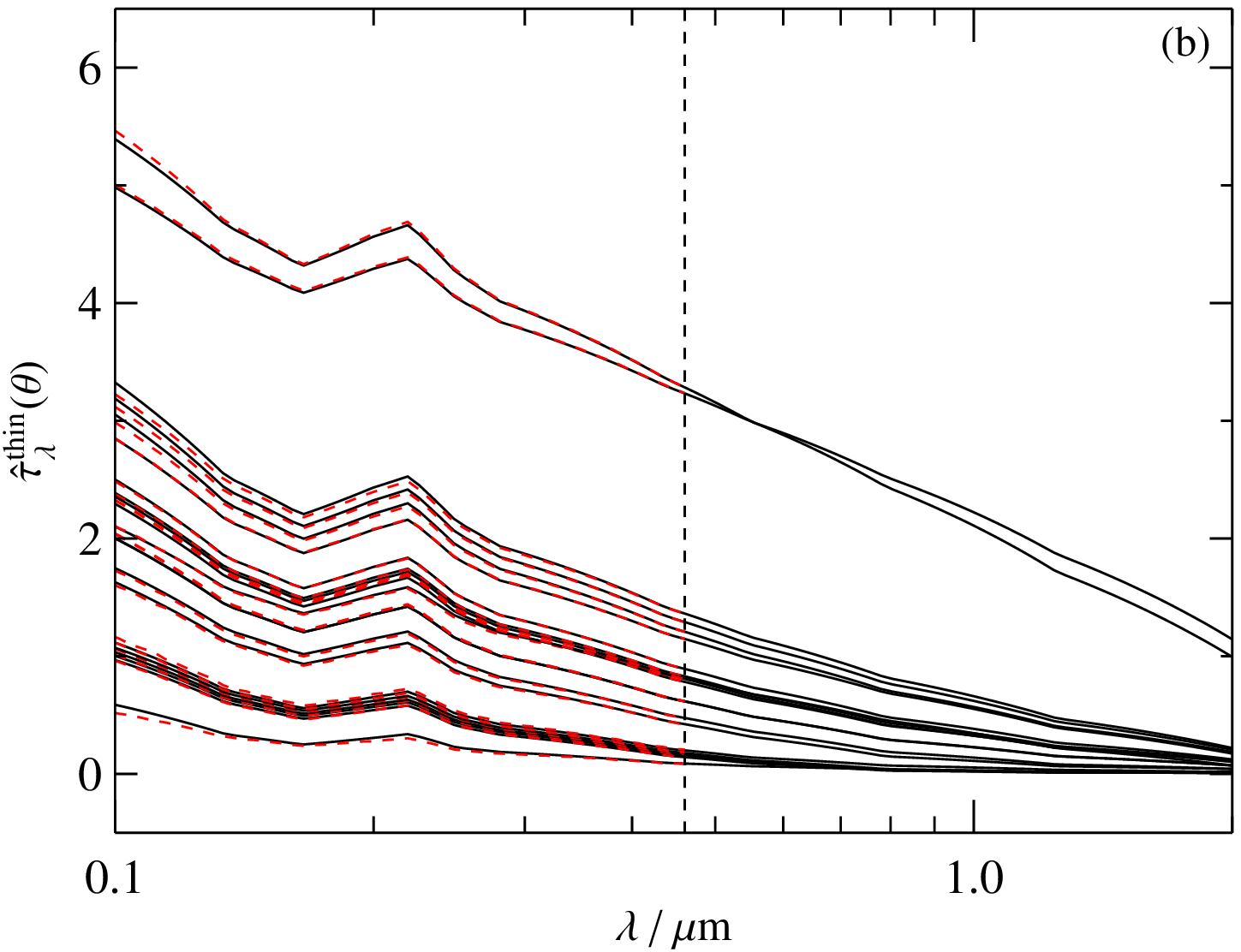}}}
	\caption{Test of the accuracy of the Gaussian Random Process method (section~\ref{app:GRP}) by means of the T04 thin disc model. ({\it a}) `Training' library of \numprint{2500} T04 thin disc attenuation curve. Solid points indicate the mean curve, along with the 100 log-spaced wavelength bins. ({\it b}) Test curves from the T04 thin disc model. Solid lines indicate the `true' curves in the whole range $0.1 < \lambda < 2.1$ \micron , dashed red lines the curves inferred in the range $0.1 < \lambda < 0.45$ \micron\ from the range $0.45 < \lambda < 2.1$ \micron\ through equation~(\ref{eq:GRF_retrieve}). In both panels, the vertical dashed line separates the ranges $0.1 < \lambda < 0.45$ and $0.45 < \lambda < 2.1$ \micron .}
	\label{fig:GRF_T04}
\end{figure*} 

For the first test, we compute 20 attenuation curves from the T04 thin disc model in the range $0.1 < \lambda < 2.1$ \micron\  by randomly drawing 20 pairs of $\theta$ and \taubP\ in the ranges $0 \leq \cos{\theta} \leq 1$ and $0.0 \leq $ \taubP $\leq 8$, respectively. We then use the training library described in Section~\ref{sec:GRP_method} to extrapolate the curves in the range $0.1 \leq \lambda \leq 0.45$ \micron\ from the region $0.45 < \lambda \leq 2.1$ \micron\ by means of equation~(\ref{eq:GRF_retrieve}). Fig.~\ref{fig:GRF_T04}b shows a remarkable agreement between the inferred and `true' curves, though the prediction becomes less accurate for curves very close to zero, where there is little information to constrain the extrapolation.

\subsubsection{Test on the \citet{Gordon2003} model}

\begin{figure*}
	\centering
	\subfigure
	{\resizebox{.45\hsize}{!}{\includegraphics{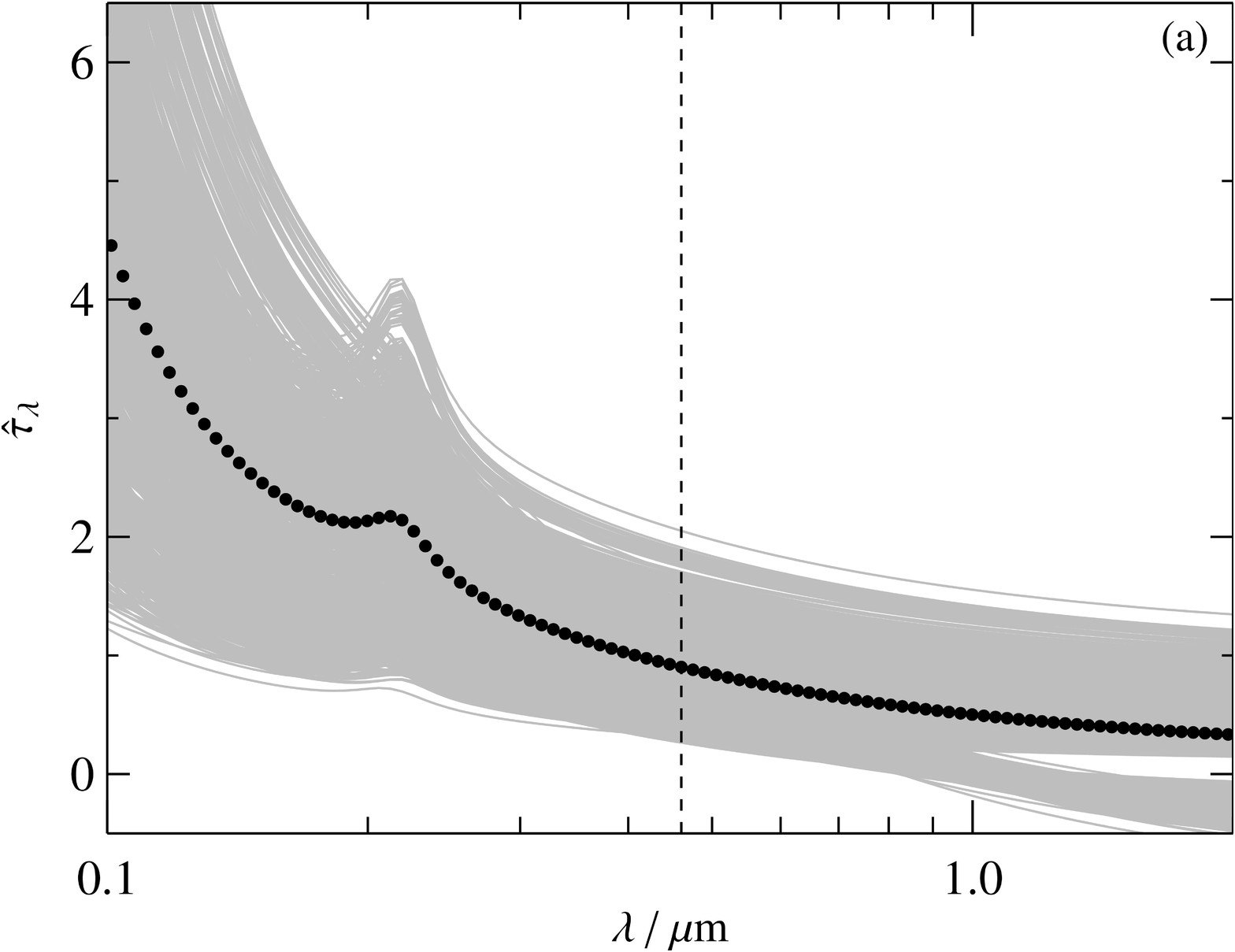}}}
	\hspace{.05\hsize}	
	\subfigure
	{\resizebox{.45\hsize}{!}{\includegraphics{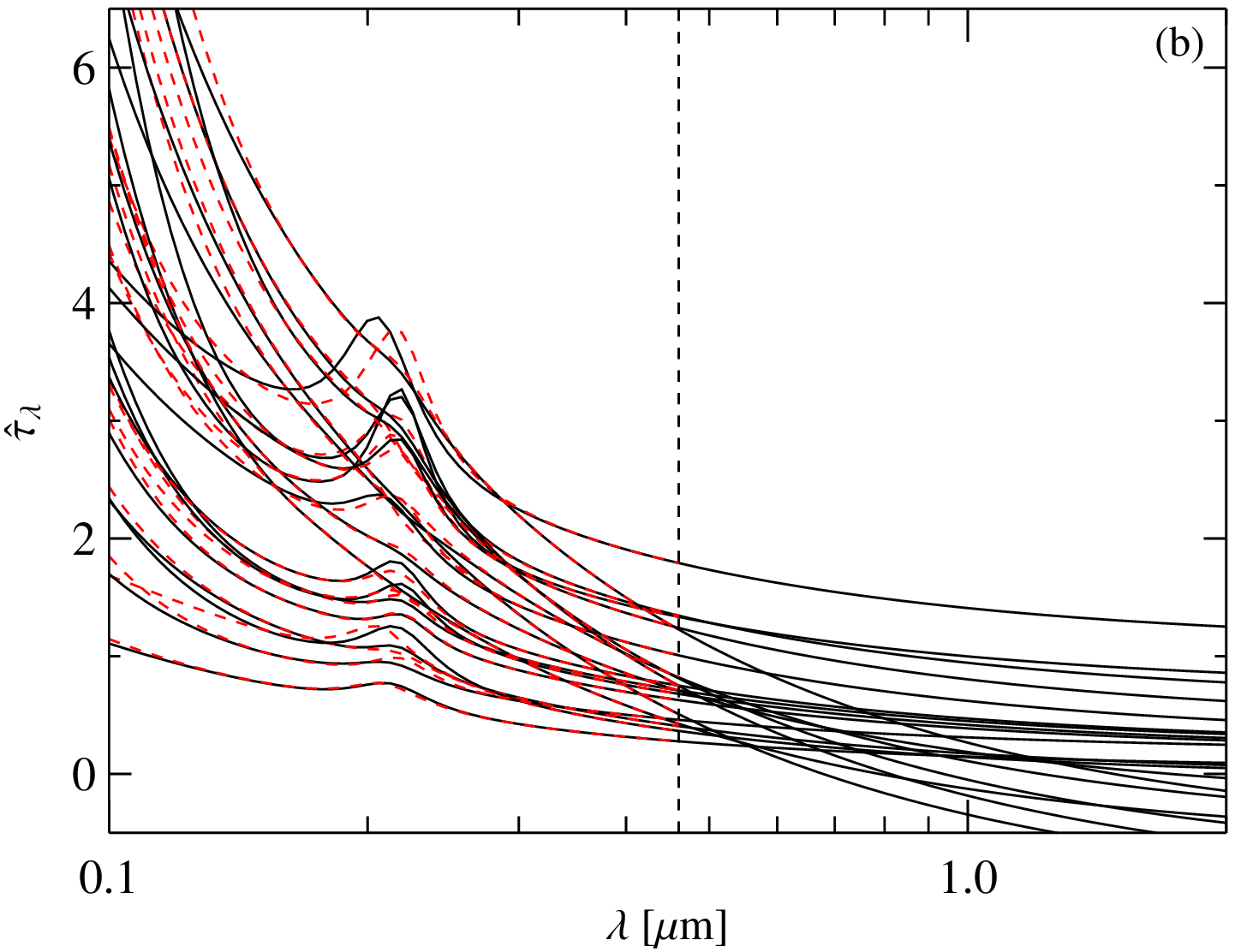}}}
	\caption{Test of the accuracy of the Gaussian Random Process method (section~\ref{app:GRP}) by means of the \citet{Gordon2003} model. ({\it a}) `Training' library of \numprint{1000} extinction curves from the parametrization of \citet{Gordon2003}. The solid points indicate the mean curve, along with the 100 log-spaced wavelength bins. ({\it b}) Test curves. Solid lines indicate the `true' curves in the whole range $0.1 < \lambda < 2.1$ \micron , dashed red lines the curves inferred in the range $0.1 < \lambda < 0.45$ \micron\ from the range $0.45 < \lambda < 2.1$ \micron\ through equation~(\ref{eq:GRF_retrieve}).  In both panels, the vertical dashed line separates the ranges $0.1 < \lambda < 0.45$ and $0.45 < \lambda < 2.1$ \micron .}
	\label{fig:GRF_Gordon}
\end{figure*}

This first test restricts the accuracy of the method in that we use the same model to generate the test sample and to build the training set. Hence, to better determine the accuracy of the method, we use a more flexible model, which allows us to generate test curves for model parameters exploring a larger range than the training set.
We build a library of extinction curves in the whole range $0.1 < \lambda < 2.1$ \micron\ following the parametrization of \citet{Gordon2003}, and, as in the previous test, try to infer the curves in the range $0.1 < \lambda < 0.45$ \micron\ from the region $0.45 < \lambda < 2.1$ \micron , eventually comparing the predicted curves with the `true' ones. \citet{Gordon2003} consider a comprehensive set of measurements of ultraviolet-to-near infrared extinction curves along different line of sights of the Large and Small Magellanic Clouds, adopting the following 6 parameters function to fit the observed curves
\begin{IEEEeqnarray}{rCl}\label{eq:gordon2003}
1.086 \, \hat{\tau}_\lambda & = & \big ( K_1 + K_2x + K_3D(x,\gamma,x_o) + K_4F(x) \big ) E(B-V)  \nonumber \\
	&& +\: R_V E(B-V)\, ,
	\label{eq:gordon}
\end{IEEEeqnarray}
where $x = \lambda^{-1}$, and
\begin{equation*}
D(x,\gamma,x_o) = \frac{x^2}{(x^2 - x_o^2)^2 + x^2 \gamma^2} \, ,
\end{equation*}
and
\begin{equation*}
F(x) = 0.5392(x - 5.9)^2 + 0.05644(x - 5.9)^3 \, .
\end{equation*}
We use equation~(\ref{eq:gordon2003}) to compute a library of 1000 extinction curves (i.e. the `training library'), shown in Fig.~\ref {fig:GRF_Gordon}a, sampled uniformly in $\log{\lambda}$ in the range $0.1 < \lambda < 2.1$ \micron . We consider four sets of parameters corresponding to the average `SMC Bar Sample', `SMC Wing Sample', `LMC/LMC2 Supershell Sample' and `LMC Average Sample' given in table 3 of \citet{Gordon2003}. To accomplish this, we randomly select one of the four sets of parameters and randomly draw each parameter from a Gaussian distribution centred on the parameter with variance equal to the error quoted in table 3 of \citet{Gordon2003}. Next, as for the library of T04 thin disc curves, we compute the mean curves $\mu_X$ and $\mu_Y$ in the ranges $0.1 < \lambda < 0.45$ and $0.45 < \lambda < 2.1$ respectively, and the sample covariance matrices $\mathbfss{C}_{YY}$ and $\mathbfss{C}_{XY}$ as defined in equations~(\ref{eq:cyy}) and (\ref{eq:cxy}), respectively. After, we compute 20 test curves by randomly drawing the parameters $K_1, K_2, K_3, K_4, x_0, \gamma$ from Gaussian distributions with variance equal three times the error quoted in table 3 of \citet{Gordon2003}. Finally, we calculate the curves in the region $0.1< \lambda < 0.45$ \micron\ from the range $0.45< \lambda < 2.1$ \micron\ through equation~(\ref{eq:GRF_retrieve}). We highlight that in this case the test curves span a wider range in the 6-dimensional parameter space than the library used to infer the curves, since the parameters $K_1, K_2, K_3, K_4, x_0, \gamma$ can reach more extreme values being drawn from Gaussian distributions with a larger variance than the one adopted to compute the training library.
Fig.~\ref{fig:GRF_Gordon}b shows the results of the Gaussian Random Process inference: it is noteworthy how well the curves in the range $0.1< \lambda < 0.45$ \micron\ are predicted from the region $0.45< \lambda < 2.1$ \micron , though the accuracy decreases at $\lambda \lesssim 0.25$ \micron . This has no impact on our analysis, since the bluer filter used in this work is the SDSS $u$, whose transmission drops to zero at 0.3 \micron . Note also that predictions obtained with a single Gaussian Random Process are accurate for both curves with and without the ultraviolet bump at 2175 \AA .

\subsection{Application of the method to the T04 thick disc and bulge models}
 
\begin{figure*}
	\centering
	\subfigure
	{\resizebox{.45\hsize}{!}{\includegraphics{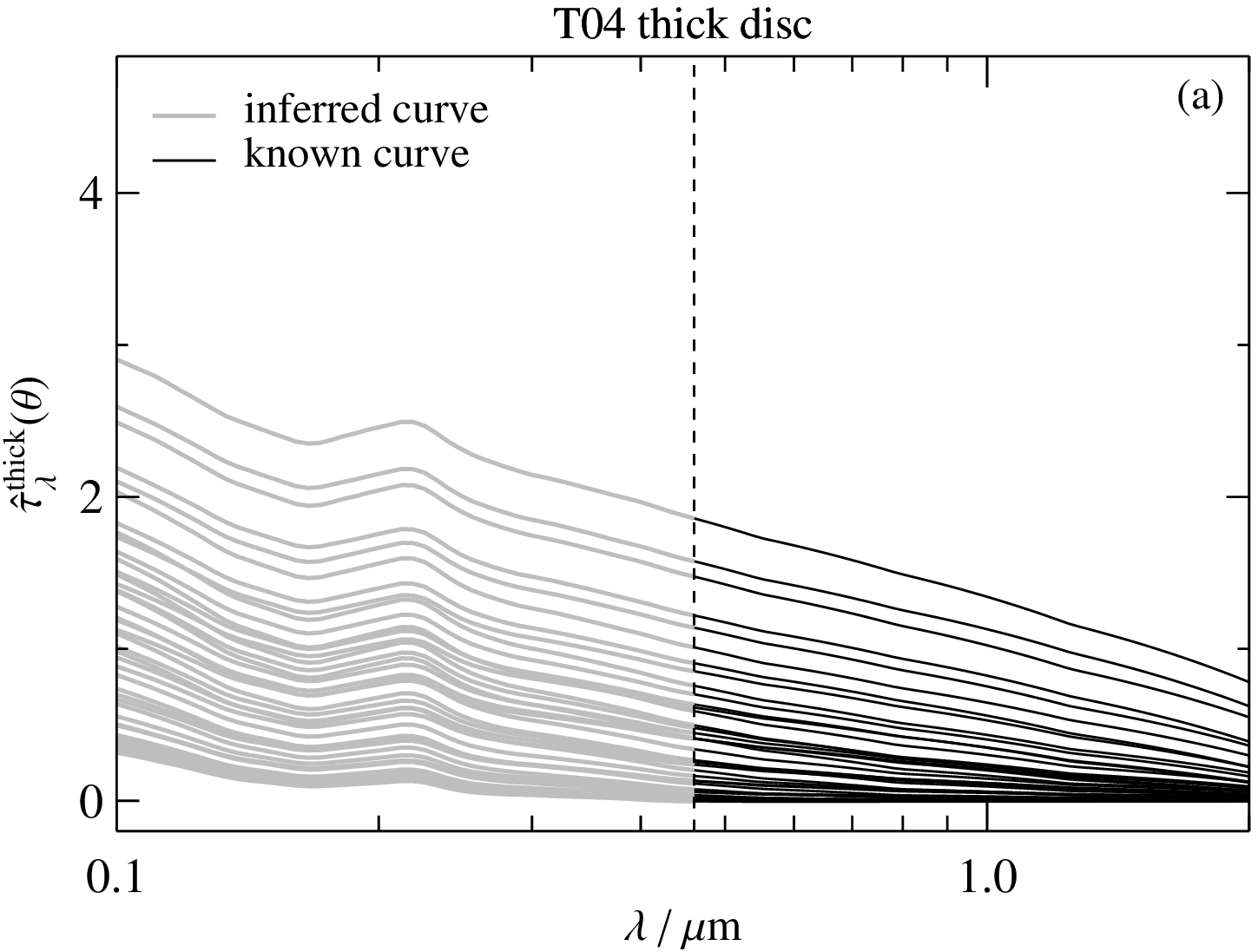}}}
	\hspace{.05\hsize}	
	\subfigure
	{\resizebox{.45\hsize}{!}{\includegraphics{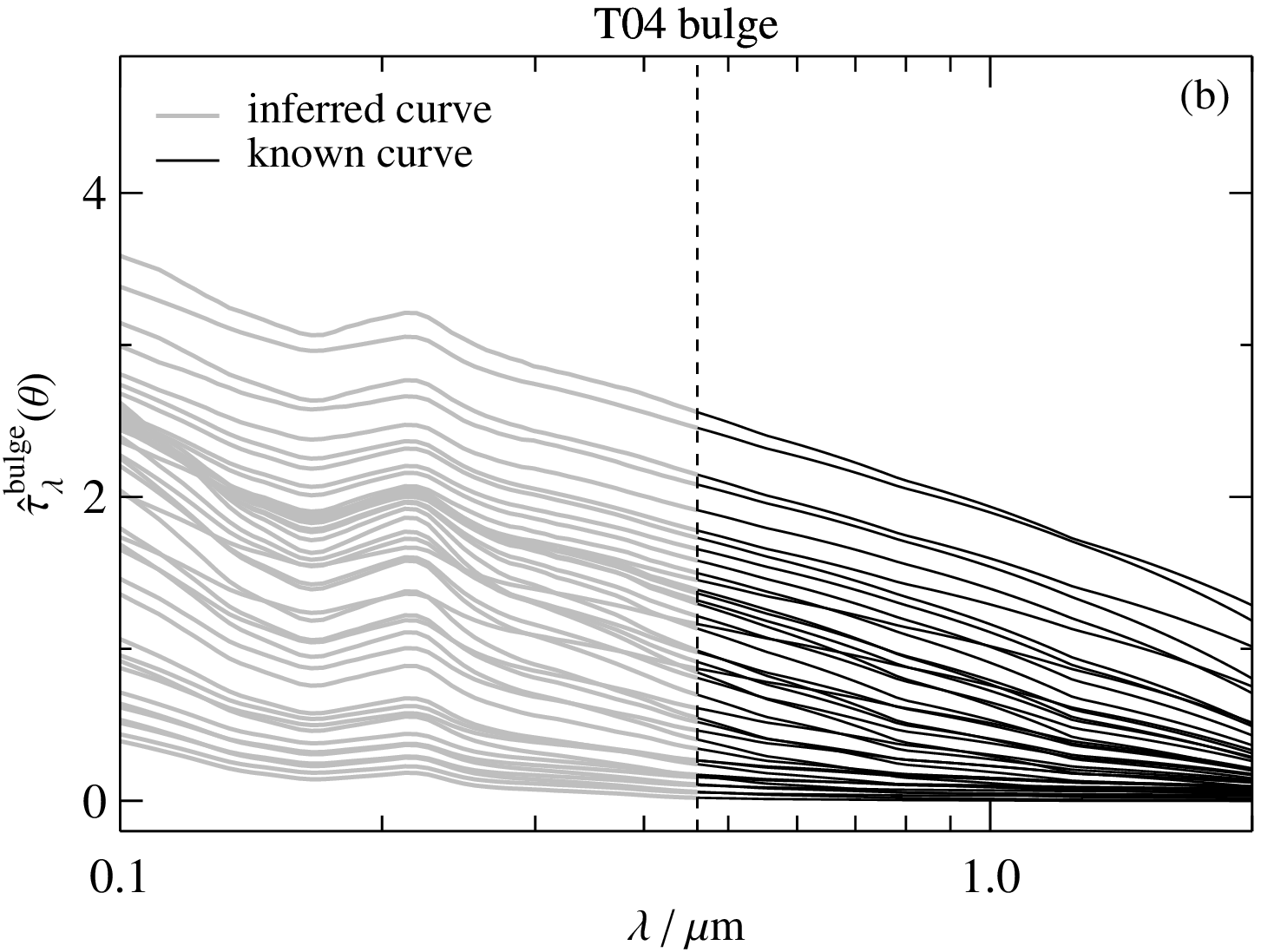}}}
	\caption{Results of the extrapolation of the T04 thick disc and bulge models by means of a Gaussian Random Process. ({\it a}) Solid lines indicate the T04 thick disc model computed by \citet{Tuffs2004} in the range $0.45< \lambda < 2$ \micron, dashed lines the curves inferred in the range $0.1< \lambda < 0.45$ \micron\ through equation~(\ref{eq:GRF_retrieve}). ({\it b})~Same as~({\it a}), but for the T04 bulge model.}
	\label{fig:GRF_disc_bulge}
\end{figure*}

The previous tests demonstrate that the extrapolation method described above, based on a Gaussian Random Process, is a reliable way to infer attenuation and extinction curves in the range $0.1 < \lambda < 0.45$ \micron\ from the region $0.45 < \lambda < 2.1$ \micron , exploiting a training sample and the correlation existing among different wavelength bins. We therefore apply the algorithm to the set of thick disc and bulge curves in the ranges $0 \leq \cos{\theta} \leq 1$ and $0.0 \leq $ \taubP $\leq 8$ computed by T04. We show in Figs.~\ref{fig:GRF_disc_bulge}a,b the curves predicted in the range $0.1 < \lambda < 0.45$ \micron\ from the region $0.45 < \lambda < 2.1$ \micron\ by means of equation~(\ref{eq:GRF_retrieve}).

\section{An importance sampling-based method to treat smooth systematics}\label{app:importance}

\subsection{Methods: Monte Carlo and Importance Sampling integration}

Any method to measure dust attenuation based on the ratio of the spectral energy distributions of two objects (e.g. stars, galaxies) assumes that the intrinsic spectral properties (i.e. un-attenuated) are equal. In this work, we measure the effect of dust from the ratio of the (normalised) mean spectral energy distribution of galaxies in different axis ratio bins to the face-on one. This traces dust attenuation if the mean stellar populations properties of the galaxies are constant across the axis ratio bins, since this guarantees that the un-attenuated spectral energy distributions are also constant. We check this hypothesis by comparing the bin-averaged and sample-averaged values of \logOH , \psiS, $z$ and $R_{90}$, finding that they systematically vary as a function of the axis ratio, hence biasing the measurement of dust attenuation.
To correct the biases, we develop an original method which uses the concepts and formalism of importance sampling. While in usual computations of average quantities each object has the same weight, this method allows us to assign a different weight to each object in each axis ratio bin accordingly to how likely it is to be drawn from a distribution common to all the bins. In this way we are able to substantially decrease the impact of biases on the attenuation measurements.
We recall here the basic concepts of Monte Carlo and importance sampling (a Monte Carlo technique) in the general case of $n$-dimensional distributions.

\begin{figure}
	\centering
	\resizebox{\hsize}{!}{\includegraphics{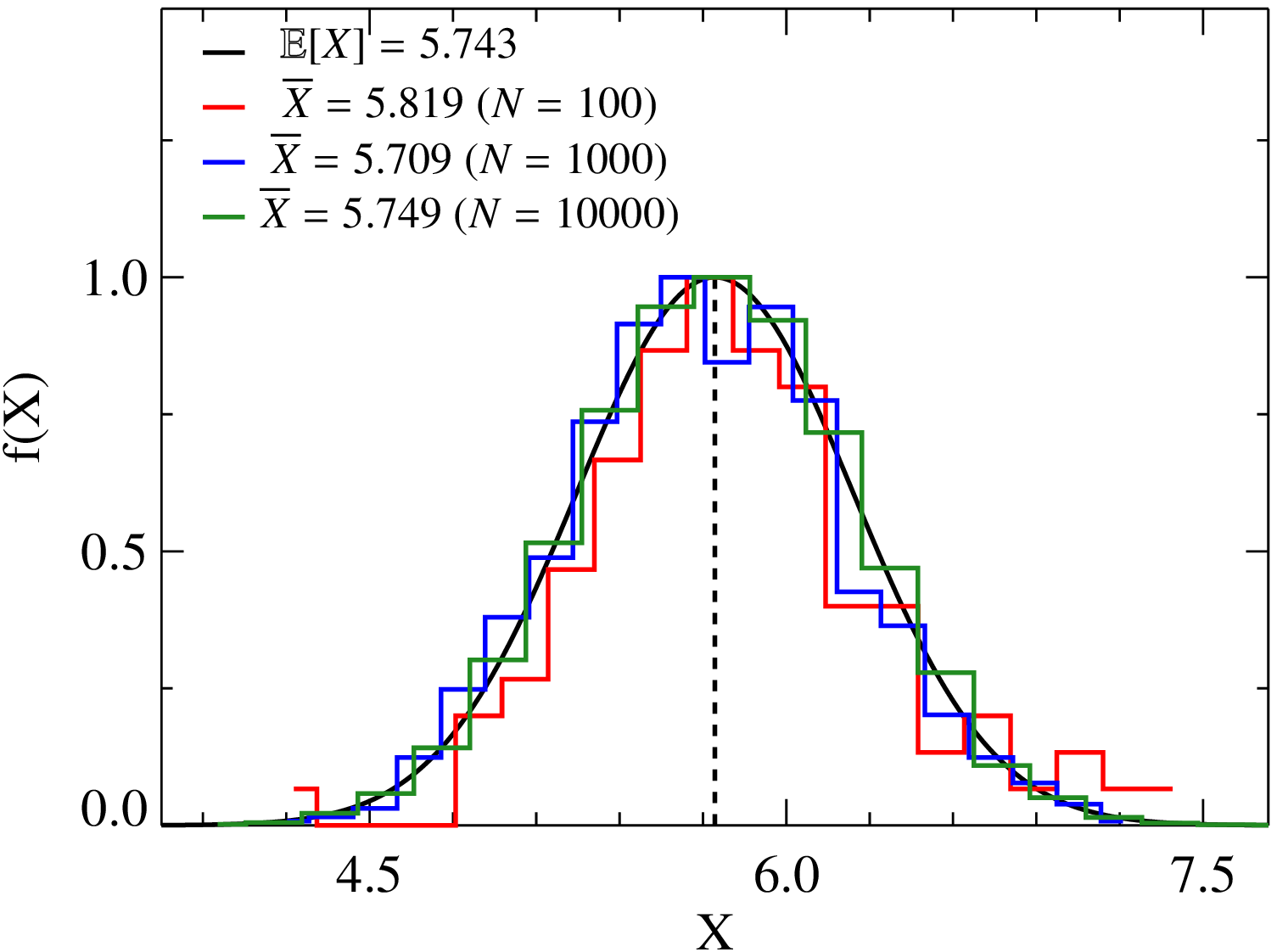}}
	\caption{Example of Monte Carlo integration of the univariate Gaussian function $f(x) = \mathcal{N}(5.743,0.248)$ (black line). Red, blue and dark green lines are histograms of 100, 1000 and $10\,000$ samples drawn from the Gaussian distribution $\mathcal{N}(5.743,0.248)$, used to compute the integral of $f(x)$ by means of equation~(\ref{eq:MonteCarloSamp}).}
	\label{fig:MonteCarloSamp}
\end{figure}

We start by recalling basic Monte Carlo integration, following the nomenclature of \citet{MonteCarloIntegration}. Suppose we want to evaluate the integral
\begin{equation} \label{eq:integral}
\mathbb{E}[h(\bmath{X})] = \int_{\mathcal{X}} d\bmath{x} \, h(\bmath{x}) \, f(\bmath{x}) \, ,
\end{equation}
over the domain $\mathcal{X}$, where $h(\bmath{x})$ is a function, $f(\bmath{x})$ a probability distribution, $\mathbb{E}$ indicates the expected (i.e. mean) value, and $\mathcal{X} \in \mathbb{R}^n$. In practical problems, it is often impossible to solve the above integral analytically, and numerical techniques which are effective in low dimensions (i.e. quadrature rules) become inefficient in highly dimensional problems (the so-called `curse of dimensionality'). The Monte Carlo method allows the estimation of the integral~(\ref{eq:integral}) through the simple expression 
\begin{equation} \label{eq:MonteCarloSamp}
\overline{h}_N = \frac{1}{N} \sum_{i=1}^N h(\bmath{x}_i) \, ,
\end{equation}
where $\overline{h}_N$ is the Monte Carlo estimate of $\mathbb{E}[h(\mathbi{X})]$ and $N$ the number of random variates $\bmath{x}_i$ drawn from the distribution $f(\bmath{x})$. The power of this method is that it transforms the operation of integration into summation. To apply equation~(\ref{eq:MonteCarloSamp}), we only need to draw random numbers from the multi-variate distribution $f(\bmath{x})$, then, by means of the Strong Law of Large Numbers, the above expression converges to the `true' value $\mathbb{E}[h(X)]$, with variance 
\begin{equation} \label{eq:MonteCarloSampVar}
var(\overline{h}_N) = \frac{1}{N^2} \sum_{i=1}^N \left [ h(\bmath{x}_i) - \overline{h}_N \right ] ^ 2 \, .
\end{equation}
Fig.~\ref{fig:MonteCarloSamp} shows an example of (uni-variate) Monte Carlo integration for $h(\bmath{x})=x$ with Gaussian probability distribution $ f(x) = \mathcal{N}(\mu,\sigma^2), \, x \in \mathbb{R}$, and $\mu=5.743$, $\sigma^2=0.248$. In this case one can easily solve the integral~(\ref{eq:integral}) analytically, obtaining the solution $\mathbb{E}[X]=5.743$. To compute the Monte Carlo estimate of the integral $\mathbb{E}(x)$, we draw $N$ random numbers from the probability distribution $f(x) = \mathcal{N}(5.743,0.248)$, then compute $\overline{x}_N$ with equation~(\ref{eq:MonteCarloSamp}). We show in Fig.~\ref{fig:MonteCarloSamp} the random samples for $N=100,1000$ and $10000$, along with the corresponding $\overline{x}_N$. Note that the error on $\overline{x}_N$ decreases approximately as $1/\sqrt{N}$, as expected from equation~(\ref{eq:MonteCarloSampVar}).

\begin{figure}
	\centering
	\resizebox{\hsize}{!}{\includegraphics{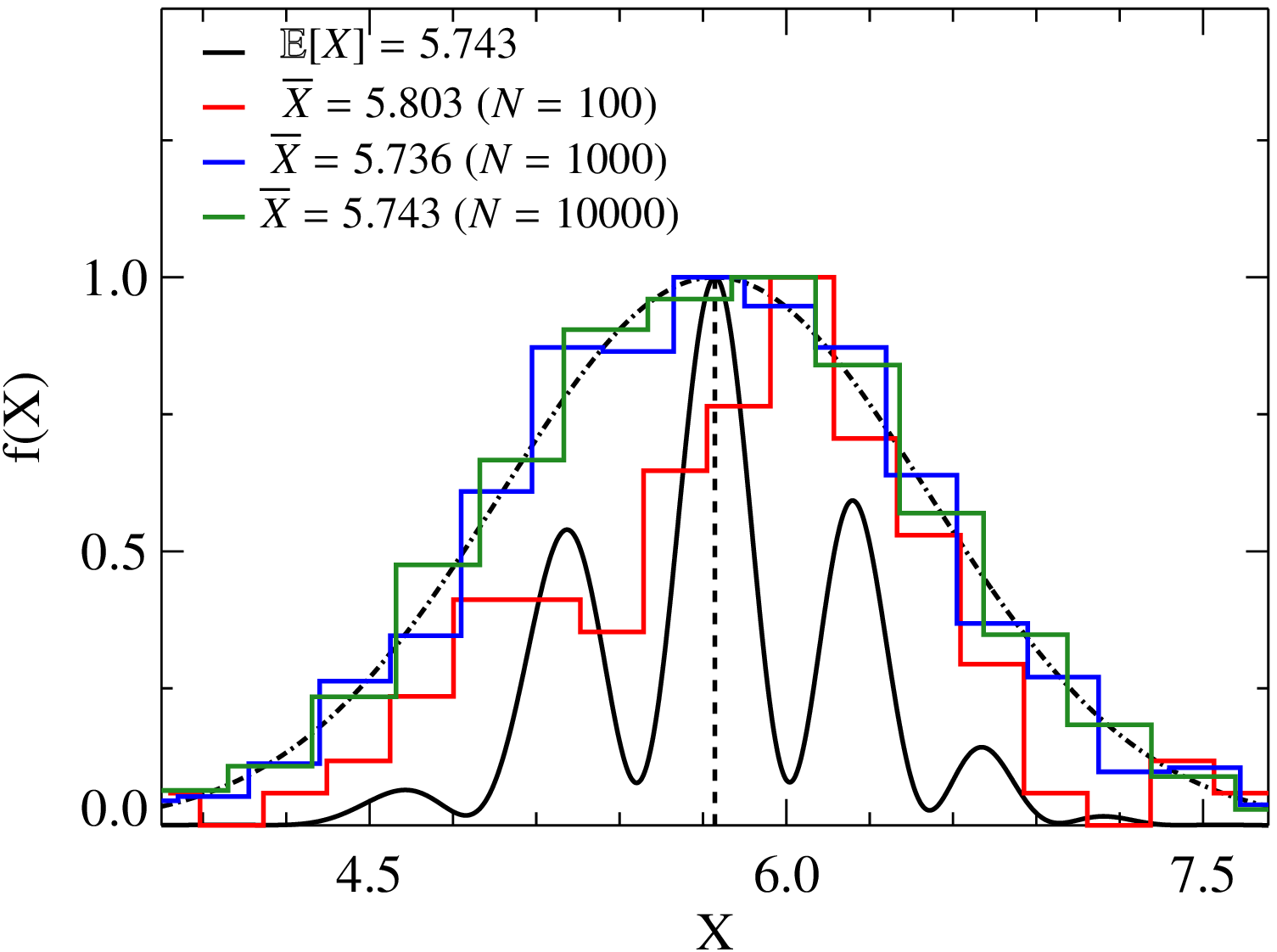}}
	\caption{Example of Importance Sampling integration of the univariate function $f(x) = \mathcal{N}(5.743,0.498) 0.2 \sin{x^2} + \mathcal{N}(5.743,0.498)$ (black line). Dot-dashed line is the importance function $g(x) = \mathcal{N}(5.743,0.77)$, red, blue and dark green lines are histograms of 100, 1000 and $10\,000$ samples drawn from the importance function $g(x)$ used to compute the integral of $f(x)$ by means of equation~(\ref{eq:ImpSampInt}).}
	\label{fig:MonteCarloImpSamp}
\end{figure}

In the previous example, we have solved the integral~(\ref{eq:integral}) in the simple case of $x$ following a Gaussian probability distribution$f(x)=\mathcal{N}(\mu,\sigma^2)$, exploiting the fact that we can efficiently draw samples from a Gaussian distribution, using, for instance, the Box-Muller transform \citep{Box_Muller}. In many situations it is not feasible, or computationally efficient, to draw samples from $f(\bmath{x})$, so we can exploit the fact that the integral~(\ref{eq:integral}) is equivalent to 
\begin{equation} \label{eq:ImpSampInt}
\mathbb{E}[h(\mathbi{X})] = \int_{\mathcal{X}} dx \, h(\bmath{x}) \, \frac{f(\bmath{x})}{g(\bmath{x})} g(\bmath{x}) \, ,
\end{equation}
where we have introduced the `importance function' $g(\bmath{x})$, to rewrite the Monte Carlo estimate~(\ref{eq:MonteCarloSamp}) as
\begin{equation}\label{eq:ImpSamp}
%\overline{h}_N = \frac{1}{N} \sum_{i=1}^N h(x_i) \frac{f(x_i)}{g(x_i)}  \, .
\overline{h_N^{IS}} =  \frac{\sum_{i=1}^N h(\bmath{x}_i) \, f(\bmath{x}_i)/g(\bmath{x}_i)}{\sum_{i=1}^Nf(\bmath{x}_i)/g(\bmath{x}_i)}  \, .
\end{equation}
Equation~(\ref{eq:ImpSamp}) expresses the importance sampling estimate of the integral~(\ref{eq:integral}), and it differs from the Monte Carlo estimate~(\ref{eq:MonteCarloSamp}) since it allows the evaluation of properties of a distribution (e.g. the expected value) by drawing samples from a different distribution. 
Also, importance sampling is a variance reduction technique, since, while equation~(\ref{eq:MonteCarloSamp}) implies a uniform sampling from the probability distribution $f(\bmath{x})$, equation~(\ref{eq:ImpSamp}) allows one to concentrate the sampling on the most `important' regions of $f(\bmath{x})$ (i.e. those that contribute the most to the integral), by drawing samples from a different distribution $g(\bmath{x})$. Equation~(\ref{eq:ImpSamp}) is similar to (\ref{eq:MonteCarloSamp}), except for the weight factor $w(\bmath{x}_i)=f(\bmath{x}_i)/g(\bmath{x}_i)$, which accounts for the fact that we draw samples from the `wrong' distribution. Here we normalise the weights to 1, so the functions $f(\bmath{x})$ need not to be normalised. In this case the quantity $\overline{h_N^{IS}}$ is biased with bias \citep{Geweke1989}
\begin{equation}\label{eq:ImpSampBias}
\Delta(\overline{h_N^{IS}}) =  \frac{\sum_{i=1}^N h(\bmath{x}_i) \, f(\bmath{x}_i)/g(\bmath{x}_i)}{\sum_{i=1}^Nf(\bmath{x}_i)/g(\bmath{x}_i)}  \, ,
\end{equation}
and variance
\begin{equation} \label{eq:ImpSampVar}
var(\overline{h_N^{IS}}) = \frac{N}{N-1} \frac{ \sum_{i=1}^N \left ( h(\bmath{x}_i)-\overline{h}_N \right )^2 w_i^2 }{\left ( \sum_{j=1}^N w_j \right )^2} \, .
\end{equation}
To be a valid importance function, $g(\bmath{x})$ has to fulfil the following condition
\begin{equation}\label{eq:ConditImpSamp}
g(\bmath{x}) = 0  \implies f(\bmath{x}) = 0 \, ,
\end{equation}
which guarantees the equality of the integrals~(\ref{eq:integral}) and (\ref{eq:ImpSampInt}).
This means that the function $g(\bmath{x})$ has to be non-null for every $\bmath{x}$ in the support of $f(\bmath{x})$, or, in other words, that all the values of $\bmath{x}$ which have non-null probability in $f(\bmath{x})$ must have non-null probability also in $g(\bmath{x})$. Though this is a sufficient condition for an importance function, in practical applications we need a further constraint, that is an estimator with finite variance. This translates in a further condition 
\begin{equation}\label{eq:ConditImpSamp2}
\txn{for } M > 0 \txn{ and }  \forall x \txn{, } \frac{f(\bmath{x})}{g(\bmath{x})} < M  \implies var(\overline{h}_N) < \infty \, ,
\end{equation}
which means that, for every $\bmath{x}$ in the domain $\mathcal{X}$, the tails of $g(\bmath{x})$ must be at least as heavy as the ones of $f(\bmath{x})$.
Fig.~\ref{fig:MonteCarloImpSamp} exemplifies the importance sampling integration of the uni-variate function $f(x)=\mathcal{N}(5.743,0.498)\, 0.2 \sin{x^2} + \mathcal{N}(5.743,0.498)$. Since we can not easily draw samples from $f(x)$, we adopt a Gaussian distribution $g(x) = \mathcal{N}(\mu_x,\sigma_x)$ as importance function and draw samples from it. We then compute the importance sampling estimate of $f(x)$ with equation~(\ref{eq:ImpSamp}), obtaining the estimates reported in Fig.~\ref{fig:MonteCarloImpSamp}.

\subsection{Application of the method to the biases of the W11 sample}

We now describe how we make use of the importance sampling formalism to correct for the biases in the dust attenuation measurements. We can arrange the gas-phase metallicity \logOH , specific star formation rate \psiS , redshift $z$ and radius which contains 90 per cent of the $r$-band Petrosian flux $R_{90}$, in a vector $\bmath{x} = [\logOH, \psiS,z,R_{90}]$. Each galaxy is then associated to a vector $\bmath{x}_i$, and we can write the systematic variation of the bin-averaged values of \logOH, \psiS, $z$ and $R_{90}$ with the axis ratio $b/a$ as
\begin{equation*}
\mathbb{E}[ p(\bmath{x}|b/a) ] = \mathbb{E}[p(\bmath{x})] + \Delta(\bmath{x}|b/a) \, ,
\end{equation*}
where $\Delta(\bmath{x}|b/a)$ indicates the axis ratio dependent bias.
We can re-write the above expression by using the definition of expected value~(\ref{eq:integral}) in the case of a discrete distribution
\begin{equation*}
\sum_i \bmath{x_i} p[\bmath{x_i}|(b/a)_i] = \sum_i \bmath{x_i} p(\bmath{x_i}) + \Delta[\bmath{x}_i|(b/a)_i]\, ,
\end{equation*}
where we sum over al possible values of $\bmath{x}$, and $p(\bmath{x_i})=p(\bmath{x_j})$ if we assume that the observations are independent and equiprobable when accounting separately for the effect of the axis ratio dependent  bias.
We can now introduce a new probability distribution $p^\prime(\bmath{x})$, which we assume independent of the axis ratio $b/a$, so as $p^\prime(\bmath{x}|b/a) = p^\prime(\bmath{x})$. We can interpret this as the probability distribution $f(\bmath{x})$ of equation~(\ref{eq:ImpSamp}), and $p(\bmath{x}|b/a)$ as the importance function $g(\bmath{x})$ of the same equation~(\ref{eq:ImpSamp}).
If we apply equation~(\ref{eq:ImpSampInt}), replacing $f(\bmath{x})$ with $p^\prime(\bmath{x}|b/a)$ and $g(\bmath{x})$ with $p(\bmath{x}|b/a)$, we obtain 
\begin{equation}\label{eq:w_mean}
\overline{\bmath{x}} = \frac{ \sum_i \bmath{x_i} p^\prime(\bmath{x_i})/p[\bmath{x_i}|(b/a)_i] } {\sum_i p^\prime(\bmath{x_i})/p[\bmath{x_i}|(b/a)_i] }\, ,
\end{equation}
which is by construction constant across the axis ratio bins.

To apply this method we must be able to compute the probability distribution $p(\bmath{x}|(b/a))$ in each axis ratio bin. To accomplish this, we can not rely on multi-dimensional histograms, since the probability distributions so obtained are non-smooth and dependent on the choice of the bin widths, hence we appeal to a \emph{kernel density estimation}. This uses a kernel to assign to each point in the space defined by the parameters $\bmath{x}$ a density (i.e. a probability). A commonly adopted kernel is the Gaussian kernel, which weights each data point with a multi-variate Gaussian function centred on the point itself, with diagonal covariance. This allows a smooth estimate of the joint probability distribution, though the choice of the diagonal covariance is critical. Fortunately, it is possible to find an optimal covariance, also called filter bandwidth, through the mean square error minimisation, by adapting the Gaussian kernel covariance to the density of the points at each location \citep[`adaptive Gaussian kernel density estimator', see][]{Silverman1986}. 
We use the \LibAGF\ library \citep{Mills2011} to compute for each axis ratio bin $k$ a smooth probability distribution $p_k(\bmath{x}|(b/a)_k)$, which can be evaluated at arbitrary points in the parameter space defined by the vector $\bmath{x}$. Next, we need to find a suitable probability distribution $p^\prime(\bmath{x})$, so as criteria~(\ref{eq:ConditImpSamp})--(\ref{eq:ConditImpSamp2}) are valid. We therefore adopt the function
\begin{equation*}
p^\prime(\bmath{x}) = \prod_k p_k(\bmath{x}|(b/a)_k)^{0.1} \, ,
\end{equation*}
which satisfies condition~(\ref{eq:ConditImpSamp}), since a null value in any of the functions $p_k(\bmath{x}|(b/a)_k)$ make null the corresponding value of $p^\prime(\bmath{x})$.  The choice of the exponent of the product of the densities $p_k(\bmath{x}|(b/a)_k)$ determines, for each axis ratio bin, the fraction of galaxies with significant weights; an exponent equal to 1 produces a very peaked probability distribution $p^\prime(\bmath{x})$, so as less than 1 \% of the objects within each bin have significant weights. This produces an almost perfect erase of the biases, assuming that the physical quantities $\bmath{x} = [\logOH, \psiS,z,R_{90}]$ are perfectly measured. Otherwise, this result would be dominated by biases and errors on the measurements of the physical parameters $\bmath{x}$. After some experimentation, we find that a value of the exponent of 0.1 produces a decrease of the biases at the level of $ < 10$ percent, while keeping the fraction of objects with significant weights above 10 percent, and hence decreasing the risk of being dominated by measurement biases and errors. To check the validity of condition~(\ref{eq:ConditImpSamp2}), we examine the behaviour of the tails of the distribution $p^\prime(\bmath{x})$ through the histograms of the weights $w_k(\bmath{x_i}) = p^\prime(\bmath{x})/p[\bmath{x_i}|(b/a)_k]$.
In the end, we obtain for each galaxy in the low- and high-\muS\ samples a weight $w_k(\bmath{x_i})$, which we use to compute the expectations (i.e. the mean values) of the \HaHb\ ratio and $ugrizYJH$ attenuation in each axis ratio bin.  

\section{Markov Chain Monte Carlo}\label{app:MCMC}

In a Bayesian framework, the degree of belief in a set of parameters, which define a model, given a set of data, is quantified by the posterior probability distribution defined by Bayes' Theorem. Evaluating the posterior probability distribution from a multi-dimensional grid of parameters can be computationally demanding because, at fixed accuracy, computational time scales exponentially with the number of parameters. To overcome this limitation, the posterior probability distribution can be represented through a set of samples drawn from it. In practice, a sampled representation allows an easy computation of one-dimensional (or two-dimensional) marginal posterior probabilities, which provide a quantitative summary of the knowledge about individual parameters (or different pairs of parameters) given a set of data.

Markov Chain Monte Carlo (MCMC) is a general method to generate random samples from a large class of probability distributions for which direct sampling algorithms are not known. Several MCMC algorithms exist \citep [see][and references therein]{MCMC_handbook}, some based on a random walk through the parameter space. The power of random walk-based MCMC algorithms is that the set of all points visited by the random walk converges towards the desired target distribution. In this work, we consider as the target distribution the posterior probability distributions of the five adjustable parameter of the dust model \tauVbc ,\taubP ,\Tthin, \Tthick\ and \Tbulge\ (see Section~\ref{sec:dust}).
In the following section we will give details of the specific random walk method adopted, and discuss the diagnostics that we use to asses the convergence of the MCMC to the target distribution. 

\subsection{The Metropolis-Hastings algorithm}\label{subsec:MH}

The Metropolis-Hastings sampler \citep{Metropolis1953,Hastings1970} is one of the most widely used MCMC algorithms.The algorithm iteratively explores the parameter space by means of a random walk, eventually converging towards the target distribution. It needs a set of starting parameters $\bmath{X}_{j=0}$, which are usually randomly drawn within the priors range. Then, at each iteration $j$, a new set of parameters $\bmath{X}'_j$ is drawn from a proposal distribution $q(\bmath{X}'_j|\bmath{X}_{j-1})$, and the relative probabilities of the parameters $\bmath{X}'_j$ and $\bmath{X}_{j-1}$ are compared by means of the Metropolis ratio, defined as
\begin{equation}\label{eq:metrop_ratio}
r =\frac{p(\bmath{X}'_j|D,I)}{p(\bmath{X}_{j-1}|D,I)}\cdot \frac{q(\bmath{X}_{j-1}|\bmath{X}_{j})}{q(\bmath{X}'_j|\bmath{X}_{j-1})}
\end{equation}
Next, the Metropolis ratio $r$ is compared with a random number $n$ uniformly drawn between 0 and 1, 
and the new set of parameters is accepted if $n\le r$ (setting $\bmath{X}_j=\bmath{X}'_j$), otherwise is rejected (setting $\bmath{X}_j = \bmath{X}_{j-1}$).
In the case of a symmetric proposal distribution, the second factor of equation~(\ref{eq:metrop_ratio}) is 1, therefore the Metropolis ratio reduces to the likelihood ratio for the two set of parameters $\bmath{X}'_j$ and $\bmath{X}_{j-1}$.
To obtain a stationary target distribution, these steps must be repeated for a long enough time, which depends on the way the parameter space is explored. This exploration is set by the adopted proposal distribution, which determines the moves in the parameter space. The proposal distribution has to satisfy the condition that any point in the support of the target distribution can be reached from any other point through a set of proposal moves. The commonly adopted Gaussian proposal distribution fulfils this condition, since it has infinite support, though in the case of pathological posteriors it can fail in the exploration of some regions of the parameter space. The  optimal proposal distribution, which minimizes the number of moves and the computational time, has the same shape of the posterior distribution. Though, in many situations this information has little relevance, since the posterior distribution is not known {\it a priori}, being the goal of the algorithm to compute it. A solution is to use an adaptive proposal distribution, which has the property of learning the shape of the posterior distribution as the chain moves through the parameter space \citep[see][and references therein]{Haario2001,Brooks2011}.\footnote{Chapter 4 of \citet{Brooks2011} is also freely available online at \url{http://citeseerx.ist.psu.edu/viewdoc/download?doi=10.1.1.161.2424&rep=rep1&type=pdf}} 
The package \CosmoMC\ implements an adaptive Gaussian proposal distribution, whose covariance is computed from the posterior samples obtained so far. 
In our work we have no previous knowledge of the posterior probability distribution, hence such an adaptive algorithm needs to (inefficiently) obtain a lot of samples before obtaining a good estimate of the posterior covariance, which is in turn used to set the covariance of the Gaussian proposal distribution. Hence, we use the nested sampling algorithm \citep{Skilling2004,Skilling2006} to rapidly explore the parameter space and sample the unknown posterior distribution, then implementing in \CosmoMC\ the posterior covariance obtained with such an algorithm. Also, this allows us to check the consistency of the posterior distribution obtained with two different algorithms.

\subsection{Convergence diagnostics of the MCMC run}\label{sec:convergence}

The main difficulty of the Metropolis-Hastings algorithm, and in general of random-walk based MCMC, is the determination of the convergence of the chain to the target distribution \citep[see][and references therein]{Kass1997,Cowles1996}. 
A way to accomplish this is to run multiple parallel chains, comparing on-the-fly the samples obtained from the different chains. \CosmoMC\ implements such a convergence algorithm, comparing the between- and within-chain variances from the last half of the samples of each parameter, by means of the \citet{Gelman1992} $\hat{R}$ parameter.
In this work, we run $m=6$ chains, hence after $n$ steps we have, for each parameter, $n \times m$ samples, that we indicate as $X_{ji}$, $j=1,...,m;\; i=1, ...,n$. To compute $\hat{R}$, we consider, for each parameter $X$ separately, the between-chain variance $B$ defined as
\begin{equation*}
B = \frac{n}{m-1}\sum_{j=1}^m \left( \bar{X_j} - \bar{X} \right)^2 \, ,
\end{equation*}
where
\begin{align*}
\begin{split} 
\bar{X_j} = \frac{1}{n}\sum_{i=1}^n X_{ji} 
\end{split} 
\begin{split} 
\bar{X} = \frac{1}{m} \sum_{j=1}^m \bar{X_j} 
\end{split} 
\end{align*}
are the mean of the parameter $X$ within any chain $j$ and across all the chains, respectively.
We then define the within-chains variance $W$ as
\begin{equation*}
W = \frac{1}{m}\sum_{j=1}^m {s_j}^2 \, ,
\end{equation*}
where
\begin{equation*}
{s_j}^2 = \frac{1}{n-1} \sum_{1=1}^n \left ( X_{ji} - \bar{X_j} \right) ^2
\end{equation*}
is the variance of the parameter $X$ in the chain $j$.
Finally, we can compute the convergence parameter $\hat{R}$ as
\begin{equation*}
\hat{R} = \sqrt{ \frac{n-1}{n} + \frac{1}{n}\frac{B}{W} } \, .
\end{equation*}
A value of $\hat{R} < 1.1$ for each parameter of interest usually indicates good convergence \citep[see Section 11.6 of ][]{Gelman2004}. In our case, we set in \CosmoMC\ the converge criterion $\hat{R} \leq 1.01$.

\begin{figure*}
	\centering
	\subfigure
	{\resizebox{.45\hsize}{!}{{\includegraphics{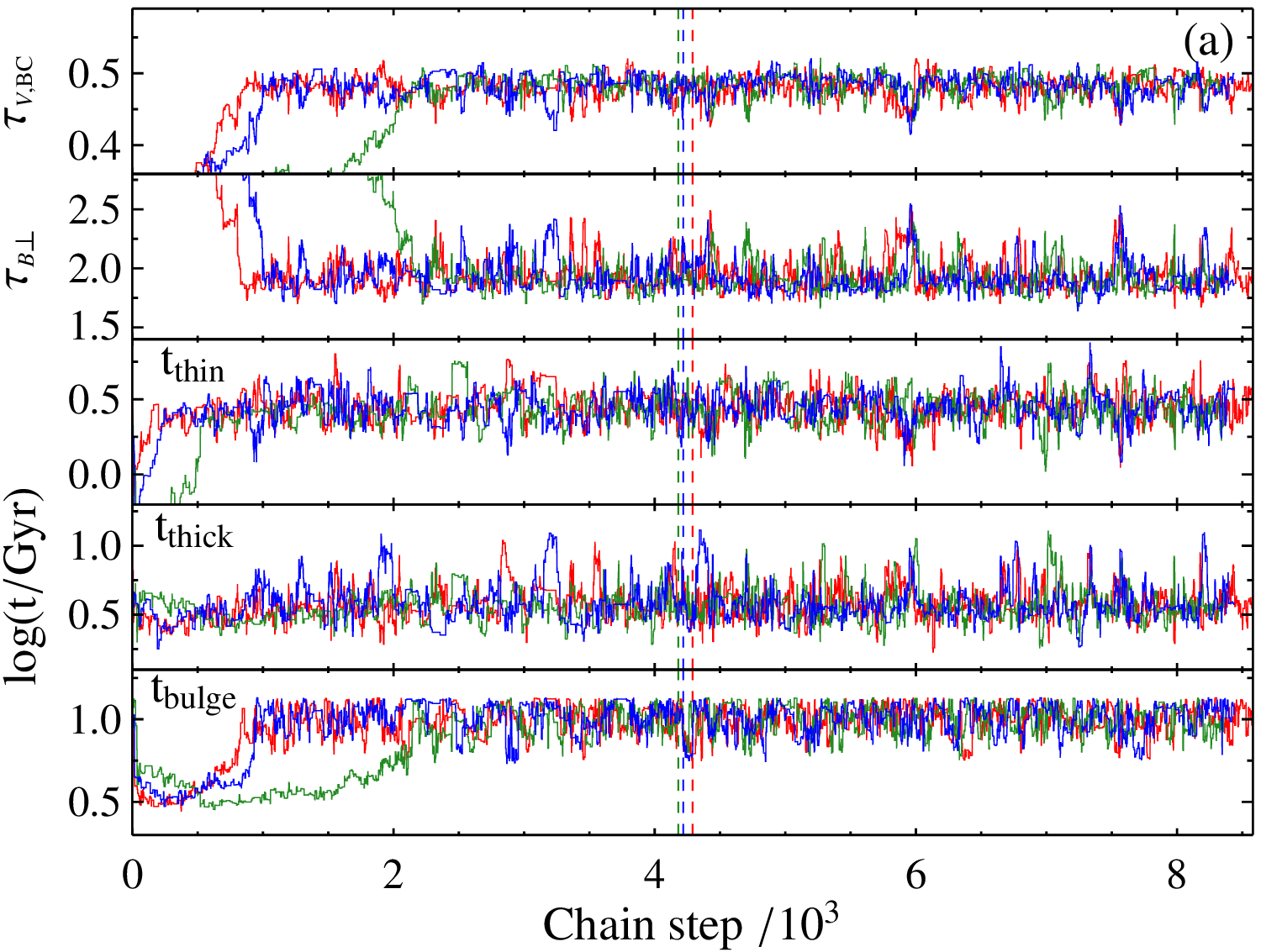}}}}
	\hspace{.05\hsize}
	\subfigure
	{\resizebox{.45\hsize}{!}{{\includegraphics{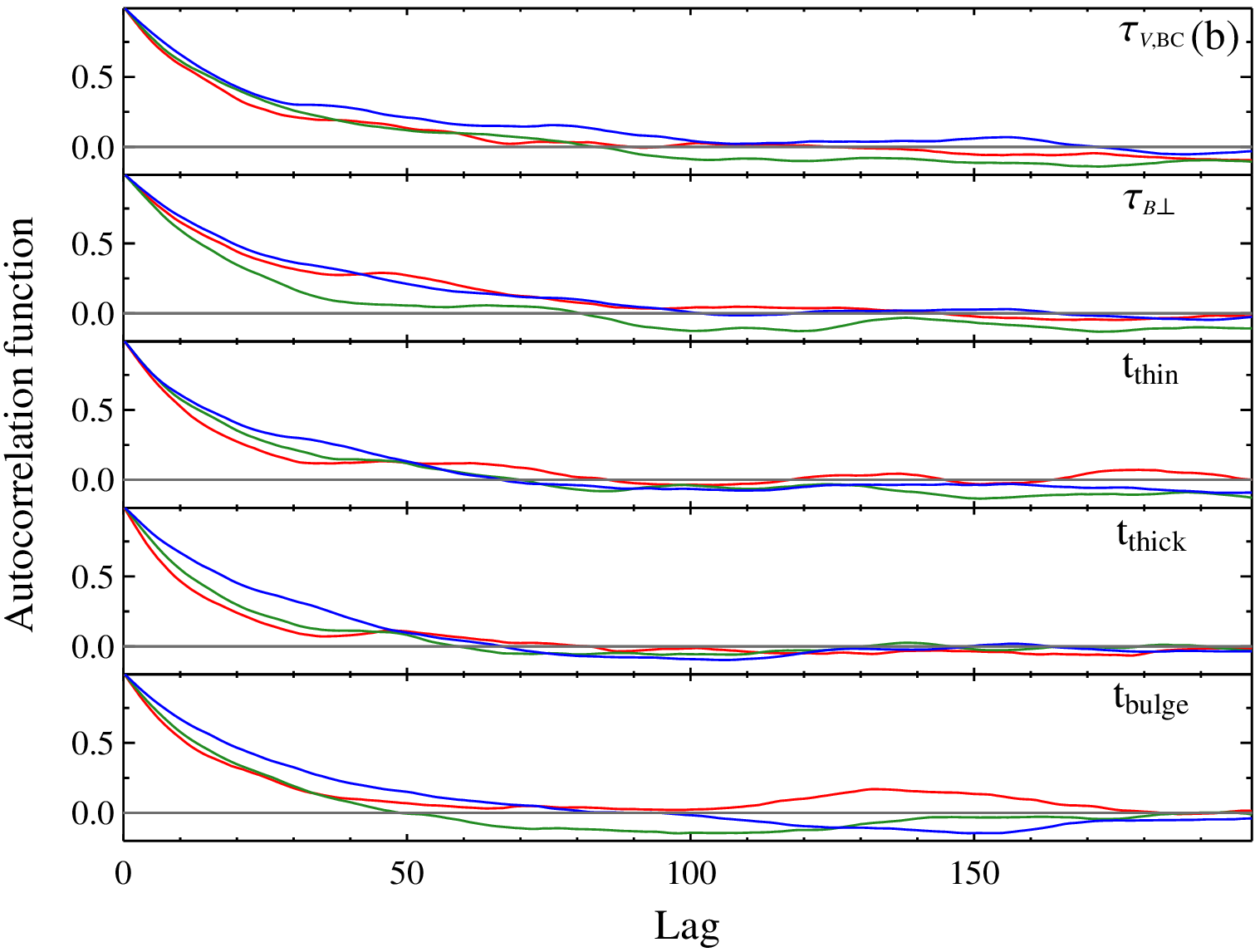}}}}
	\caption{Exploration of the five-dimensional parameter space defined by \tauVbc, \taubP, \Tthin, \Tthick\ and \Tbulge\ by the MCMC described in Section~\ref{sec:bayes_approach}. ({\it a}) Iteration number against value of the draw of each parameter, at each iteration (i.e. `trace-plot') for three randomly selected chains among the six chains run (indicated by the different colours), for the low-\muS\ galaxies. Vertical dashed lines separate the samples discarded (first half of each chain) from those used to compute the posterior probability distribution (second half of each chain). ({\it b}) Auto-correlation function as a function of the lag (i.e. the distance between two iterations) for each parameter, for the last half of the same three randomly selected chains of ({\it a}) (indicated by the different colours).}
	\label{fig:conv_low}
\end{figure*}

\begin{figure*}
	\centering
	\subfigure
	{\resizebox{.45\hsize}{!}{{\includegraphics{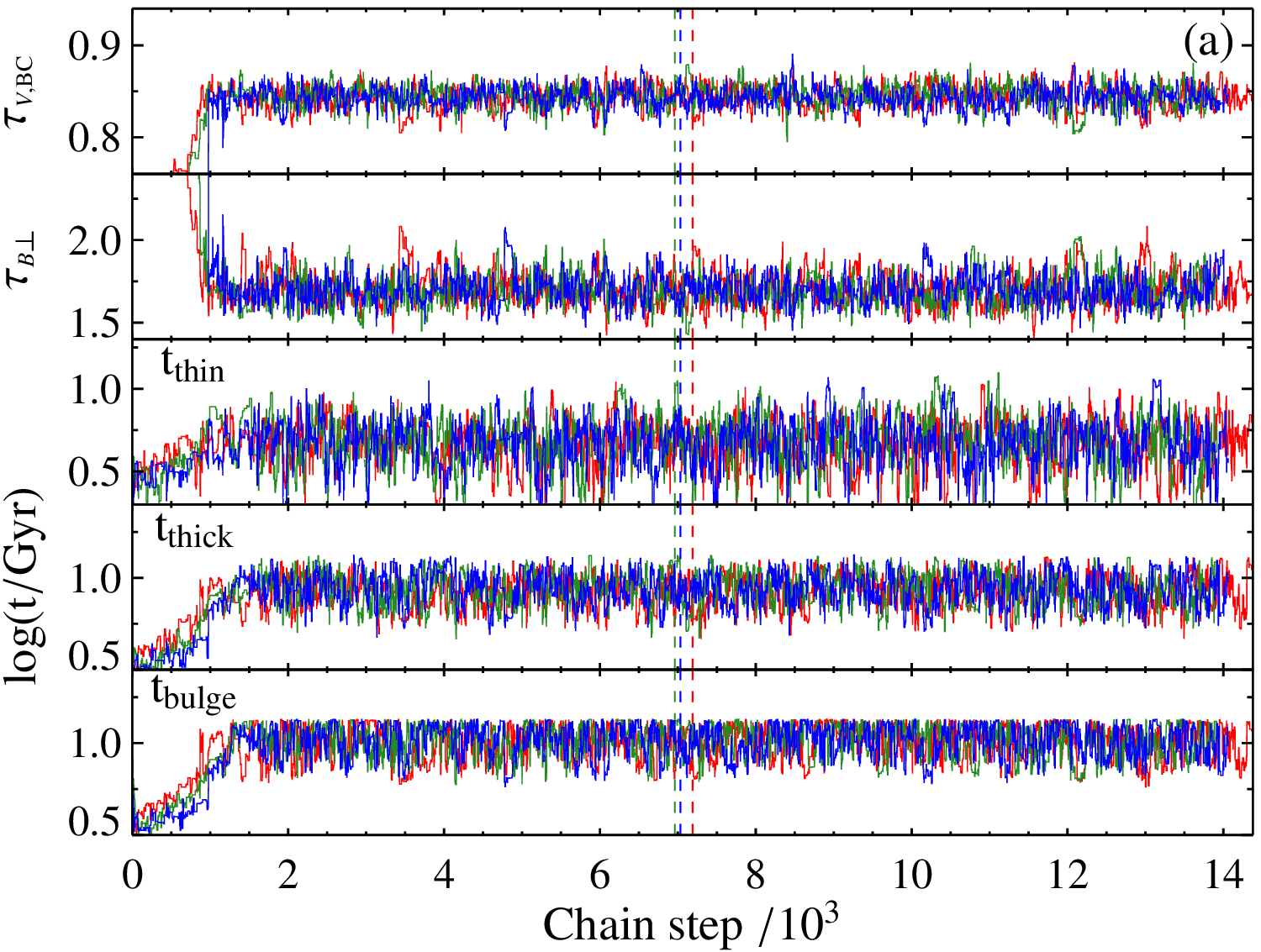}}}}
	\hspace{.05\hsize}
	\subfigure
	{\resizebox{.45\hsize}{!}{{\includegraphics{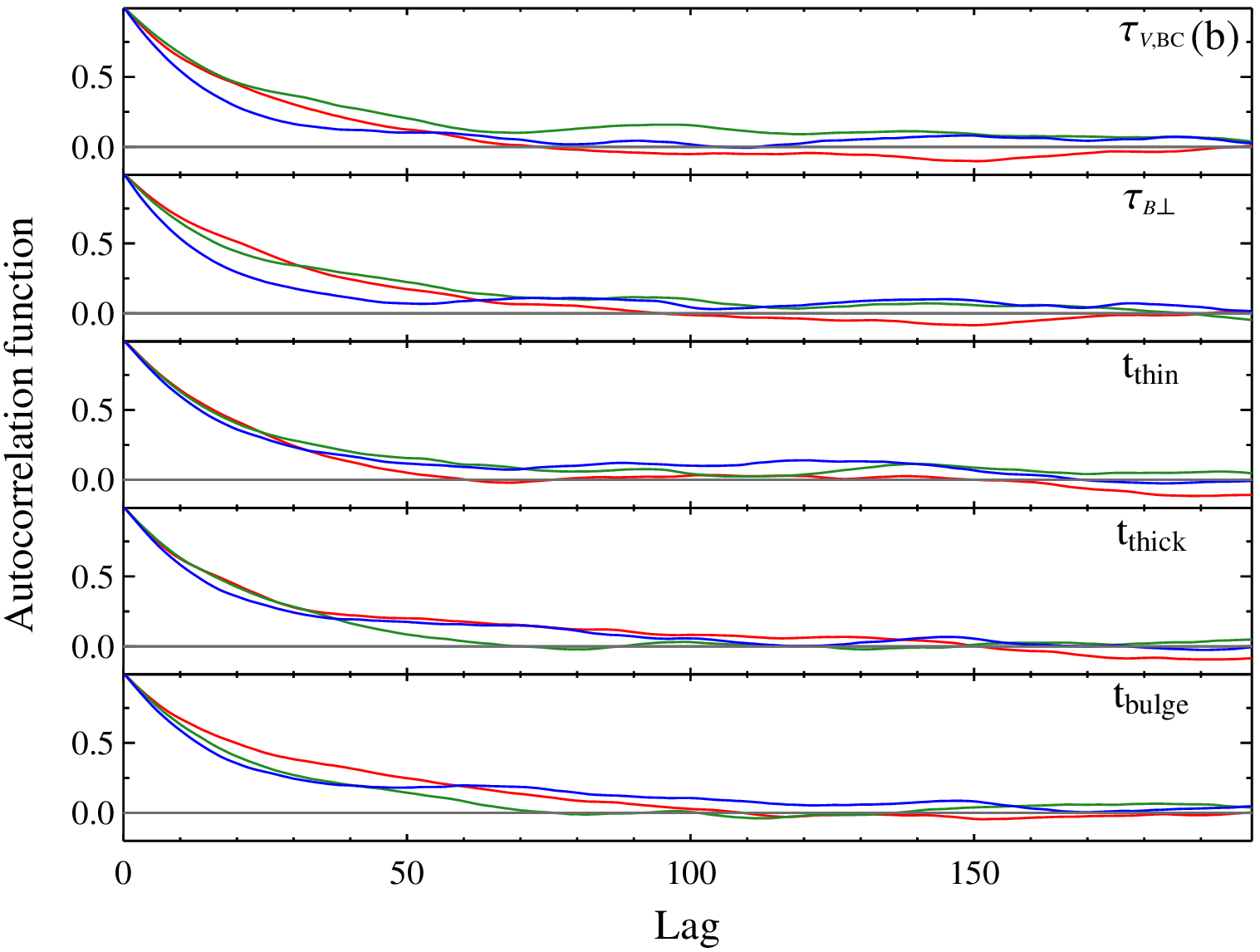}}}}
	\caption{Exploration of the five-dimensional parameter space defined by \tauVbc, \taubP, \Tthin, \Tthick\ and \Tbulge\ by the MCMC described in Section~\ref{sec:bayes_approach}. ({\it a}) Iteration number against value of the draw of each parameter, at each iteration (i.e. `trace-plot') for three randomly selected chains among the six chains run (indicated by the different colours), for the high-\muS\ galaxies. Vertical dashed lines separate the samples discarded (first half of each chain) from those used to compute the posterior probability distribution (second half of each chain).  ({\it b}) Auto-correlation function as a function of the lag (i.e. the distance between two iterations) for each parameter, for the last half of the same three randomly selected chains of ({\it a}) (indicated by the different colours).}
	\label{fig:conv_high}
\end{figure*}

To further check the exploration of the parameter space, we plot in Figs. \ref{fig:conv_low}a--\ref{fig:conv_high}a the value of the draw of each parameter at each iteration against the iteration number (i.e. the so called `trace-plot') for three randomly selected chains among the six chains run. Figs. \ref{fig:conv_low}a--\ref{fig:conv_high}a show that both for the low- and high-\muS\ data sets the chains move from the starting points to the region of the parameter space of highest probability, without showing any large-scale trend which could suggest a bad exploration of the parameter space. This also indicates that discarding the first half of any chain, as suggested in \CosmoMC , is a conservative choice which allows us to keep only draws that are likely sampling the target distribution.

Finally, we plot in Figs. \ref{fig:conv_low}b--\ref{fig:conv_high}b the value of the auto-correlation function between any pair of draws as a function of the separation between the draws (i.e. the `lag'), for each parameter and for the same three randomly selected chains shown in Figs. \ref{fig:conv_low}a--\ref{fig:conv_high}a. The auto-correlation quantifies the amount of correlation between subsequent draws, and starts from 1 at lag $l=1$, since each move depends on the previous one in a deterministic way through the proposal distribution. In the case of a good exploration of the parameter space, the auto-correlation function drops to $\approx 0$ after a lag $l^\prime$, meaning that samples drawn at iterations $i$ and $i+l^\prime$ are independent. Figs. \ref{fig:conv_low}b--\ref{fig:conv_high}b show that the auto-correlation function  drops and starts oscillating around zero at $l^\prime \approx 70$ for both the low- and high-\muS\ data sets, as expected in the case of good exploration of the parameter space. 

These tests indicate that the chains are well exploring the parameter space, and that the drawn samples have likely converged to the stationary target distribution, which in this case is the posterior probability distribution of the five parameters \tauVbc ,\taubP ,\Tthin, \Tthick\ and \Tbulge .


\section{Nested sampling}\label{app:nested}

As discussed in Appendix \ref{app:MCMC}, the Metropolis-Hastings algorithm with Gaussian proposal distribution has two major limitations: the tuning of the proposal distribution, which is crucial to guarantee a good exploration of the parameter space, and the assessment of the convergence of the chain, which is non-trivial in the general case of multi-modal distributions. These two caveats become even more serious during the very first exploration of a model against a set of data, since the posterior distribution is unknown and the proposal distribution can not be tuned in an optimum way. This usually makes the first runs of a MCMC computationally demanding, since many likelihood evaluations are needed before one can start tuning the proposal distributions using the samples drawn from the posterior. Moreover, a single MCMC can fail in the exploration of multi-modal posterior distributions with well separated peaks, so multiple chains and/or annealing (or tempering) schemes are needed to ensure the exploration of all modes. Another limitation arises from the nature of Markov Chains (i.e. the markovian property), which makes each step in the chain dependent on the previous one, making impossible to parallelise single MCMC. 
Motivated by the need of a fast and reliable algorithm to compute the Bayesian evidence, \citet{Skilling2004,Skilling2006} proposed a novel Monte Carlo approach, which allows the computation of the multi-dimensional evidence integral in Bayes's theorem by transforming it into a one-dimensional integral. The algorithm, called nested sampling, explores the parameter space in a stochastic way by means of set of `active points', shrinking exponentially at each iteration the prior volume from which the samples are drawn around the regions of highest probability. Besides the evidence, the drawn samples allow, as a by-product, the computation of the posterior probability distribution. In practice, this is accomplished by computing, at each iteration, the likelihood on $N$ points uniformly drawn from the prior distribution $\pi(\bmath{\Theta})$, then selecting the point with the lowest likelihood $\mathcal{L}_{min}$ and, at the next iteration, drawing points from $\pi(\bmath{\Theta})$ conditioned to have $\mathcal{L}(\bmath{\Theta})>\mathcal{L}_{min}$. The technical difficulty is to efficiently draw points from the conditional prior distribution $\pi(\bmath{\Theta}|\mathcal{L}(\bmath{\Theta})> \mathcal {L}_{min})$, since the volume occupied by parameters for which $\mathcal{L}(\bmath{\Theta})>\mathcal{L}_{min}$ shrinks also exponentially with the iterations. The \MultiNest\ algorithm faces this problem by appealing to the 'simultaneous ellipsoidal nested sampling method' of \citet{Feroz2008}. This allows the approximation of contours with constant likelihood with ellipsoids, whose optimal number and bounds are determined by an expectation-maximisation algorithm. This allows an efficient sampling from the conditional density $\pi(\bmath{\Theta}|\mathcal{L}(\bmath{\Theta})> \mathcal{L}_{min})$, hence solving the major technical obstacle of the nested sampling algorithm.
Also, \MultiNest\ exploits the possibility of parallelisation of the nested sampling algorithm, strongly reducing the wall-clock time required for inference on a mulit-core computer.

In this work, following authors recommendations given in the manual of the package, we run \MultiNest\ with $N=500$ active points, setting the tolerance of the evidence, which is the stopping criterion, to 0.5, and the sampling efficiency, which determines the balance among the accuracy of the posterior and the length of the run, to 0.7.
The algorithm converges after about $\numprint{15000}$ likelihood evaluations. We then compute the covariance among of all pairs of parameters, and include it in the \CosmoMC\ run.
As a further consistency check, we compare the marginal and joint posterior probability distributions obtained with \MultiNest\ and \CosmoMC\, finding an excellent agreement among the two.

\label{lastpage}

\end{document}